\documentclass[pre,showpacs,twocolumn,preprintnumbers,amsmath,amssymb]{revtex4}

\usepackage{graphicx}
\usepackage{dcolumn}
\usepackage{bm}
\usepackage{parskip}
\usepackage{subfigure}
\usepackage{amssymb}
\usepackage{amsmath}
\usepackage{amssymb}
\usepackage{graphicx}
\newcommand{\bare}[1]{\mathaccent"7017{#1}}
\def\be{\begin{eqnarray}}
\def\ee{\end{eqnarray}}
\def\be{\begin{equation}}
\def\ee{\end{equation}}
\begin{document}
\title{Crossover from low-temperature to high-temperature fluctuations. \\ II.  Nonuniversal thermodynamic Casimir forces of anisotropic systems}

\author{Volker Dohm}

\affiliation{Institute for Theoretical Physics, RWTH Aachen
University, D-52056 Aachen, Germany}

\date {21 August 2017}

\begin{abstract}
The finite-size renormalization-group approach for isotropic O$(n)$-symmetric systems introduced previously [V. Dohm, Phys. Rev. Lett. {\bf 110}, 107207 (2013)] is extended to weakly anisotropic O$(n)$-symmetric systems. Our theory is formulated within the  $\varphi^4$ model with lattice anisotropy in a $d$-dimensional block geometry with periodic boundary conditions. It describes the crossover from low-temperature to high-temperature fluctuations including Goldstone-dominated and critical fluctuations for $1\leq n \leq \infty$ in $2<d<4$ dimensions. An exact representation is derived for the large-distance behavior of the bulk correlation function of anisotropic systems in terms of the principal correlation lengths and an anisotropy matrix  ${\bf \bar A}$. This includes the long-ranged correlations with an anisotropic algebraic decay at low temperatures due to the Goldstone modes for $n>1$. We calculate the finite-size scaling functions of the excess free energy and thermodynamic Casimir force. Exact results are derived in the large-$n$ limit.  Applications are given for $L_\parallel^{d-1} \times L$ slab geometries with a finite aspect ratio $\rho=L/L_\parallel$ as well as for the film limit $\rho \to 0$. For
weakly anisotropic systems two-scale-factor universality is replaced by multiparameter universality since the scaling functions depend on $d(d+1)/2-1$ nonuniversal anisotropy parameters, in addition to two nonuniversal thermodynamic length scales. The latter depend also on the anisotropy parameters.
This implies a substantial reduction of the predictive power of bulk and finite-size theory for anisotropic systems as compared to isotropic systems.
The validity of multiparameter universality is confirmed analytically for a nontrivial example of the $d=2,n=1$ universality class. Anisotropy-dependent minima of the Casimir force scaling function are found below $T_c$ for $\rho \ll 1$. Both the sign and magnitude of the Casimir amplitude in the Goldstone and critical regimes are affected by the lattice anisotropy. Also a nonuniversal  shift of the film critical temperature is shown to be caused by anisotropy in the large-$n$ limit. Quantitative predictions are made that can be tested by Monte Carlo simulations for $\varphi^4$ models and for spin models in the Ising, $XY$, and Heisenberg universality classes.
\end{abstract}
\maketitle

\newpage
\renewcommand{\thesection}{\Roman{section}}
\renewcommand{\theequation}{1.\arabic{equation}}
\setcounter{equation}{0}
\section{ Introduction and Summary}
In the preceding paper \cite{dohm2017I} we have developed an analytic theory of the crossover from low-temperature to high-temperature fluctuations including critical fluctuations in confined O$(n)$-symmetric systems in $2<d<4$ dimensions
with periodic boundary conditions (BC) and short-range interactions. A brief account was given in \cite{dohm2013}.
The cornerstone of this theory is the simultaneous description of two different types of long-range fluctuations: critical fluctuations of the $(d,n)$ universality class at a finite critical temperature $T_c$ \cite{fish-1} and classical fluctuations due to massless Goldstone modes \cite{goldstone,wagner} at low temperatures. Both types of fluctuations exist in $O(n)$-symmetric systems undergoing a second-order phase transition which opens up the opportunity of studying the crossover between the two different types of fluctuation-induced forces  \cite{kardar}: the critical Casimir force $F_{\text Cas,c}$ \cite{fisher78,krech} and the Casimir force $F_{\text Cas,G}$ induced by Goldstone fluctuations \cite{zandi2004}. The theory was formulated within the O$(n)$-symmetric $\varphi^4$ model but owing to universality the results are applicable to a large class of O$(n)$-symmetric systems.  A restriction of this theory \cite{dohm2017I,dohm2013} was the assumption of spatial isotropy which is a characteristic of fluids and superfluids.

In this paper we extend the theory to O$(n)$-symmetric systems with weak spatial anisotropy which represent an important class of systems with cooperative phenomena such as superconductors and magnetic materials. These phenomena can be described by spin models (e.g., Ising, $XY$, or Heisenberg models) or by lattice and continuum $\varphi^4$ theories. The advantage of the latter is the analytic tractability of lattice anisotropy for all $n$ in terms of a $d \times d$ symmetric anisotropy matrix ${\bf A}$ \cite{cd2004,dohm2006,chen-zhang,dohm2008,dohm2009PJ,kastening-dohm} whereas
anisotropic spin models are amenable to an analytic treatment only in special cases \cite{hout,Wu1966,WuCoy,CoyWu,Vaidya1976,Indekeu,Berker,kastening2012,pri,car-1,night1983,Abra,Yurishchev,shenoy1995},
primarily for $d=2$ Ising models.
 Weakly anisotropic systems with a critical point constitute a
subclass of systems that belong to the same $(d,n)$ bulk universality class as isotropic systems. Such anisotropic systems have been shown \cite{Indekeu,cd2004,dohm2006,chen-zhang,dohm2008,dohm2009PJ,kastening-dohm,DG,bruce}
to have finite-size scaling functions and  critical bulk amplitude relations and correlation functions that differ from those of isotropic systems in that they depend on the matrix elements of ${\bf A}$.  The latter represent parameters that depend on system-dependent properties such as the lattice structure, coupling constants, and correlation lengths of anisotropic systems. According to the usual terminology
\cite{pri,priv,pelissetto,dohm2005}, a quantity is {\it nonuniversal} if it depends on such system-dependent parameters.

Due to spatial anisotropy there exists no unique bulk correlation length but rather $d$ different correlation lengths
in the directions of the $d$ principal axes which, within the $\varphi^4$ theory, are determined by the nonuniversal eigenvectors of ${\bf A}$. Weakly anisotropic systems near criticality still have a single correlation-length exponent $\nu$ which is the same as for isotropic systems. We do not consider {\it strongly} anisotropic systems (see, e.g., \cite{tonchev,diehl-2,diehl2010}) with critical exponents different from those of the usual $(d, n)$ universality classes, and we do not study phenomena that may arise from the interplay between spatial and spin anisotropy \cite{lin}.

For anisotropic confined systems, nonuniversality manifests itself by the fact that the principal correlation lengths and  principal directions are unrelated to the orientation of the surfaces of the confining geometry.  This is a system-dependent physical source of nonuniversality that one cannot get rid of by a formal transformation to an isotropic system (Sec. II). An additional source of nonuniversality may enter through spatial anisotropies at the confining surfaces. So far the potential complexity of nonuniversal finite-size effects due to the interplay between confinement, surface orientation, and weak anisotropy has remained  unexplored in the literature on the Casimir force  \cite{krech}, on finite-size theory  \cite{brankov}, and on boundary critical phenomena  \cite{diehl1997}. In particular no theory describing the crossover from weak to strong anisotropy has been developed.

In a confined system with a characteristic size $L$, the fundamental quantity from which the Casimir force per unit area $F_{{\text Cas}}=-\partial [Lf^{{\text ex}}]/\partial L$
can be derived is the excess free energy density (divided by $k_BT$) $f^{{\text ex}}=f-f_b$ where $f$ and  $f_b$ are the free energy densities of the confined and the bulk system, respectively. In the early discussion on the universality properties in finite systems near criticality \cite{pri} the picture of "two-scale-factor universality" was put forward. As an extension of two-scale-factor universality for bulk systems \cite{aharony1974,ger-1,hohenberg1976,weg-1}, it was hypothesized that the singular part $f_s$  of $f$ can be described (for large $L$, small $t=(T-T_c)/T_c$ and small ordering field ${\bf h}= h\;{\bf e}_h$ with a unit vector ${\bf e}_h$) by
\begin{equation}
\label{1b}
f_s (t, h, L) = L^{-d} \;  F (C_1 t L^{1/\nu},
C_2 h L^{\beta \delta/\nu})
\end{equation}
with universal critical exponents $\nu, \beta, \delta$ and the
universal scaling function $F(x,  y)$, where the two constants $C_1$ and
$C_2$  are universally related to the bulk constants $A_1$ and $A_2$ of  the singular bulk part
\begin{equation}
\label{1ax} f_{b,s} (t, h)= \lim_{L\to \infty} f_s (t,  h, L)= A_1 |t|^{d \nu} \; W_\pm (A_2 h|t|^{-\beta \delta})
\end{equation}
with the universal scaling function $W_\pm (z)$ above $(+)$ and below $(-)$
$T_c$. The definition was made quite precise in that it was stated
by Privman and Fisher \cite{pri} that the metric factors $C_l$ and $C_2$ are the {\it only} nonuniversal, system-dependent parameters entering (\ref{1b}), and that no further nonuniversal prefactor $C_0$ is required. It was asserted that, for given geometry and BC,  $F(x, y)$ is "the same for all systems in a given universality class" \cite{priv}. Subsequently the same universality properties have been attributed to the critical Casimir force \cite{Indekeu1984,Indekeu,KrDi92a,krech,brankov}.

It is the hallmark of two-scale-factor universality that the critical Casimir force $F_{\text Cas,c}$ at $t=0,h= 0$  for large $L$ is predicted to have the simple size dependence
\be
\label{casimirTc}
F_{\text Cas,c}= X_c\; L^{-d}
\ee
without any nonuniversal factor, i.e., where the Casimir amplitude  $X_c(d,n)$ is a universal number.
It was noted \cite{pri,priv,aharony}, however,  that lattice anisotropy is a marginal perturbation in the renormalization-group (RG) sense, thus it was not obvious a priori to what extent two-scale-factor universality is valid in the presence of anisotropic couplings \cite{priv}. Correspondingly, no proof was given in the literature for the validity of the universal form of (\ref{casimirTc}) for weakly anisotropic systems.

As far as the experimental observability of weak anisotropy is concerned, it was claimed \cite{toldin,diehl2009,diehl2010} that critical Casimir forces are active only in isotropic fluids where the ordering degrees of freedom can move in and out of the system. This is in disagreement with the arguments presented in \cite{wil-1,comment}, that a Casimir force, in the form of an electrical potential difference, appears at the junction between an anisotropic high-$T_c$ superconducting film and the bulk superconductor due to the transfer of Cooper pairs from the film to the bulk. Quantitative estimates indicated that this effect should be directly measurable. Furthermore, it was argued in \cite{wil-1} that the Casimir force in superconductors should be similar to that observed in isotropic superfluid $^4$He \cite{garcia}. This theory based on vortex-loop fluctuations \cite{wil-1} did not succeed, however, in describing a finite Casimir force
\be
\label{casimirGold}
F_{\text Cas,G}= X_0 L^{-d}
\ee
in the Goldstone-dominated low temperature region, unlike the experimental \cite{garcia}, theoretical \cite{zandi2004,biswas2010}, and numerical \cite{hucht,vasilyev2009,hasenbusch2010} findings of $F_{\text Cas,G}$ well below the superfluid transition of $^4$He and of $XY$ and  $\varphi^4$  models. Essentially the same result (\ref{casimirTc}) for $F_{\text Cas,c}$ was predicted \cite{wil-1} at the superconducting phase transition as for the $\lambda$ transition of $^4$He, without an effect of the lattice anisotropy of the superconducting system.
An earlier theory for an anisotropic XY model was invoked \cite{wil-1} where lattice anisotropy was claimed to be "irrelevant" \cite{shenoy1995}.

By contrast, it was found in analytical calculations \cite{Indekeu,cd2004,dohm2006,dohm2008,chen-zhang,kastening-dohm,DG}
and Monte Carlo (MC) simulations \cite{selke2005,selke2006,selke2009} that, although critical exponents are unchanged by weak anisotropy, finite-size effects in the subclass of weakly anisotropic systems must be distinguished from those in the subclass of isotropic systems within the same universality class. The crucial distinguishing feature is the absence of two-scale-factor universality \cite{cd2004,dohm2006,dohm2008,dohm2009PJ,chen-zhang,kastening-dohm,DG}
in the subclass of weakly anisotropic systems. It was shown for several geometries and boundary conditions that the nonuniversal anisotropy matrix ${\bf A}$ affects the finite-size scaling properties of $f_s$, of $F_{\text Cas,c}$, of the Binder cumulant, of the susceptibility, of the correlation length, and of the helicity modulus. In particular, as an  exact result in the large-$n$ limit, it was found  for film geometry with periodic BC  \cite{cd2004} that (\ref{casimirTc}) is to be replaced by
\be
\label{casimirTcaniso}
F_{\text Cas,c}({\bf \bar A})= X_c({\bf \bar A}) L^{-d}
\ee
with an nonuniversal amplitude $X_c({\bf \bar A})$  that depends on the reduced anisotropy  matrix ${\bf \bar A}= {\bf A}/(\det{\bf  A})^{1/d}$. A nonuniversal amplitude $X_c(\xi_\perp/\xi_\parallel)$ was also found in the exactly solvable mean-spherical \cite{DG} and Gaussian \cite{kastening-dohm} models in film geometry for several BC where $X_c$ depends on the ratio of the bulk correlation lengths $\xi_\perp$ and $\xi_\parallel$  perpendicular and parallel to the boundaries.

A serious lack of knowledge remained, however, with regard to a theory of finite-size effects in anisotropic $O(n)$ symmetric systems with {\it finite} $1<n< \infty$ that covers the region {\it below} $T_c$  including the crossover to the Goldstone regime.
So far this was achieved in \cite{dohm2013} only for isotropic systems, as derived in detail in the preceding paper \cite{dohm2017I} in terms of a crossover scaling function for $f_s$.

The goal of this paper is to study the effect of anisotropy
in both the critical and low-temperature regions including the Goldstone regime for $n> 1$ as well as on the crossover from low to high temperatures for general $1 \leq n \leq \infty$ in block, slab and film geometries with periodic BC.
The predictions of our theory can be tested by MC simulations for $n$-component $\varphi^4$ lattice models \cite{hasenbusch2010,Hasenbuschgesamt} or Ising ($n=1$), $XY$ ($n=2$), and Heisenberg ($n=3$) models. The concept of our theory should be applicable also to the case of Dirichlet BC \cite{dohm2014} which are relevant to Casimir forces in anisotropic superconductors \cite{wil-1,schneider2004} and to finite-size effects in magnetic materials.

A judgement about the validity or violation of two-scale-factor universality in bulk and confined systems requires the combined analysis of {\it both} the free energy {\it and} the bulk correlation function \cite{hohenberg1976,pri,priv}.
Thus, before developing finite-size theory in Sec. III
we lay the groundwork of anisotropic {\it bulk} theory near $T_c$ in Sec. II where a prediction is derived for the scaling form of the bulk correlation function $G_b$ of weakly anisotropic systems satisfying {\it multiparameter universality} {\cite{dohm2008,dohm2009PJ,kastening-dohm}. At $h=0$ our exact asymptotic (large ${\bf x}$, large ${\bar \xi_\pm(t)}$, finite $|{\bf x}|/{\bar \xi_\pm}\geq 0 $) result is for general $n$ above $T_c$ $(+)$ and for $n=1$ below $T_c$ $(-)$
\begin{eqnarray}
\label{3nbarAhNull} G_b ({\bf x}, t) = \frac{D_1}{ [{\bf x}\cdot ({\bf \bar A}^{-1}{\bf x})]^{( d -2 + \eta)/2}}\;
 \Phi_\pm \Big(\frac{[{\bf x}\cdot ({\bf \bar A}^{-1}{\bf x})]^{1/2}}{\bar \xi_\pm(t)}\Big),\;\;\;
\end{eqnarray}
 and the asymptotic result for the transverse correlation function $G_{b,{\rm T}}$ for general $n>1$ below $T_c$ is
\begin{eqnarray}
\label{xtransBx}
G_{b,{\rm T}} ({\bf  x}, t) &=&
{\cal B}_{\rm T}\;|t|^{2\beta}\Big(\frac{\bar \xi_{\rm T}(t)}{ [{\bf x}\cdot ({\bf \bar A}^{-1}{\bf x})]^{1/2} }\Big)^{d-2}.\;\;\;\;\;\;
\end{eqnarray}
Here $\eta$ and $\beta$ are the critical exponents of the isotropic system, $\bar \xi_\pm(t)$ and $\bar \xi_{\rm T}(t)$
are the geometrical mean of the corresponding principal correlation lengths,
$\Phi_\pm$ is the universal scaling function of the isotropic theory \cite{dohm2017I}, and the dimensionless reduced anisotropy matrix ${\bf \bar A}$ is expressed in terms of ratios of principal correlation lengths, as defined in Sec. II. This matrix depends on $d(d+1)/2-1$ anisotropy parameters. The nonuniversal constants $D_1$ and ${\cal B}_{\rm T}$ are universally related to the amplitudes $A_1$ and $A_2$ of the bulk free energy (\ref{1ax}) as determined by the relations (\ref{3n3}) and (\ref{3n4}). The universal constants appearing in these relations are the same as those for isotropic systems in the same universality class. Although the results (\ref{3nbarAhNull}) and (\ref{xtransBx}) are derived within the $\varphi^4$ lattice model, our hypothesis of multiparameter universality predicts that (\ref{3nbarAhNull}) and (\ref{xtransBx}) are valid for all systems in the same universality class including, e. g., fixed-length spin models. Thus $G_b$ and $G_{b,{\rm T}}$ have a universal structure but depend, in general, on $d(d+1)/2+1$ independent nonuniversal parameters, thus violating two-scale-factor universality defined in \cite{hohenberg1976,priv,pri}. In particular $G_b$ and $G_{b,{\rm T}}$ exhibit a nonuniversal directional dependence in space, in contrast to isotropic systems. The validity of multiparameter  universality is confirmed analytically in Sec. II for a nontrivial example of the $d=2,n=1$ universality class.

For {\it confined} anisotropic systems, we briefly summarize some aspects of our approach of this paper and present the structure of the ensuing scaling function of the Casimir force. We consider the  $\varphi^4$ model with lattice anisotropy in a $d$-dimensional rectangular $ L_1 \times L_2 \cdot \cdot \cdot \times L_d$ block geometry with finite aspect ratios
\begin{eqnarray}
\label{aspect}
\rho_\alpha=L/L_\alpha, \;\;\; \alpha= 1, 2, ..., d-1,\;\; \rho_d\equiv1,\;\;\;\;\;\;
\end{eqnarray}
where we have chosen $L\equiv L_d$ as a reference length. A shear transformation is performed such that the interaction becomes isotropic (in the long-wavelength limit) which implies that the block geometry is transformed, in general, into a non-rectangular parallelepiped. The isotropic form of the transformed interaction permits one to perform the field-theoretic renormalizations which are characteristic for the bulk $(d,n)$ universality class but does not eliminate the nonuniversal anisotropy matrix  ${\bf A}$ which now enters the shape of the parallelepiped and the skew orientation of the transformed nonrectangular lattice, i.e., the transformed system has a nonuniversal shape, nonuniversal wave-vectors ${\bf k}'$ describing the transformed lattice structure, and nonuniversal BC. Thus the shear transformation restores isotropy without restoring universality. An important ingredient of our theory is then the separation of the lowest mode of the order-parameter fluctuations and an approximation with regard to the higher modes such that a simultaneous treatment of the critical and the Goldstone modes in a finite geometry is achieved. A crucial advantage is provided by the minimal renormalization approach at fixed dimension \cite{dohm1985,dohm2008} whose renormalization constants are the same above and below $T_c$. This makes possible to derive in Secs. III and IV a single finite-size crossover scaling function between the low- and high-temperature regions including the critical region. We find that, for anisotropic systems in a block geometry, (\ref{1b}) is replaced for $2<d<4$ and general $n$ by
\begin{eqnarray}
\label{1c} f_s (t, h, \{L_\alpha\}, {\bf A}) \; = \; L^{-d} \;
F (\tilde x, \; \tilde x_h,\{\rho_\alpha\}, {\bf \bar A}),\\
%
\label{xhtilde}
\tilde x=t\;(L/\bar \xi_{0+})^{1/\nu}\;,
\;\tilde x_h= h\;( L/\bar \xi_c)^{\beta\delta/\nu} ,
\end{eqnarray}
where $\bar \xi_{0+}$ and $\bar \xi_c$ are the geometrical mean of the amplitudes of the principal correlation lengths for $T> T_c$, $h= 0$ and for $T=T_c$, $ h\neq 0$ of the anisotropic system as defined in (\ref{ximean}) and (\ref{meancorampcx}), respectively. For cubic geometry $(\rho_\alpha=1)$,
(\ref{1c}) agrees with Eq. (1.3) of \cite{dohm2008} through the exact relations (\ref{tildexx}) and (\ref{hprime}) for the scaling variables $\tilde x$ and $ \tilde x_h$. The matrix $\bar {\bf A}$ in (\ref{1c}) is the same as in (\ref{3nbarAhNull}) and (\ref{xtransBx}).
Two-scale factor universality is replaced by multiparameter universality,
with a nonuniversal critical amplitude $F (0, 0,\{\rho_\alpha\}, {\bf \bar A})$.
For the  explicit results of $F (\tilde x, \;0,\{\rho_\alpha\}, {\bf \bar A})$ at $h= 0$ see (\ref{scalfreeaniso}) for finite $n$ and (\ref{3.9xx1})-(\ref{corrinf})
in the limit $n\to \infty$. They exhibit nonuniversal finite-size effects caused by the long-ranged anisotropic correlations  at low temperatures due to the Goldstone modes for $n>1$ and by the anisotropic critical correlations near $T_c$ for $n\geq 1$.

The same structure is obtained for the Casimir force of the anisotropic system at $ h= 0$
\begin{equation}
\label{Fcas}
F_{\text Cas} (t, \{L_\alpha\},{\bf  A})  \; = \; L^{-d} \;X(\tilde x,\{\rho_\alpha\}, {\bf \bar A}).
\end{equation}
The scaling function $X$ includes a description of the crossover from the critical ($t=0$) Casimir amplitude
\begin{equation}
\label{Xc}
X_c({\bf \bar A})=X(0,\{\rho_\alpha\}, {\bf \bar A})
\end{equation}
to the low-temperature $(t\to -\infty)$ Casimir amplitude
\begin{equation}
\label{XNULL}
X_0({\bf \bar A})= X(-\infty,\{\rho_\alpha\}, {\bf \bar A})
\end{equation}
for $n>1$. The dependence on ${\bf \bar A}$ is, in general, quite complicated [see (\ref{Kd})-(\ref{calJ3x})]. It does not appear, e.g., merely in the simple form of a prefactor that could be absorbed in the factor $L^{-d}$. Even in the case of $\infty^{d-1} \times L$ film or strip geometries [see item (c) below] considered earlier  \cite{Indekeu,cd2004,DG,kastening-dohm,kastening2012}, anisotropy changes the prefactor and the scaling argument. As shown in Sec. IV, both the magnitude and sign of $X_c({\bf \bar A})$ and $X_0({\bf \bar A})$ are affected by the anisotropy matrix ${\bf \bar A}$, and both amplitudes $X_c({\bf \bar A})$ and $X_0({\bf \bar A})$ differ significantly from their counterparts  $X_c({\bf 1})$ and $X_0({\bf 1})$ of isotropic systems in the same universality class and with the same geometry and BC. This demonstrates that the picture of two-scale-factor universality for finite systems described in \cite{pri,priv} is oversimplified and that the corresponding claims in the literature with regard to the universality of the Casimir force  scaling function (see, e.g., \cite{krech,vasilyev2009,GrDi07,hucht2011,toldin2013,diehl2014}) are not correct for the subclass of weakly anisotropic systems.

More specifically, since all critical exponents and thermodynamic bulk scaling functions are the same in isotropic and weakly anisotropic systems of the same universality class \cite{dohm2008}, one can immediately predict these bulk quantities for, e.g., anisotropic Ising-like  $(n=1)$ magnets on the basis of the knowledge of these quantities for ordinary $(n=1)$ fluids. This is not the case, however, for their bulk correlation functions and finite-size scaling functions (even if the two different systems have the same shape and the same boundary conditions). Since $\bar {\bf A}$ has $d(d+1)/2-1$ independent matrix elements, a quantitative analysis of finite-size effects in anisotropic systems requires, in general, significantly more nonuniversal information than in isotropic systems, namely up to five additional nonuniversal parameters in three dimensions.  An identification of $\bar {\bf A}$ for real anisotropic systems is a nontrivial experimental task. Thus the violation of two-scale-factor universality in the subclass of weakly anisotropic systems is not a formal property  but constitutes a substantial reduction of the predictive power of the finite-size theory  as compared to the simpler situation in the subclass of isotropic systems. A summary of further results and predictions of this paper is given below.

(a) {\it Bulk properties of anisotropic systems.} In Sec. II a derivation is given for the exact representation of ${\bf \bar A}$ in terms of ratios of principal correlation lengths within the $\varphi^4$ theory.  The ellipsoidal form of surfaces with constant $G_b({\bf x},t)$ and $G_{b,{\rm T}}({\bf x},t)$ in ${\bf x}$ space is identified analytically in terms of $\bar {\bf A}$ through (\ref{quadratic}). A conjecture is made for determining the principal axes of anisotropic fixed-length spin models. The range of validity in the parameter space of ${\bf A}$ is discussed. Quantitative predictions are made for the ratio of bulk correlation functions.

(b) {\it Predictions for slab geometry}. In Sec. IV, the predictions of the Casimir force scaling function for $n\geq1$ are specified  for two models  with a diagonal and a non-diagonal anisotropy matrix ${\bf \bar A}$ in three dimensions. These results demonstrate the effect of lattice anisotropy on (i) the crossover from far below to far above $T_c$ including anisotropy-dependent minima near $T_c$ (Fig. 1), (ii) the change of sign of the critical Casimir amplitude (Fig. 2), (iii) the change of sign of the Casimir amplitude at low temperature (Fig. 3). The prediction (ii) demonstrates that the proof \cite{hucht2011} for the vanishing of the Casimir force at $T_c$ for cubic geometry is not valid for anisotropic systems. It also shows that  microscopic details such as the presence of a next-nearest-neighbor (NNN) interaction do show up in nonuniversal macroscopic effects on the Casimir force. Some of our approximate results for finite $n$ have large-$n$ limits that agree with the exact result derived in Sec. VI.

(c) {\it Predictions for film geometry}. In Secs. V and VI we find the scaling form of the excess free energy density of the $d$-dimensional anisotropic film system at $ h=0$
\begin{eqnarray}
\label{ffilmscalinganisox}
&&f^{\text ex}_{\text film}(t,L,{\bf A})= L^{-d}\; F^{\text ex}_{\text film}(C_1tL^{1/\nu},{ \bf \bar A}), \;\;\;\;\;\;\;\;\\
\label{Ffilmanisot}
 &&F^{\text ex}_{\text film}(C_1tL^{1/\nu},{ \bf \bar A})=C_0({ \bf \bar A})\;F^{\text ex}_{\text film,iso}(C_1tL^{1/\nu}), \;\;\;\;\;\;\;\;\\
\label{Cnullani}
 &&C_0({ \bf \bar A})=[({ \bf \bar A^{-1}})_{dd}]^{-d/2},\;\;\;\;\\
 \label{Ceinsani}
 &&C_{1}({\bf A})= [1/ \xi^\perp_{0+}({\bf A})]^{1/\nu},
\end{eqnarray}
where $F^{\text ex}_{\text film,iso}$ is the scaling function of the isotropic film system which here, however, has a different scaling argument containing the nonuniversal $\bf A$-dependent amplitude $\xi^\perp_{0+}({\bf A})$, (\ref{scalfilmxixneux}), of the bulk correlation length
perpendicular to the film boundaries. The nonuniversal prefactor $ C_0({ \bf \bar A})$ violates two-scale-factor universality as defined through the scaling form (\ref{1b}).
Thus both the overall amplitude and the argument of the scaling function are affected by anisotropy. The structure of (\ref{ffilmscalinganisox}) is consistent with previous results \cite{cd2004,DG,kastening-dohm,kastening2012} and with multiparameter universality for film geometry. It is exact for $n\to \infty, d\leq 3$ (Sec. VI). For finite $n$, no shift of the film transition temperature is captured, as in previous work. The progress achieved here is the derivation of the scaling functions $F^{\text ex}_{\text film}$ and $X_{\text film}$
for general $n$ below $T_c$ and the quantitative description of anisotropy effects above and below $T_c$ (Fig. 4). Our results are at variance with an earlier result for anisotropic superconducting films below $T_c$ \cite{wil-1} without an anisotropy effect. We predict that the "universal" Casimir force scaling function $X_{\text film}$ observed in isotropic superfluid $^4$He films \cite{garcia} and calculated in isotropic $XY$ and $\varphi^4$ models with Dirichlet BC \cite{hucht,vasilyev2009,hasenbusch2010,dohm2014,biswas2010,KrDi92a,GrDi07} is different from the scaling function of weakly anisotropic models with the same BC in the same bulk universality class.

(d)  {\it Large-$n$ limit}. In this limit exact finite-size scaling functions
are derived in Sec. VI for block, slab, and film geometries. Unlike the case of finite $n$, they are valid for arbitrary ${\bf \bar A}$ provided that $\det {\bf A} >0$. They describe the anisotropy effects on the complete crossover from the Goldstone regime at low temperatures to high temperatures far above bulk $T_c$.
For $d>3$, the nonuniversal anisotropy effect on the shift of the finite film critical temperature $T_{cf}(L)$ is determined. This demonstrates that superconducting films
should exhibit an anisotropy-dependent fractional shift of the Kosterlitz-Thouless transition temperature different from that in isotropic superfluid $^4$He films.

(e) {\it Other finite-size scaling functions}.
Our result for the finite-size scaling function of $f_s$ is derived from an order-parameter distribution function $\propto \exp [- H^{eff}]$, (\ref{deltafprime6}), with an exponential form whose exponent can be interpreted as an effective Hamiltonian. As shown in \cite{Esser}, the same distribution function determines the finite-size scaling functions of, e.g., the specific heat, the susceptibility, and the order parameter. Thus these scaling functions can be calculated parallel to the calculation of $f_s$ presented in this paper.
If appropriate BC are employed, the corresponding nonuniversal finite-size effects are well measurable in real anisotropic systems such as  superconductors \cite{schneider2004}, magnetic materials \cite{alpha}, alloys \cite{onukiBook}, and solids with structural phase transitions \cite{bruce-1} and in compressible anisotropic systems \cite{dohm2011}. As a further example we mention the critical Binder cumulant \cite{priv,Binder,Esser,dohm2008}
\begin{equation}
\label{1f} U^* ({\bf \bar A}) \; =
\frac{1}{3}\;\Big[\frac{\partial^4 F(0,y,\{\rho_\alpha\}, {\bf \bar A})/ \partial y^4}{(\partial^2  F(0,y,\{\rho_\alpha\}, {\bf \bar A})
/\partial y^2)^2}\Big]_{y=0}
 \;
\end{equation}
defined through $F$, (\ref{1c}).
MC data for an anisotropic $d=2$ Ising model \cite{kam,selke2005,selke2009} and results of the $\varphi^4$ theory \cite{dohm2008,dohm2006,cd2004,kastening2013} support multiparameter universality but violate two-scale-factor universality (Secs. II and VII).
\renewcommand{\thesection}{\Roman{section}}
\renewcommand{\theequation}{2.\arabic{equation}}
\setcounter{equation}{0}
\section{Shear transformation and bulk theory}
In the following we discuss the shear transformation relating isotropic and anisotropic bulk correlation functions \cite{dohm2008,dohm2006,cd2004}. The latter
provide the basis for identifying the principal axes and  correlation lengths as measurable quantities.
Crucial steps have been performed in \cite{cd2004,dohm2006,dohm2008}
for general $n$ above $T_c$ and for $n=1$ below $T_c$.
Here we extend the analysis
to general $n$ below $T_c$ in order to demonstrate the nonuniversal algebraic decay of correlation functions that is the origin for nonuniversal low-temperature Casimir forces in anisotropic systems with $n>1$. The feature of
multiparameter universality \cite{dohm2008,dohm2009PJ,kastening-dohm} is further discussed and tested analytically for the $d=2,n=1$ universality class.

\subsection{ Shear transformation}

We briefly recall relevant elements of previous work \cite{cd2004,dohm2006,dohm2008}
on the basis of the $\varphi^4$ lattice Hamiltonians I (2.1)
and I (2.13). (Equation numbers preceded by "I" are those of \cite{dohm2017I}.)
Instead of the isotropic interaction I (2.14) we assume
an anisotropic short-range interaction with the long-wavelength form
\begin{eqnarray}
\label{2h}
\delta \widehat K ({\bf k}) = {\bf k}\cdot{\bf A}{\bf k} +
\sum^d_{\alpha, \beta, \gamma, \delta} B_{\alpha \beta \gamma
\delta} \; k_\alpha k_\beta k_\gamma k_\delta +
O (k^6)\;\;
\end{eqnarray}
where the symmetric $d\times d$ anisotropy matrix ${\bf A}$ depends on the microscopic
lattice structure and on the couplings $K_{i,j}$ through the dimensionless second moments
\begin{equation}
\label{2i} A_{\alpha \beta} = A_{\beta \alpha} = N^{-1} \sum^N_{i,
j = 1} (x_{i \alpha} - x_{j \alpha}) (x_{i \beta} - x_{j \beta})
K_{i,j}.
\end{equation}
Its real eigenvalues $\lambda_\alpha \; ,
\alpha = 1, 2, ..., d,$ and eigenvectors ${\bf e}^{(\alpha)}$ are
determined by ${\bf A
e}^{(\alpha)}=\lambda_\alpha {\bf e}^{(\alpha)}$ with ${\bf
e}^{(\alpha)} \cdot {\bf e}^{(\beta)} = \delta_{\alpha
\beta}$.
A necessary condition for {\it weak} anisotropy is $\lambda_\alpha> 0$
which implies corresponding restrictions for the couplings $K_{i,j}$ (see also Sec. II. D).

We bring the $O(k_\alpha k_\beta)$ part of (\ref{2h}) into an isotropic form
in order to make possible the use of renormalizations in Secs. II. C and III
that are the same as those of the isotropic $\varphi^4$ theory.
This is achieved by a shear transformation that consists of a rotation and rescaling of lengths in the direction of ${\bf e}^{(\alpha)}$.
The rotation is provided by the orthogonal matrix
${\bf U}={\bf U}\big(\{{\bf e}^{(\alpha)}\}\big)$ with matrix elements
$U_{\alpha \beta} = e_\beta^{(\alpha)}$
where $e^{(\alpha)}_\beta $ denote the
Cartesian components of the eigenvectors ${\bf e}^{(\alpha)}$. The
rescaling is provided by the diagonal matrix
\begin{eqnarray}
\label{matrixlambda}
&& {{\mbox {\boldmath$\lambda$}} = \bf U AU}^{-1},\\
 \label{det}
&&\det{{\mbox {\boldmath$ \lambda$}}= \det {\bf  A} }= \prod^d_{\alpha = 1}
\lambda_\alpha> 0 ,
\end{eqnarray}
with diagonal elements $\lambda_\alpha $. The reduced diagonal matrix ${\bf \bar{\mbox
{\boldmath$\lambda$}}}$ with diagonal elements $\bar \lambda_\alpha$ is defined as
\begin{eqnarray}
\label{reducedlambda}
&&{\bf \bar{\mbox
{\boldmath$\lambda$}}}={{\mbox
{\boldmath$\lambda$}} / (\det{{\mbox {\boldmath$ \lambda$}}})^{1/d}}= {\bf U \bar AU}^{-1},\\
\label{lambdaalpha}
&&\bar \lambda_\alpha= \frac{\lambda_\alpha}{ (\det {\bf {\mbox
{\boldmath$\lambda$}}})^{1/d}}
=\prod^d_{\beta=1,\; \beta \neq\alpha}\Big(\frac{\lambda_\alpha}{\lambda_\beta}\Big)^{1/d},
\end{eqnarray}
with $\det{\bar{\mbox {\boldmath$ \lambda$}}}= 1 $ and the reduced anisotropy matrix ${\bf \bar A}={\bf A}/(\det {\bf A})^{1/d}$, $\det {\bf \bar A}=1$.
So far both the eigenvectors ${\bf e}^{(\alpha)}$ and the reduced rescaling matrix ${\bf \bar{\mbox {\boldmath$\lambda$}}}$ are defined as a function of ratios of second moments of the microscopic couplings $K_{i,j}$ which we call parametrization (i). In ${\bf x}$ space and ${\bf k}$ space the transformation is
\begin{eqnarray}
\label{xj}
{\bf x'_j} &=& {\mbox {\boldmath$\lambda$}}^{-1/2} {\bf U}{\bf x_j}\:,\\
\label{kprime}
{\bf k}'&=&{\mbox {\boldmath$\lambda$}}^{1/2} {\bf U k},\\
\label{phiprime}
\varphi'({\bf x'_j})& =& (\det{\bf A})^{1/4}\varphi({\bf U}^{-1}
{\mbox{\boldmath$\lambda$}}^{1/2} {\bf x'_j})\;, \\
\label{uprime}
u'_0&=&(\det{\bf A})^{-1/2}u_0,\\
\label{2oox}
h'&=& (\det {\bf A})^{1/4}  h,
\end{eqnarray}
which leads to the
isotropic form of (\ref{2h}) with ${\bf A}'={\bf 1}$,
\begin{eqnarray}
\label{Aprimematrix}
 \label{2ox}
&&\delta \widehat K ({\bf k})  = \delta \widehat K ({\bf
U}^{-1}{\mbox {\boldmath$\lambda$}}^{-1/2} {\bf k'})
\equiv \delta \widehat K' ({\bf k'})
= {\bf k'}^2 + O(k'^4).\nonumber \\
\end{eqnarray}
This yields an exact relation between the anisotropic and isotropic bulk correlation functions $G_b ({\bf x}, t, h)$ and  $G'_b({\bf x}', t,  h')$, respectively, for arbitrary ${\bf x}, t, h$
\begin{eqnarray}
\label{2y} &&G_b ({\bf x}, t, h) = (\det {\bf A})^{- 1/2} G'_b
({\mbox {\boldmath$\lambda$}}^{-1/2} {\bf U} {\bf x}, t, (\det{\bf
A})^{1/4} h).\nonumber\\
\end{eqnarray}
Outside the asymptotic critical region, $G_b$ and $G'_b$ depend on all details of the couplings and the lattice structure.
The isotropic correlation function $G_b'$ is the basis of defining the second-moment bulk correlation length
\begin{equation}
\label{3b} \xi_\pm'(t, h') = \left(\frac{1} {2d} \frac{\sum_{\bf
x'}\; {\bf x}'^2 \;G_b' ({\bf x}', t, h')}{\sum_{\bf x'}\; G_b'
({\bf x}', t, h')} \right)^{1/2} \;
\end{equation}
for general $n$ above and at $T_c$ and for $n = 1$ below $T_c$.
In the asymptotic region (large ${\bf x}'$, large $\xi'_\pm$, finite $|{\bf x}'|/ \xi'_\pm\geq 0 $, small $h'$) $G_b'$ has the {\it isotropic} scaling form
\cite{dohm2008,pri}
\begin{eqnarray}
\label{3c} &&G'_b  ({\bf x'}, t, h') = D'_1 | {\bf x'} |^{- d + 2 -
\eta} \Phi_\pm (|{\bf x'}| / \xi'_{\pm}, D'_2  h'
t|^{-\beta\delta} ), \;\;\;\;\;\;\;\;\;\\
%
\label{3d}
&&\xi'_\pm(t, h') = \xi'_{0+} |t|^{- \nu} X_\pm (D'_2  h'
|t|^{-\beta\delta}) \;
\end{eqnarray}
with the universal scaling functions $\Phi_\pm(z,y)$ and $X_\pm( y)$.
Application of (\ref{2y}) to (\ref{3c})  yields an analytic identification of the principle axes and principal correlation
lengths $\xi^{(\alpha)}_\pm(t,h)$
in terms of the $T$-independent eigenvectors ${\bf e}^{(\alpha)}$ and eigenvalues $\lambda_\alpha$
\cite{dohm2008},
\begin{eqnarray}
\label{xilambda}
&&\xi_\pm^{(\alpha)} (t, h) = \lambda_\alpha^{1/2}
\xi_\pm'(t, h'),\\
\label{principalx}
&&\xi^{(\alpha)}_\pm(t,0)= \xi^{(\alpha)}_{0\pm}|t|^{-\nu},\;\;\xi_{0 \pm}^{(\alpha)} = \lambda_\alpha^{1/2} \xi_{0 \pm}',
\end{eqnarray}
with $\xi_\pm'(t,0)= \xi'_{0\pm}|t|^{-\nu}$.
Note that $\xi_\pm'(t, h')$ is not measurable since $G_b'$ is not observable, in contrast to $G_b$ of the original anisotropic system. In particular, the correlation lengths $\xi^{(\alpha)}_\pm(t, h)$ are measurable quantities
which have the nonuniversal ratios \cite{dohm2008}
\begin{eqnarray}
\label{3qx}
\frac{\xi_+^{(\alpha)}}{\xi_+^{(\beta)}}=\frac{\xi_-^{(\alpha)}}{\xi_-^{(\beta)}}
=\frac{\xi_{0+}^{(\alpha)}}{\xi_{0+}^{(\beta)}} =\frac{\xi_{0-}^{(\alpha)}}{\xi_{0-}^{(\beta)}}=
\Big(\frac{\lambda_\alpha}{\lambda_\beta}\Big)^{1/2}=\Big(\frac{\bar\lambda_\alpha}{\bar\lambda_\beta}\Big)^{1/2}.\;\;\;\;\;\;
\end{eqnarray}
According to (\ref{lambdaalpha}) and (\ref{3qx}), the diagonal matrix ${{\mbox {\boldmath$\bar \lambda$}}}$ has the diagonal elements
\begin{eqnarray}
\label{lambdaalphaxix}
&&\bar \lambda_\alpha= \prod^d_{\beta=1,\; \beta \neq\alpha}\big(\xi_{0\pm}^{(\alpha)}/\xi_{0\pm}^{(\beta)}\big)^{2/d}
=\big(\xi_{0\pm}^{(\alpha)}/\bar \xi_{0\pm}\big)^2,\;\;\;\\
\label{ximean}
&&\bar \xi_{0\pm}=\big[\prod^d_{\alpha = 1} \xi_{0\pm}^{(\alpha)}\big]^{1/d}=(\det {\bf A})^{1/(2d)}\xi_{0\pm}'
\end{eqnarray}
with $\bar \xi_\pm(t, 0) =\bar \xi_{0 \pm} |t|^{- \nu}$ where  the characteristic length
\begin{eqnarray}
\label{meancorx}
\bar \xi_\pm(t, h)&=&\big[\prod^d_{\alpha = 1} \xi_\pm^{(\alpha)}(t,  h)\big]^{1/d}=
\big[V^\pm_{corr}(t,  h)\big]^{1/d}\;\;\;\;\;\;
\end{eqnarray}
is the geometric mean of the principal correlation lengths determining the
ellipsoidal correlation volume $V^\pm_{corr}$ \cite{cd2004}. Thus Eq. (\ref{lambdaalphaxix}) determines the reduced rescaling matrix
\begin{equation}
\label{lambdaquerdef}
{{\mbox {\boldmath$\bar \lambda$}}}={\bf \bar{\mbox {\boldmath$\lambda$}}} \big(\{\xi_{0 \pm}^{(\alpha)}\}\big) \;\;\;
\end{equation}
as a function of the observable ratios $\xi_{0\pm}^{(\alpha)}/ \bar\xi_{0\pm}$ for general $d$ which we call parametrization (ii), with ${\bf e}^{(\alpha)}$ remaining the same as in parametrizatíon (i). Both ${\bf \bar{\mbox {\boldmath$\lambda$}}} \big(\{\xi_{0 \pm}^{(\alpha)}\}\big)$ and $V^\pm_{corr}$ are nonuniversal quantities and the vectors ${\bf e}^{(\alpha)}$ have a nonuniversal orientation.

It is remarkable that the exact relations (\ref{3qx}) and (\ref{lambdaalphaxix}) are quite simple for general $d$ and for arbitrary short-range interactions. They have the same form as for the Gaussian model (where $u_0 =0$).  Since $\bar \lambda_\alpha$ depends on the couplings
$K_{i,j}$ through (\ref{2i}),
Eqs. (\ref{3qx}) and (\ref{lambdaalphaxix}) provide $d-1$
relations
between the couplings $K_{i,j}$ and the ratios  $(\xi_{0\pm}^{(\alpha)}/\bar \xi_{0\pm})^2$ or $(\xi_{0\pm}^{(\alpha)}/(\xi_{0\pm}^{(\beta)})^2$ that are specific for the $\varphi^4$  theory and the Gaussian model.
We note that all relations given above are exact within the $\varphi^4$ theory and are valid for both the continuum and the lattice version. These relations  can be extended to general $n>1$ below $T_c$, see (\ref{ratioT}) and (\ref{ratios}).

It is well established that $f'_{b,s}(t,  h')$ of the {\it isotropic} system has the asymptotic (small $t$, small $ h'$) scaling form
\cite{pri}
\begin{equation}
\label{1aiso} f'_{b,s} (t, h') = A'_1 |t|^{d \nu} \; W_\pm (A'_2 h'|t|^{-\beta \delta})
\end{equation}
with the universal scaling function $W_\pm (z)$ above $(+)$ and below $(-)$
$T_c$ and the nonuniversal amplitudes $A'_1$ and  $A'_2$.
The shear transformation with (\ref{2oox}) implies the exact relation \cite{dohm2008}
\begin{eqnarray}
\label{bulkrelationsing}
&&f_{b,s}(t,h)= (\det {\bf A})^{- 1/2} f'_{b,s}\big(t,(\det {\bf A}\big)^{ 1/4}h)\nonumber\\&&=A'_1(\det {\bf A})^{- 1/2} |t|^{d \nu} \; W_\pm (A'_2(\det {\bf A})^{1/4} h|t|^{-\beta \delta})\;\;\;\;\;\;\;\;\;\;
\end{eqnarray}
which is just the scaling form (\ref{1ax}) of the {\it anisotropic} system with the nonuniversal amplitudes
\begin{equation}
\label{3mm}
A_1 = A'_1 (\det {\bf A})^{- 1/2}, \;\;\;A_2 = A'_2 (\det
{\bf A})^{ 1/4} \;.
\end{equation}
Thus the simple result is that the thermodynamic scaling function $W_\pm $ of the bulk free energy is the same for both isotropic and anisotropic systems for general $n$, only the amplitudes $A_i$ and $A'_i$ are different.
The situation is more complicated for the bulk correlation function and the finite-size scaling functions of anisotropic systems.
\subsection{Bulk correlation function and multiparameter universality in anisotropic systems}
Using the  parametrization (ii) of ${\bf \bar{\mbox {\boldmath$\lambda$}}} \big(\{\xi_{0 \pm}^{(\alpha)}\}\big)$, (\ref{lambdaquerdef}),
we have derived \cite{dohm2006,dohm2008} critical bulk relations involving the principal correlation lengths and the large-distance behavior of the critical correlation function $G_b ({\bf x}, 0, 0)$ of  weakly anisotropic systems
\begin{eqnarray}
\label{3u}&& f_{b,s}(t,0) \bar\xi_+(t,0)^d= A_1 (\bar\xi_{0+})^d
= Q_1 =\text {universal},\;\;\;\;\\
\label{3y}
&&\left(\Gamma_+ / \Gamma_c \right)
\left(\bar \xi_c / \bar\xi_{0 +} \right)^{2 - \eta}
 = Q_2  = \text {universal} \; , \quad \\
\label{3yx}
&&\left(\Gamma_+ / \Gamma_c \right)
\left(\xi_c^{(\alpha)} / \xi_{0 +}^{(\alpha)} \right)^{2 - \eta}
 = Q_2  = \text {universal} \; , \quad \\
\label{3z}
&& G_b ({\bf  x}^{(\alpha)}, 0, 0) = (\bar\xi_{0+})^{-d}{\Gamma_+}\;\widetilde Q_3\left[ \xi_{0 +}^{(\alpha)} /| {\bf  x}^{(\alpha)} | \right]^{d - 2 +
\eta},\;  \qquad \\
\label{principalratio}
&&\xi^{(\alpha)}_{0-}/\xi^{(\alpha)}_{0+}= \xi_{0-}'/\xi_{0+}'= X_-( 0)=  \text {universal},\;\;\;
\end{eqnarray}
with universal constants $Q_1(d,n)$, $Q_2(d,n)$, $\widetilde Q_3(d,n)$, and $X_-(0)$ that are the same for all isotropic and weakly anisotropic systems (not only $\varphi^4$ models)
within the same  $(d,n)$ universality class (for isotropic systems see Tables 6.1 and 6.3 of \cite{priv} and Table I of \cite{tarko}). The same statement holds for the universal constants $P_2(d,n)$, $P_3(d,n)$, and $W_1(d,n)$ of the amplitude relations given in (\ref{3n2})-(\ref{3n4}) below, as noted in \cite{dohm2008}. These universal features involving $d(d+1)/2+1$ independent nonuniversal parameters were called {\it  multiparameter universality} for weakly anisotropic systems \cite{dohm2008}.
In (\ref{3y})-(\ref{3z}),  ${\Gamma_+}$ is the amplitude of the bulk susceptibility $\chi_b ={\Gamma_+}t^{-\gamma}$ of the anisotropic system at $ h= 0$ above $T_c$, and $\bar \xi_c$
denotes the mean correlation-length amplitude at $T_c$
\begin{eqnarray}
\label{meancorampcx}
\bar \xi_c=\Big[\prod^d_{\alpha = 1} \xi_c^{(\alpha)}\Big]^{1/d}
=(\det{\bf A})^{1/(2d)-\nu/(4\beta\delta)}\xi_c'
\end{eqnarray}
where $\xi^{(\alpha)}_c$ is defined by
$\xi_\pm^{(\alpha)} (0, h) = \xi^{(\alpha)}_c| h|^{-\nu/(\beta\delta)}$
for small $h\neq 0$. In (\ref{meancorampcx}) we have used (\ref{2oox}), (\ref{xilambda}), and \cite{dohm2008}
$\xi'_\pm(0,h') = \xi'_c\; |  h'|^{-\nu/(\beta\delta)}$.
In (\ref{3z}), $G_b ({\bf  x}^{(\alpha)}, 0, 0)$ describes the critical large-distance behavior along the principal direction $\alpha$.
In the following we generalize it to arbitrary directions of ${\bf x}$ and to finite $t$ and $ h$ above and below $T_c$.

From (\ref{det}), (\ref{reducedlambda}), (\ref{xilambda}), (\ref{principalx}), and (\ref{meancorx}) we obtain
\begin{eqnarray}
\label{relationlambda}
{\mbox
{\boldmath$\lambda$}}^{1/2} \xi'_\pm(t,h)={\bf \bar{\mbox
{\boldmath$\lambda$}}}^{1/2} \bar \xi_\pm(t, h).
\end{eqnarray}
Substituting
(\ref{relationlambda})
into the first argument of $\Phi_\pm$ in (3.19) of \cite{dohm2008} yields the correlation function for an arbitrary direction of ${\bf x}$ in terms of
${\bf \bar{\mbox {\boldmath$\lambda$}}}$ and $\bar \xi_\pm(t, h)$,
\begin{eqnarray}
\label{3nbar} G_b ({\bf x}, t, h) &=& D_1 |{\bf \bar{\mbox
{\boldmath$\lambda$}}}^{-1/2} {\bf U}{\bf x}|^{- d + 2 - \eta}
\nonumber\\ &\times& \Phi_\pm (|{\bf \bar{\mbox
{\boldmath$\lambda$}}}^{-1/2} {\bf U}{\bf x}| /\bar \xi_\pm, D_2  h
|t|^{-\beta\delta}),\;\;\;\;\;\\
\label{3dbar}
\bar \xi_\pm(t, h) &=& \bar \xi_{0+} |t|^{- \nu} X_\pm (D_2  h
|t|^{-\beta\delta}) \; ,\\
\label{3nn}D_1& = &D'_1 \;(\det {\bf A})^{(- 2 + \eta) / (2 d)} \; ,\\
\label{3n1} D_2 &=& D'_2\; (\det {\bf A})^{1/4} \;,
\end{eqnarray}
with the universal scaling functions $\Phi_\pm(z,y)$ and $X_\pm( y)$
defined by the {\it isotropic} scaling form (\ref{3c}), (\ref{3d}).
We use the normalization $X_+ (0) = 1$ which implies \cite{dohm2008}
$\bar \xi_{0 -} /\bar \xi_{0 +} = X_-(0) = \text {universal}.$
The spatial argument $|{\bf \bar{\mbox
{\boldmath$\lambda$}}}^{-1/2} {\bf U}{\bf x}|$
can be rewritten as
\begin{eqnarray}
\label{xierechnungx}
&&|{\bf \bar{\mbox
{\boldmath$\lambda$}}}^{-1/2} {\bf U}{\bf x}|
= [({\bf \bar{\mbox
{\boldmath$\lambda$}}}^{-1/2} {\bf U}{\bf x})\cdot ({\bf \bar{\mbox
{\boldmath$\lambda$}}}^{-1/2} {\bf U}{\bf x})]^{1/2}\nonumber\\
&&=[{\bf x}\cdot ({\bf U}^{-1}{\bf \bar{\mbox
{\boldmath$\lambda$}}}^{-1} {\bf U}{\bf x})]^{1/2}=[{\bf x}\cdot ({\bf \bar A}^{-1}{\bf x})]^{1/2},
\end{eqnarray}
with
\begin{eqnarray}
\label{reducedA}
{\bf \bar A} ={\bf U}^{-1} {\bf \bar{\mbox {\boldmath$\lambda$}}}{\bf U}
\end{eqnarray}
[compare (\ref{matrixlambda}) and (\ref{reducedlambda})]. This leads to
\begin{eqnarray}
\label{3nbarA} G_b ({\bf x}, t,  h) &=& D_1 [{\bf x}\cdot ({\bf \bar A}^{-1}{\bf x})]^{(- d + 2 - \eta)/2}
\nonumber\\ &\times& \Phi_\pm ([{\bf x}\cdot ({\bf \bar A}^{-1}{\bf x})]^{1/2} /\bar \xi_\pm, D_2  h
|t|^{-\beta\delta}).\;\;\;\;\;\;\;\;\;\;
\end{eqnarray}
Due to the two parametrizations for ${\bf \bar{\mbox {\boldmath$\lambda$}}}$ given above, the matrix ${\bf \bar A}$, (\ref{reducedA}), can be represented (i) in terms of the second-moments of the couplings $K_{i,j}$, or (ii) in terms
of the correlation-length amplitudes $\xi_{0+}^{(\alpha)}$ through (\ref{lambdaquerdef}),
\begin{eqnarray}
\label{Anonunix}
&&{\bf \bar A}\big(\{\xi_{0\pm}^{(\alpha)},{\bf e}^{(\alpha)}\}\big)={\bf U}\big(\{{\bf e}^{(\alpha)}\}\big)^{-1}\;{\bf \bar{\mbox {\boldmath$\lambda$}}} \big(\{\xi_{0\pm}^{(\alpha)}\}\big)\;{\bf U}\big(\{{\bf e}^{(\alpha)}\}\big).\;\;\;\;\;\,\;\;\;
\end{eqnarray}
Within the $\varphi^4$ theory, both parameterizations (i) and (ii) are equivalent near $T_c$.   Consistency with the relations (\ref{3u})-(\ref{principalratio}) requires us to employ the parametrization (ii)  with the representation (\ref{Anonunix}) for applications beyond $\varphi^4$ theory (see the examples discussed below). In particular,
choosing ${\bf  x}^{(\alpha)}= x^{(\alpha)} {\bf e}^{(\alpha)}$ along the principal direction $\alpha$ and using (\ref{lambdaalphaxix}) we obtain from
(\ref{3nbarA}) and (\ref{Anonunix})
\begin{eqnarray}
\label{3oxi} &&G_b ({\bf  x}^{(\alpha)}, t, h)= D_1 \Big \{\big[\xi_{0 \pm}^{(\alpha)}/\bar \xi_{0 \pm}\big]/|
x^{(\alpha)} | \Big\}^{ d - 2 + \eta} \nonumber \\&& \;\;\;\;\;\;\;\;\;\;\;\;\;\;\;\;\;\;\;\times\;\;\Phi_\pm \left(| x^{(\alpha)}|/
\xi_\pm^{(\alpha)}(t,  h), D_2  h |t |^{- \beta \delta}\right).\;\;\;\;\;\;\;\;\;
\end{eqnarray}
Comparison with (\ref{3z}) yields the identification of $D_1$ in terms of the observable quantities $\Gamma_+$ and $\bar \xi_{0+}$ and the universal constants $\widetilde Q_3$ and $\Phi_\pm(0,0)$,
\begin{eqnarray}
\label{3oxiD} && D_1= (\bar \xi_{0+})^{-2+\eta}\Gamma_+\widetilde Q_3\;\Phi_\pm(0,0)^{-1}.
\end{eqnarray}

At $h\neq 0$, the results presented for $G_b$ are valid above, at, and below $T_c$ for general $n$. In the limit $ h \to 0$, however, they remain applicable below $T_c$ only for $n=1$, and the following separate discussion is necessary for the case $n>1$ for $ h \to  0$ below $T_c$. We choose the coordinate system in the $n$ dimensional ${\bm \varphi}'$ space such that ${\bf h}' = (h', 0,..., 0)$ and $<{\bm \varphi}'>'\equiv{\bf M}' = (M', 0,..., 0)$ and define the longitudinal and transverse bulk correlation functions of the transformed {\it isotropic} system as \cite{priv}
\begin{subequations}
\label{correlation functions}
\begin{align}
\label{long correl}
G'_{b,{\rm L}} ({\bf x}_i'-{\bf x}_j', t, h') =
\langle \varphi_i'^{(1)} \varphi_j'^{(1)}\rangle' - M'^2,
\\
\label{trans correl}
G'_{b,{\rm T}} ({\bf x}_i'-{\bf x}_j', t, h') =
\langle \varphi_i'^{(\kappa)} \varphi_j'^{(\kappa)}\rangle'  \;\;
\end{align}
\end{subequations}
for each $\kappa = 2,...,n$. For $h' \to  0 $ below $T_c$, where the correlations are long-ranged with an algebraic decay due to the Goldstone modes for $n>1$, no second-moment correlation length can be defined. Following \cite{hohenberg1976,priv} we define the bulk transverse correlation length $\xi'_{\rm T} $ by the large-distance behavior of $G'_{b,{\rm T}}$ for $d>2$
\begin{eqnarray}
\label{trans correl large}
G'_{b,{\rm T}} ({\bf x}', t, 0) &\approx & {\cal C}_{\rm T}  M'_b(t, 0)^2 \big[\xi'_{\rm T}(t)/|{\bf x}'|\big]^{d-2},\\
\label{constT}
{\cal C}_{\rm T}& = &\Gamma(d/2)/[2\pi^{d/2}(d-2)],
\end{eqnarray}
for $|{\bf x}'|\to \infty$ where $ M'_b(t, 0)$ is the bulk order parameter.
The asymptotic  form of $\xi'_{\rm T}$ near $T_c$ is
$\xi'_{\rm T}(t) = \xi'_{0\rm T} |t|^{- \nu}$,
with the  nonuniversal amplitude $\xi'_{0 \rm T}$. As shown within the isotropic theory \cite{priv}, the amplitude ratio
\begin{equation}
\label{3exT} \xi'_{0\text T} / \xi'_{0 +} = \text {universal} \;
\end{equation}
is a universal quantity, thus no new nonuniversal parameter is introduced below $T_c$ for $n>1$. The large-distance behavior of the longitudinal correlation function is related to that of $G'_{b,{\rm T}}$  by \cite{fisher-priv}
\begin{eqnarray}
\label{long correl large}
G'_{b,{\rm L}} ({\bf x}', t,0) \approx (1/2)(n-1)[G'_{b,{\rm T}} ({\bf x}', t,0)/ M'_b(t,0)]^2 .  \;\;\;\;\;
\end{eqnarray}
Thus $\xi'_{\rm T} $ governs the large-distance behavior of both $G'_{b,{\rm T}}$ and $G'_{b,{\rm L}}$ below $T_c$.
The experimentally measurable correlation functions are not $G'_{b,{\rm L}}$ and $G'_{b,{\rm T}}$ of the transformed isotropic system but $G_{b,{\rm L}}$ and $G_{b,{\rm T}}$ of the original anisotropic system. At $ h= 0$, $G_b$ and $G'_b$  are related by the shear transformation
\begin{eqnarray}
\label{2y prime} &&G_{b,{\rm L}} ({\bf x}, t,  0) = (\det {\bf A})^{- 1/2} G'_{b,{\rm L}}
({\mbox {\boldmath$\lambda$}}^{-1/2} {\bf U} {\bf x}, t,  0),\;\;\;\;\;\;\;\;
\\
\label{3y prime}&& G_{b,{\rm T}} ({\bf x}, t, 0) = (\det {\bf A})^{- 1/2} G'_{b,{\rm T}}
({\mbox {\boldmath$\lambda$}}^{-1/2} {\bf U} {\bf x}, t, 0).\;\;\;\;\;\;\;\;
\end{eqnarray}
Along the direction of the
principal axis $\alpha$, the large-distance behavior of $G_{b,{\rm T}} ({\bf x^{(\alpha)}}, t,0)$  is, similar to (\ref{3oxi}),
\begin{eqnarray}
\label{3oxtransB} G_{b,{\rm T}} ({\bf x}^{(\alpha)}, t, 0) \approx {\cal C}_{\rm T} B^2|t|^{2\beta}\big[\xi^{(\alpha)}_{\rm T}(t)/|  x^{(\alpha)}|\big]^{d-2}
\end{eqnarray}
where  $\xi_{\rm T}^{(\alpha)}(t)$ is the principal transverse correlation length in the direction of ${\bf e}^{(\alpha)}$,
$ \xi_{\rm T}^{(\alpha)}(t)=\xi_{0{\rm T}}^{(\alpha)}|t|^{-\nu},\;\; \xi_{0{\rm T}}^{(\alpha)}=\lambda_\alpha^{1/2} \xi_{0{\rm T}}'$,
with the nonuniversal ratio
\begin{eqnarray}
\label{ratioT}
\xi_{0 {\rm T}}^{(\alpha)} / \xi_{0 {\rm T}}^{(\beta)} =
(\lambda_\alpha / \lambda_\beta)^{1/2}.
\end{eqnarray}
In (\ref{3oxtransB}) we have used
$ M_b(t,0) =(\det {\bf A})^{-1/4}M'_b(t,0)\approx B|t|^\beta$ for $|t|\ll1$. Similar to (\ref{3nbar}) and (\ref{3nbarA}), we obtain for an arbitrary direction of ${\bf x}$
\begin{eqnarray}
\label{xtransB}
G_{b,{\rm T}} ({\bf  x}, t, 0) &\approx&
{\cal C}_{\rm T} B^2|t|^{2\beta}\Big(\frac{\bar \xi_{\rm T}}{ [{\bf x}\cdot ({\bf \bar A}^{-1}{\bf x})]^{1/2} }\Big)^{d-2},\;\;\;\;\;\;\;\;\\
\label{corrvolT}
\bar \xi_{\rm T}(t)&=&\Big[\prod^d_{\alpha = 1} \xi_{\rm T}^{(\alpha)}(t)\Big]^{1/d},
\end{eqnarray}
with the mean transverse correlation length $\bar \xi_{\rm T}$. The result for the longitudinal correlation function  $G_{b,{\rm L}}$ follows from (\ref{long correl large})-(\ref{3oxtransB}). In  (\ref{xtransB}) the representation (ii) of
\begin{eqnarray}
\label{AnonunixlT}
{\bf \bar A}\big(\{\xi_{0{\rm T}}^{(\alpha)},{\bf e}^{(\alpha)}\}\big)={\bf U}\big(\{{\bf e}^{(\alpha)}\}\big)^{-1}\;{\bf \bar{\mbox {\boldmath$\lambda$}}} \big(\{\xi_{0{\rm T}}^{(\alpha)}\}\big)\;{\bf U}\big(\{{\bf e}^{(\alpha)}\}\big)\;\;\;\;\;\;
\end{eqnarray}
in terms of $\xi_{0 {\rm T}}^{(\alpha)}$  and $\bar \xi_{0{\rm T}}=\Big[\prod^d_{\alpha = 1} \xi_{0\rm T}^{(\alpha)}\Big]^{1/d}$ can be employed
using (\ref{lambdaquerdef}) and (\ref{ratioT}).
According to (\ref{3qx}) and (\ref{ratioT}), $\varphi^4$ theory predicts the relations
\begin{eqnarray}
\label{ratios}
\xi_{0+}^{(\alpha)}/\xi_{0+}^{(\beta)} =\left\{
\begin{array}{r@{\quad \quad}l}
                         \; \xi_{0-}^{(\alpha)}/\xi_{0-}^{(\beta)}\quad          & \mbox{for} \;n=1\;, \\
                         \; \xi_{0 {\rm T}}^{(\alpha)} / \xi_{0 {\rm T}}^{(\beta)}& \mbox{for} \;n>1 .\;
                \end{array} \right.
\end{eqnarray}
Eqs. (\ref{3nbarA}),  (\ref{3oxi}), (\ref{3oxtransB}), and (\ref{xtransB})  are the central results of this section which are asymptotically exact within the  $\varphi^4$ theory. They are valid for $n=1,d>1$ and for $n\geq 2, d>2$ including the large-$n$ limit. They show that anisotropy changes the dominant power-law behavior for general $n$. In particular they show that the matrix ${\bf \bar A}^{-1}$ governs the large distance behavior of the bulk correlation function for general $n$ above and below $T_c$ via the quadratic form (\ref{xierechnungx}), and similarly in ${\bf k}$ space, determined by the shear transformation
\begin{eqnarray}
\label{quadratic}
{\bf x}\cdot (\bar \xi_\pm^{-2}{\bf \bar A}^{-1}{\bf x}) &=&{\bf x}'\cdot (\xi_\pm'^{-2}{\bf x}'),
\\
 {\bf k}\cdot (\bar \xi_\pm^{2}\;{\bf \bar A}{\bf k})&=&{\bf k}'\cdot (\xi_\pm'^{2}\;{\bf k}'),
\end{eqnarray}
and similarly with $\xi_{\rm T}'$ and $\bar \xi_{\rm T}$.
Equation (\ref{quadratic}) describes the ellipsoidal and spherical forms of constant $G_b$ and $G'_b$ surfaces in ${\bf x}$ and ${\bf x}'$ space, respectively (compare Fig. 1 of \cite{Vaidya1976}), and similarly in ${\bf k}$ and ${\bf k}'$ space, expressed in terms of the correlation lengths $\bar \xi_\pm$ and $\xi_\pm'$ in $d$ dimensions [compare (\ref{3nbarA}) and (\ref{3c})] at fixed $t,h$. At criticality these ellipsoidal surfaces are  described by
${\bf x}\cdot ({\bf \bar A}^{-1}{\bf x})=\text{const}$
 as follows from the power-law part of (\ref{3nbarA}).
After substituting ${\bf \bar A}\big(\{\xi_{0+}^{(\alpha)},{\bf e}^{(\alpha)}\}\big)$ and ${\bf \bar A}\big(\{\xi_{0{\rm T}}^{(\alpha)},{\bf e}^{(\alpha)}\}\big)$,
$G_b$ and $G_{b,{\rm T}}$ are expressed in terms of (a) the thermodynamic amplitudes $D_1$, $D_2$ and $B$, (b) the eigenvectors ${\bf e}^{(\alpha)}$, and (c) the amplitudes $\xi_{0+}^{(\alpha)}$, $\xi_{0-}^{(\alpha)}$, and $\xi_{0 {\rm T}}^{(\alpha)}$ of the principal correlation lengths. The former amplitudes (a) are universally related to $A_1$ and  $A_2$ \cite{dohm2008},
\begin{eqnarray}
\label{3n2} &&A_2 / D_2 \; = P_2= \text {universal} \; ,\\
\label{3n3} &&D_1 A_2^{-2} A_1^{- 1 - \gamma / (d \nu)} \; = P_3= \text {universal},\\
\label{3n4} &&B/(A_1A_2) = - W_1 = \text {universal},
\end{eqnarray}
where $W_1\equiv \lim_{y\to 0}\partial W_-(y)/\partial y$. Such relations hold also for the isotropic system with primed amplitudes $A'_1,A'_2,D'_1,D'_2,B'$ (see I (3.51) - I (3.53) and Eqs. (3.10), (3.11), and (A12) of \cite{dohm2008,correctA12}) where the universal bulk constants $P_2$, $P_3$, and $W_1$ of the isotropic system are the same as for  the anisotropic system in the same universality class.
This is an important ingredient of multiparameter universality.
Among  the amplitudes (c), there are only $d-1$ independent amplitudes because of the universal relations (\ref{3u}), (\ref{principalratio}), and (\ref{3exT}).
In addition, the knowledge of $d(d-1)/2$ nonuniversal parameters determining the directions of ${\bf e}^{(\alpha)}$ is necessary. Altogether there are
$d(d+1)/2 +1$ independent nonuniversal parameters determining the bulk correlation function. The hypothesis of two-scale-factor universality states \cite{hohenberg1976} that the correlation function near $T_c$ is {\it fully determined} once the two thermodynamic amplitudes $A_1$ and $A_2$ have been chosen. Our results show that the knowledge of $A_1$ and $A_2$ and of the universal quantities  $Q_1,P_2,P_3,W_1$ is not sufficient: According to (\ref{3u}), the amplitude $A_1$ determines only the product $\prod^d_{\alpha = 1} \xi_{0 +}^{(\alpha)}$ but not each factor $ \xi_{0 +}^{(\alpha)}$ separately that would be needed in (\ref{3oxi}) above $T_c$, thus the hypothesis is not valid for the subclass of weakly anisotropic {\it bulk} systems. A natural consequence is the violation of  two-scale-factor universality also for {\it confined} systems (Sec. IV). Our conclusion is not in conflict with the early proofs of two-scale-factor universality for bulk systems \cite{aharony1974,ger-1,hohenberg1976,weg-1} since these proofs were given only for isotropic systems. Also the derivation in \cite{pri,priv} was based on the assumption of a single bulk correlation length $\xi_\infty$ that does not exist in anisotropic systems.

Nevertheless, as pointed out in \cite{dohm2006,dohm2008}, some degree of universality is maintained for the bulk properties of anisotropic systems: The critical exponents
and the functions $W_\pm(z)$, $\Phi_\pm(z,y)$, and $X_\pm( y)$ are the same as those of isotropic $\varphi^4$ theory and are independent of the coupling $u_0$, of the lattice spacing, and of the higher-order moments $B_{\alpha\beta \gamma \delta}$ etc. of the $\varphi^4$ lattice theory. We anticipate that they would also remain independent of higher-order couplings, such as those of $\varphi^6$ terms if such terms were included in the $\varphi^4$ Hamiltonian.
Furthermore the same principal axes and correlation lengths can be obtained from $\varphi^4$ models on various lattices with a large variety of different couplings.  (For a few examples see Sec. IX of \cite{dohm2008}.) This means that
a large number of members in the subclass of
anisotropic $\varphi^4$ models have the same asymptotic bulk correlation functions and amplitude relations near $T_c$ if they have the same nonuniversal parameters (a)-(c) specified above. This prediction can be tested by MC simulations for $\varphi^4$ lattice models \cite{hasenbusch2010,Hasenbuschgesamt}.

We hypothesize that this kind of multiparameter universality is valid not only for all $\varphi^4$ lattice models but quite generally for all other systems in the subclass of weakly anisotropic systems of the $(d,n)$ universality classes provided that the parametrization (ii) is used for reasons of consistency with (\ref{3u})-(\ref{principalratio}), i.e., that a reduced anisotropy matrix of such systems is constructed according to (\ref{Anonunix}),
with  $ {{\mbox {\boldmath$\bar \lambda$}}}$   expressed in terms of $\xi_{0+}^{(\alpha)}/\xi_{0+}^{(\beta)}$ according to (\ref{lambdaalphaxix}),
and with ${\bf U}$ and ${\bf U^{-1}}$ determined by the unit vectors ${\bf e}^{(\alpha)}$ defining the principal axes.
Such systems include the $O(n)$-symmetric spin models with the Hamiltonian
\begin{eqnarray}
\label{Hspin}
\beta H^{\text spin} = - \sum_{i,j} K^{\text spin}_{i,j} {\bf S}_i \cdot {\bf S}_j- \sum_{i} \beta {\bf h} \cdot {\bf S}_i
\end{eqnarray}
with $\beta= 1/(k_B T)$ where the $n$-component spin variables ${\bf S}_i$ have a fixed length  ${\bf S}_i^2=1$.
For $n=1$,  ${\bf S}_i\equiv \sigma_i= \pm 1$ denote the discrete variables of the Ising model. We take this model for testing our hypothesis.

An appropriate quantity is the ratio of the bulk correlation functions at criticality along two different principal directions $\alpha$ and $\beta$ which, according to (\ref{3qx}), (\ref{3z}), (\ref{3nbar}), and (\ref{3oxi}), $\varphi^4$ theory predicts to have the large-${\bf  x}$ behavior
\begin{eqnarray}
\label{corrlambdaratiox} \frac{G_b ({\bf  x}^{(\alpha)}, 0, 0)}{G_b ({\bf  x}^{(\beta)}, 0, 0)}\;\Bigg(\frac{|{\bf  x}^{(\alpha)}|}{|{\bf  x}^{(\beta)}|}\Bigg)^{d-2+\eta}&=&\Bigg (\frac{\lambda_{\alpha}}{\lambda_{\beta}}\Bigg)^{(d-2+\eta)/2} \\
\label{3oxicratiox}
&=& \big[\xi^{(\alpha)}_{0 +}/\xi^{(\beta)}_{0 +}\big]^{d-2+\eta}\;\;\;\; \;\;\;\;\;\;\;
\end{eqnarray}
for general $d$ and $n$ and arbitrary short-range interactions. The basic question is whether this general result is valid for anisotropic systems beyond the $\varphi^4$ theory.

We consider the anisotropic $d=2$ Ising model on a square lattice
with different positive nearest-neighbor (NN) couplings in the ``horizontal'' and ``vertical'' directions, denoted by  $K_1=2 \beta J_\parallel$ and $K_2=2\beta J_\perp$ respectively.  The principal directions 1 and 2 are  parallel to the Cartesian axes.
This implies ${\bf U}_{\text Ising} = {\bf 1}$ and ${\bf \bar A}_{\text Ising}=  {{\mbox {\boldmath$\bar \lambda$}}}_{\text Ising}$.
An exact analytic result by Wu \cite{Wu1966} is available for the critical bulk correlation function $G^{\text Ising}_b$ of this Ising model.
The large-${\bf  x}$ behavior of the ratio of the critical correlation functions $G^{\text Ising}_b$
along the principal directions $1$ and $2$ is obtained
from  Egs. (1.7), (1.11), and (5.7) of \cite{Wu1966} and
from Eqs. (5.2) and (5.9) in chapter XI of \cite{CoyWu} as
\begin{eqnarray}
\label{Isingratioxx}&& \frac{G^{\text Ising}_b ({\bf  x}^{(1)}, 0,  0)}{G^{\text Ising}_b ({\bf  x}^{(2)}, 0, 0)}\Bigg(\frac{|{\bf  x}^{(1)}|}{|{\bf  x}^{(2)}|}\Bigg)^{1/4}= \Bigg [\frac{(1+z_{1c}^2)(1-z_{2c}^2)}{(1-z_{1c}^2)(1+z_{2c}^2)}\Bigg]^{1/4}  \nonumber\\\\
\label{sinhratio}
&&=\big[\sinh(4\beta_c J_\parallel)/\sinh(4\beta_c J_\perp)\big]^{1/8}
\\
\label{xiratioisingx}
&&=\big (\xi_{01}/\xi_{02}\big)^{1/4}
\end{eqnarray}
with $z_{1c}=\tanh(2\beta_c J_\parallel),z_{2c}=\tanh(2\beta_c J_\perp)$ where
$\xi_i=\xi_{0i}t^{-\nu}$, $i=1,2$ denote the
correlation lengths above $T_c$ of the Ising model for the "rectangular lattice" employed in \cite{Indekeu}.
In deriving (\ref{sinhratio}) from (\ref{Isingratioxx}) we have used the condition of criticality \cite{Indekeu}
$\sinh(4\beta_c J_\parallel)\sinh(4\beta_c J_\perp)=1$.
Eq. (\ref{xiratioisingx}) then follows from the relation \cite{Indekeu}
\begin{eqnarray}
\label{xiratiospinx}
\xi_{0 1}/\xi_{0 2}=[\sinh(4\beta_c J_\parallel)/\sinh(4\beta_c J_\perp )]^{1/2}.\;\;\;\;\;\;
\end{eqnarray}
In (\ref{xiratioisingx})  and (\ref{xiratiospinx}) the asymptotic amplitudes of the "true" correlation lengths  are employed as defined in \cite{Indekeu} through the exponentially decaying part of the correlation function \cite{fish-2,cd2000-2,Indekeu,footnotetrue}. We conjecture that the ratio of these amplitudes is identical with the ratio of the corresponding {\it second-moment} principal correlation lengths of the anisotropic Ising model.

Applying the result (\ref{3oxicratiox}) for the $\varphi^4$ theory to the ($d=2, n=1$) universality class with $\eta=1/4$  we indeed find exact structural agreement between (\ref{3oxicratiox}) and (\ref{xiratioisingx}) in the parametrization (ii) of both models. This constitutes a nontrivial analytic confirmation of our hypothesis of multiparameter universality.

We further comment on this issue by considering the $d=2$  $\varphi^4$ lattice model with the same NN couplings on the square lattice, i.e.,  $K_x=J_\parallel/\tilde a^2$,  $K_y =J_\perp/\tilde a^2$ in the ``horizontal'' and ``vertical'' directions, respectively. This corresponds to the  diagonal anisotropy matrix
${\bf A}_{(d=2)}
=
2\left(\begin{array}{ccc}
 J_\parallel & 0 \\
 0 & J_\perp \\
\end{array}\right),$
with eigenvalues $\lambda_x=2J_\parallel >0, \lambda_y=2J_\perp >0$ and $\det {\bf A}_{(d=2)}=4J_\parallel J_\perp$. The corresponding correlation-length amplitudes  are denoted by $\xi_{0 +}^\parallel$, $\xi_{0 +}^\perp$ for the $\varphi^4$ model.
The reduced anisotropy matrix is
\begin{align}
\label{2dmatrixquer1}
{\bf \bar A}_{(d=2)}\;\;\;=&&
\left(\begin{array}{ccc}
(J_\parallel/J_\perp)^{1/2} &\;\; 0 \\
0 & \;\;(J_\perp/J_\parallel)^{1/2} \\
\end{array}\right)\\
\label{2dmatrixquer2}
&&=
\left(\begin{array}{ccc}
 \xi^\parallel_{0+}/\xi^\perp_{0+} &\;\; 0 \\
 0 & \;\;\xi^\perp_{0+}/\xi^\parallel_{0+} \\
\end{array}\right)
\end{align}
where we have used
\begin{eqnarray}
\label{ratioxi}
\xi^\parallel_{0+}/\xi^\perp_{0+}=(\lambda_x/\lambda_y)^{1/2}=(J_\parallel/J_\perp)^{1/2}
\end{eqnarray}
according to (\ref{3qx}). Eqs. (\ref{2dmatrixquer1}) and (\ref{2dmatrixquer2}) correspond to the representation (i) and (ii), respectively. Eqs. (\ref{ratioxi}) and (\ref{xiratiospinx})
demonstrate that  ratios of correlation lengths have, in general, a different dependence on the couplings for different models. Such a difference was already noted in \cite{bruce,kastening-dohm} between the $\varphi^4$ and Gaussian models and the $d=2$ Ising model. This implies that Eq. (6.7) of \cite{diehl-chamati} derived from $\varphi^4$ theory is not generally valid for lattice $O(n)$ spin models with finite $n$. Likewise, the correlation functions (\ref{corrlambdaratiox})  and (\ref{Isingratioxx}) have a different dependence on the couplings $J_\parallel,J_\perp$.
Thus our example confirms that in constructing the appropriate matrix ${\bf \bar A}$ according to (\ref{Anonunix}) for applications to bulk theories beyond the $\varphi^4$ theory, representation (ii) should be used as was done previously \cite{dohm2006,dohm2008} in the context of (\ref{3u})-(\ref{principalratio}).

In summary, multiparameter universality means that all systems of a given bulk universality class having the same amplitudes $A_1$, $A_2$, the same principal axes, and the same ratios $\xi_{0+}^{(\alpha)} / \xi_{0+}^{(\beta)}$ of the principal correlation-length amplitudes should have the same bulk correlation functions (\ref{3nbarA}),  (\ref{3oxi}), (\ref{3oxtransB}), and (\ref{xtransB})
and bulk relations (\ref{3u})-(\ref{principalratio}).
We emphasize, however, that scaling forms satisfying multiparameter universality
still have a nonuniversal character as they are functions of nonuniversal correlation-length amplitudes. For example, the ratios (\ref{corrlambdaratiox})  and (\ref{Isingratioxx}) are nonuniversal quantities. In particular  $G_{b,\rm T}$ is predicted to have a nonuniversal algebraically decaying ${\bf x}$-dependence below $T_c$ due to Goldstone modes for $n>1$, (\ref{xtransB}), and $G_b$ at criticality for general $n$ \cite{footnote2008x} where, according to (\ref{3nbarA}),
\begin{eqnarray}
\label{3nbarcx} G_b ({\bf x}, 0,  0) = \frac{D_1\;\;\Phi_+ (0,  0)}{[{\bf x}\cdot ({\bf U}^{-1}{\bf \bar{\mbox
{\boldmath$\lambda$}}}^{-1} {\bf U}{\bf x})]^{(d -2 +\eta)/2}}
\end{eqnarray}
exhibits a {\it directional nonuniversality} through ${\bf U}(\{{\bf e}^{(\alpha)}\})$  no matter what kind of representation of ${\bf \bar{\mbox {\boldmath$\lambda$}}}^{-1}$ is used, with different system-dependent amplitudes along different principal axes.

The hypothesis of multiparameter universality can be tested by MC simulations of spin and $\varphi^4$ models and by measurements in real systems. For this purpose the identification of the principal axes and correlation lengths is necessary. While an analytic identification is easily done for $\varphi^4$ models owing to the tractability of the term ${\bf k}\cdot{\bf A}{\bf k} $, no general approach to an analytic construction of the principal axes and correlation lengths has been developed for fixed-length spin models with lattice anisotropy.
For example, for the case of the $d=2$ Ising model with NN interactions on an anisotropic triangular lattice \cite{Indekeu,kastening2012},  only a {\it conjecture} for the true correlation lengths $\xi_i$ along the three directions $i$ of the bonds is known.

We suggest that the scaling forms (\ref{3nbarAhNull}), (\ref{3z}), and (\ref{3nbarcx}) can be taken as the basis for identifying the principal axes ${\bf e}^{(\alpha)}$ and the correlation lengths $\xi_{0+}^{(\alpha)}$ from a comparison with measurements or MC simulations of the bulk correlation function. This information then suffices to determine the matrix ${\bf \bar A}\big(\{\xi_{0+}^{(\alpha)},{\bf e}^{(\alpha)}\}\big)$ as well as the mean correlation length $\bar\xi_{0+}$ of the corresponding system through (\ref{lambdaalphaxix})-(\ref{lambdaquerdef}) and (\ref{Anonunix}).
These quantities are needed for a comparison of MC data with the predictions of our finite-size theory (Secs. IV-VI).
It would be interesting to compare in  more detail our exact results (\ref{3nbarAhNull}), (\ref{3z}), (\ref{3nbarA}), (\ref{ratios}), and (\ref{3nbarcx})
for the $(n=1,d=2)$ $\varphi^4$ model with the exact results
for the correlation function
of the anisotropic $d=2$ Ising model \cite{Wu1966,CoyWu,WuCoy,Vaidya1976}.

It was noted in \cite{dohm2008} that a reduced anisotropy matrix ${\bf \bar A}^{\text spin}= {\bf A}^{\text spin}/(\det {\bf A}^{\text spin})^{1/d}$ can be defined for a fixed-length spin model (\ref{Hspin}) in terms of the second moments
\begin{equation}
\label{2ifixed} A^{\text spin}_{\alpha \beta}  = N^{-1} \sum^N_{i,
j = 1} (x_{i \alpha} - x_{j \alpha}) (x_{i \beta} - x_{j \beta})
K^{\text spin}_{i,j},
\end{equation}
analogous to (\ref{2i}). It would be interesting to investigate in which cases the eigenvectors ${\bf e}^{(\alpha)}_{\text spin}$ of ${\bf \bar A}^{\text spin}$ correctly describe the exact principal axes of the correlation function of fixed-length spin model at large distances. This is obviously the case for lattices and interactions with orthorhombic symmetry. It is also the case for the $d=2$ Ising model discussed in Sec. II. E which does not have orthorhombic symmetry. We conjecture that this may be the case even for a larger class of non-orthorhombic interactions. This would facilitate the analytic determination of the principal axes of such spin models. An interesting candidate for this investigation is the $d=2$ anisotropic Ising model studied in \cite{Indekeu,kastening2012} for which no  exact analytic result has been given for the principal axes in terms of the NN couplings $K_i$.
\subsection{Amplitude of the bulk free energy density}
In the following we present explicit results for the amplitude of $f_{s, b} (t, 0)\equiv f^\pm_{s, b}(t)$ of the anisotropic system for general $n$ above and below $T_c$ and identify the length $\xi'_{0+}$ within the minimally renormalized isotropic $\varphi'^4$ theory. First we consider $f'^\pm_{b,s}(t)\equiv f'_{s, b} (t, 0)$ which
can be taken from the one-loop results I (3.15) and I (3.20) for isotropic systems after the replacement $u_0 \to u'_0$,
i.e.,
\begin{eqnarray}
\label{free-bulk-above-sing}
f'^+_{b,s}(t)&=&  -n  A_d\; (r_0 - r_{0c})^{d/2}/\;(d\;\varepsilon)+ O(u'_0),\\
\label{free-bulk-below-sing}
 f'^-_{b,s}(t) &=& - [-2(r_0 - r_{0c})]^2/(64 u'_0) \;\nonumber\\ &-&
A_d \;[-2 (r_0 - r_{0c})]^{d/2}/(d\;\varepsilon) + O(u'_0),\;\,
\end{eqnarray}
with the geometrical factor $A_d$, I (3.7))
and with the critical value of $r_0$ \cite{dohm2017I,dohm2008}
\begin{eqnarray}
\label{rnullcx}
r_{0c}&=&-4(n+2) u_0 \int_{\bf k} [\delta \widehat K (\mathbf k)]^{-1}+ O(u_0^2)\\&=&-4(n+2) u'_0 \int_{{\bf k}'} [\delta \widehat K' (\mathbf k)']^{-1}+ O({u'_0}^2),\\
\label{3dd} \int_{\bf k'} &=& (\det {\bf
A})^{1/2} \int_{\bf k} \equiv  (\det {\bf
A})^{1/2} \prod^d_{\alpha = 1}  \int^{\pi / \tilde a}_{- \pi /
\tilde a}  \frac{d k_\alpha}{2\pi}.\;\;\;\;\;\;\;\;
\end{eqnarray}
We see that $r_{0c}$ is affected by the anisotropy according to (\ref{rnullcx}) but is invariant under the shear transformation. The advantage of the transformed system is that its renormalizations can be taken from  bulk theory for isotropic systems \cite{dohm2017I} provided that they are expressed in terms of the renormalized counterparts
\begin{subequations}
\label{renormalized parameters}
\begin{align}
\label{5ax} u' = \mu^{- \varepsilon} A_d Z_{u'}(u', \varepsilon)^{-1} Z_{\varphi'}(u', \varepsilon)^2
u_0',\\
\label{rr}
r' = Z_{r'}(u', \varepsilon)^{-1} (r_0 - r_{0c})
\end{align}
\end{subequations}
of the {\it transformed} four-point coupling  $u_0'$, (\ref{uprime}), and of $r_0 - r_{0c}$. The singular part $ f'^\pm_{b,s}$  given in its unrenormalized form (\ref{free-bulk-above-sing}) and (\ref{free-bulk-below-sing})  will be denoted by
$\delta f'_b(r_0 - r_{0c}, u'_0)$. Then the renormalized counterpart is defined as
\begin{eqnarray}
\label{5dx}   f'_{R,b}(r', u',\mu) &=& \delta f'_b(Z_{r'}r',
       \mu^{\varepsilon}Z_{u'}Z_{\varphi'}^{-2}A_d^{-1}u')\nonumber\\  &-& (1/8)\mu^{-\varepsilon} r'^2 A_d
       A(u',\varepsilon)\; .
\end{eqnarray}
The renormalization constants are the same as given in I (3.25) with $u$ replaced by $u'$. This implies that the $\beta$-function $\beta_{u'}(u',\varepsilon)$ yields, through the condition $\beta_{u'}(u'^*,\varepsilon)=0$, the fixed point value $u'^*= u^*$ which is identical with the known fixed point value $u^*$ of ordinary isotropic ${\bm \varphi}^4$  theory \cite{cd2004,dohm2008}. Consequently the critical exponents of the anisotropic system are those of the isotropic system.
The same renormalization constants will be employed in the finite-size theory in Sec. IV.
The subsequent steps are parallel to those of isotropic bulk theory \cite{dohm2008,dohm2017I}. The reference length $\mu^{-1}$ is  chosen as
\begin{eqnarray}
\label{my}
&&\mu^{-1} = \xi'_{0 +}(a_0, u_0,\det {\bf A})\\
\label{xinullprime}
&& = \Big[ a'^{-1} Q^*
\exp \Big(\int_{u'}^{u^*} \frac{\zeta_{r'} (u^*) - \zeta_{r'}
(u'')}{\beta_{u'} (u'', \varepsilon)}\, d u'' \Big)\Big]^{1/2}\;\;\;\;\;\;\;\;\;\;
\end{eqnarray}
with $a'=Z_{r'}(u',\varepsilon)^{-1}a_0$ where (\ref{xinullprime}) is the exact representation of the amplitude $\xi'_{0+}$ of the second-moment bulk correlation length within the minimally renormalized $\varphi'^4$ theory  \cite{dohm2017I,dohm1985}. Note that $\xi'_{0+}$ is a function of $a_0$ and $(\det {\bf
A})^{-1/2}u_0$ through $a'$ and $u'$.
Using (\ref{bulkrelationsing}) we obtain
the singular part of the {\it anisotropic} system
\begin{eqnarray}
\label{asymfreexx}
&&f^\pm_{s,b}(t)= (\det {\bf A})^{- 1/2} A'^\pm|t|^{d \nu}\;
=A^\pm|t|^{d \nu},\\
\label{6mxx}
&&A^+ \equiv A_1 = Q_1\bar\xi_{0+}^{-d}\; \\
\label{6mx}
&&=  - \; A_d \; Q^{*d\nu} \;
\left[\frac{n}{4d} +  \frac{\nu B(u^*)}{2 \alpha}  \right]
(\det {\bf A})^{- 1/2}\;\xi_{0+}'^{-d},\;\;\;\;\;\;\;\;\;\;\\
\label{6nxx} && A^-\equiv  A_1 W_-(0) =  - \; A_d \; (2Q^*)^{d\nu}  \nonumber\\
&& \times \left[\frac{1}{64 u^*} \; +
\; \frac{1}{4d}   +  \frac{\nu  B(u^*)}{8
\alpha}  \right](\det {\bf A})^{- 1/2} \; \xi_{0+}'^{-d}, \;\;\;\;\;\;\;\;\;\;\;
\end{eqnarray}
compare I (3.44) and I (3.45), with the same  universal quantity $Q_1(d,n)$ as given in I (3.56). The quantities $Q^*$ and $B(u^*)$ are the same as defined in \cite{dohm2017I}.
From (\ref{6mx}) and (\ref{6nxx}) the same universal ratio  $A^-/A^+$
is obtained as for the isotropic system as given in I (3.57).
We express $f^\pm_{s, b}(t)$ in terms of the ellipsoidal correlation volume $ V^+_{corr}(t)\equiv V^+_{corr}(t, 0)$, (\ref{meancorx}), with a $T$-independent orientation. Together with (\ref{det}) and (\ref{principalx}), the result (\ref{asymfreexx})-(\ref{6nxx})  can be written
as
\begin{eqnarray}
\label{3jjx}
&&f^\pm_{s,b}(t)= \left\{
\begin{array}{r@{\quad \quad}l}
                          Q_1/V^+_{corr}(t) ,\;\;\;t > 0,\quad          &  \\
                         (A^-/A^+)Q_1/V^+_{corr}(t),\;\;\; t <  0&
                \end{array} \right.
\end{eqnarray}
which has the same universal form as
I (3.59) for isotropic systems where, however,  the spherical volume $V^+_{corr}(t)$ depends only on a single length scale.
\subsection{Restricted range of validity }
For the applications of the $\varphi^4$ model
for finite $n$
the following reservation must be made as indicated already in Sec. VIII. E of \cite{dohm2008}.
Consider the rotation in wave-vector space, ${\bf q} = {\bf U} {\bf k}$. It yields an interaction $\delta \widetilde K ({\bf
q}) \equiv \delta \widehat K ({\bf U}^{-1} {\bf q})$ that is diagonalized at $O(q^2)$,
\begin{eqnarray}
\label{33lx} \delta \widetilde K ({\bf q}) \;=\;\sum_{\alpha=1}^d
\lambda_\alpha q_\alpha^2 +  \sum^d_{\alpha,
\beta, \gamma, \delta} \widetilde B_{\alpha \beta \gamma \delta} \; q_\alpha
q_\beta q_\gamma q_\delta + O (q^6).\;\;\;\;\;
\end{eqnarray}
On the level of mean-field theory, a
wave-vector instability occurs when the anisotropy parameters are changed such that one eigenvalue (e.g., $\lambda_1$) or more than one of the eigenvalues vanish which may correspond to a Lifschitz point \cite{diehl-2}   (for an example see Sec. IV. E). The smallness of $\lambda_1$ in the vicinity of this point implies that some of the $O(q^4)$ terms are no longer negligible, and fluctuation affects arising from the $u_0\varphi^4$ term and $O(q^4)$ terms must be taken into account via a perturbative RG treatment incorporating a renormalized {\it shifted} eigenvalue $\lambda_{1R}
\neq \lambda_1$ \cite{diehl-2}. Thus the physical instability occurs at a point $\lambda_{1R}=0$ with $\lambda_1=\lambda_{1LP}\neq 0$ where $\lambda_{1LP}$ depends on the nonuniversal details of the model. Our subsequent theory is not applicable to this point and to its vicinity since our renormalizations will be defined with respect to the ordinary critical point of the isotropic Hamiltonian $H'$ rather than with respect to the point of instability where $\lambda_{1R}=0$. This implies that, for finite $n$, the range of applicability of our theory is restricted not only by $\det {\bf A}>0$,
(\ref{det}), but also by the requirement that the matrix ${\bf A}$ is well away from the renormalized wave-vector instability described above. This is guaranteed if ${\bf A}$ is restricted to some neighborhood of isotropy (${\bf A}_{iso}=c_0{\bf 1}$, $c_0>0$). This is taken into account in the applications in Sec. IV. C.
Our discussion applies also to the exact result for the ratio of critical bulk correlation functions (\ref{corrlambdaratiox}). Its range of validity is  limited not only by the requirement $\lambda_\alpha>0, \lambda_\beta>0$ but also by the condition that the system is away from an instability of the type discussed above. The reservations made above are not necessary for the exactly solvable case $n\to \infty$ (Sec. VI).
\subsection{Predictions for ${\bf d=2}$ Ising and ${\bf \varphi^4}$ models}
In this section we present predictions for the $d=2$ Ising universality class. We consider the  $\varphi^4$ model   at $h=0$ near $T_c$ on a square lattice with NN couplings  $K_x=K_y =J/\tilde a^2>0$ and the NNN coupling $K_d=J_d/\tilde a^2$ in the diagonal $(1,1)$ direction (Fig. 2 (a) of \cite{dohm2006} and Fig. 1 of \cite{selke2005}). The anisotropy matrix
\begin{align}
\label{simplematrixx}
{\bf A}_2
&=
2\left(\begin{array}{ccc}
 J +J_d& J_d \\
 J_d & J +J_d \\
\end{array}\right)
\end{align}
has the eigenvalues $\lambda_1=2(J+2J_d) ,\lambda_2=2J$ and eigenvectors
$ {\bf e}^{(1)} = 2^{-1/2}\left(\begin{array}{c}
  1 \\
  1 \\
\end{array}\right) \; , {\bf e}^{(2)} =  2^{-1/2}\left(\begin{array}{c}
  -1 \\
  1 \\
\end{array}\right) \; ,$
directed along the diagonals \cite{dohm2006}. These eigenvectors are valid for  both $J_d>0$ and $J_d<0$. The directions 1 and 2 are parallel and perpendicular to the $(1,1)$ direction of the bonds
$J_d$. Weak anisotropy requires  $\lambda_\alpha>0$ which implies $J>0$ and $J_d > -J/2$. The principal correlation lengths
$\xi_\pm^{(\alpha)}(t)= \xi^{(\alpha)}_{0\pm}|t|^{-\nu}$ with $\xi_{0 \pm}^{(\alpha)} = \lambda_\alpha^{1/2} \xi_{0 \pm}'$
have the ratio given in Eq. (3.31) of \cite{dohm2008,footnoteratio},
\be
\label{ratioq} \frac{\xi_{0 \pm}^{(1)}}{\xi_{0 \pm}^{(2)}} = \Big(\frac{\lambda_1}{\lambda_2}\Big)^{1/2}\equiv \; q \; = \;\left\{
\begin{array}{r@{\quad\quad}l}
               > 1 & \mbox{for} \; J_d > 0\;, \\ <1 & \mbox{for} \;
                 J_d < 0\;.
                \end{array} \right.
\ee
According to (\ref{3nbarA}) with $d=2$ and $\eta=1/4$ the exact asymptotic bulk correlation function is
\begin{eqnarray}
\label{3nbarAhNull2dx} G_b ({\bf x}, t) = \frac{D_1}{ [{\bf x}\cdot ({\bf \bar A_2}^{-1}{\bf x})]^{1/8}}\;
 \Phi_\pm \Big(\frac{[{\bf x}\cdot ({\bf \bar A_2}^{-1}{\bf x})]^{1/2}}{\bar \xi_\pm(t)}\Big)\;\;\;
\end{eqnarray}
with ${\bar \xi_\pm(t)}=\big[\xi_\pm^{(1)}(t)\xi_\pm^{(2)}(t)\big]^{1/2}$ where  $\Phi_\pm(|{\bf x'}|/\xi'_\pm)$ is the universal scaling function of the isotropic system. In terms of the anisotropy parameter
\begin{eqnarray}
\label{s}
s & = & \frac{1}{1 +J/ J_d}= \frac{(\lambda_1/\lambda_2)-1}{(\lambda_1/ \lambda_2)+1}=  \frac{q^2-1}{q^2  + 1}
\end{eqnarray}
the reduced anisotropy matrix reads \cite{dohm2008}
\begin{eqnarray}
  \label{33ax}
 {\bf \bar A}_2
 &=& \;(1-s^2)^{-1/2} \left(\begin{array}{ccc}
  1 & \;\;s  \\
  s & \;\;1  \\
\end{array}\right)\\
 \label{33qx}
&=& \;\frac{1}{2} \left(\begin{array}{ccc}
  q+q^{-1} &\;\;\; q-q^{-1}  \\
  q-q^{-1} & \;\;\;q+q^{-1}  \\
\end{array}\right)\equiv {\bf  Q}_2(q).
\end{eqnarray}
Expressing $s$ and $q$ in terms of the coupling ratio $J/ J_d$ or the correlation-length ratio $\xi_{0 \pm}^{(1)}/\xi_{0 \pm}^{(2)}$
corresponds to the representations (i) or (ii), respectively. They are equivalent within the $\varphi^4$ model near $T_c$.
As a special case we obtain the ratio of the correlation functions at $T_c$  for large $|{\bf  x}^{(\alpha)}|$ along the two diagonals according to (\ref{corrlambdaratiox})
\begin{eqnarray}
\label{neucorrlambdaratiox} \frac{G_b ({\bf  x}^{(1)}, 0)}{G_b ({\bf  x}^{(2)}, 0)}\;\Bigg(\frac{|{\bf  x}^{(1)}|}{|{\bf  x}^{(2)}|}\Bigg)^{1/4}&=&\Bigg (\frac{J+2J_d}{J}\Bigg)^{1/8}.
\end{eqnarray}
The predictions (\ref{3nbarAhNull2dx})-(\ref{neucorrlambdaratiox}) can be tested by MC simulations for the $d=2$ $\varphi^4$ model \cite{Hasenbuschgesamt}.

The $d=2$ Ising model with the same couplings is called "anisotropic triangular model" in \cite{Berker,Indekeu}. Multiparameter universality predicts that the same result for the correlation function $G^{\text Ising}_b $ is valid as for (\ref{3nbarAhNull2dx}) provided that  $\xi_{0 \pm}^{(\alpha)}$ and  ${\bf e}^{(\alpha)}$  are replaced by the corresponding quantities of the Ising model.  Exact analytic results for $G^{\text Ising}_b$  are available \cite{Vaidya1976} which yield the same principal directions ${\bf e}^{(\alpha)}$ as for the $\varphi^4$ model (i.e., along the diagonals, see Fig. 1 of \cite{Vaidya1976}). The explicit results  of \cite{Vaidya1976} for $T\neq T_c$, however, are given in terms of the couplings $E_i$ rather than principal correlation lengths
$\xi_\pm^{(\alpha){\text Ising}}$. Although no exact derivation has been given for $\xi_\pm^{(\alpha){\text Ising}}$ a conjecture
can be derived \cite{kastening2013} from the conjecture for the true correlation lengths $ \xi_i$ above $T_c$ along the three directions $i$ of the bonds of an  anisotropic triangular lattice  of the Ising model \cite{Indekeu}.
Using $\xi_+^{(1){\text Ising}}/\xi_+^{(2){\text Ising}}=\big[\sinh(4\beta_c J)\big]^{-1}$ as obtained from
Eqs. (8) and (9) of \cite{kastening2013} and assuming multiparameter universality we find, in analogy to (\ref{sinhratio}) and (\ref{xiratioisingx}), for the ratio
of the correlation functions at $T=T_c$ along the  diagonals,
\begin{eqnarray}
\label{neuIsingratioxx}&& \frac{G^{\text Ising}_b ({\bf  x}^{(1)}, 0)}{G^{\text Ising}_b ({\bf  x}^{(2)}, 0)}\Bigg(\frac{|{\bf  x}^{(1)}|}{|{\bf  x}^{(2)}|}\Bigg)^{1/4}
=\big[\sinh(4\beta_c J)\big]^{-1/4}\;\;\;\;\;\;\;\;\;
\end{eqnarray}
with  the condition of criticality \cite{hout,Berker}
$\sinh^2(4\beta_c J)+\sinh(4\beta_c J)\sinh(4\beta_c J_d)=1$.
For small $\beta_c J_d$ the r.h.s of (\ref{neuIsingratioxx}) is $1+J_d/(4J)$, in agreement with (\ref{neucorrlambdaratiox}) for small $J_d$
The prediction (\ref{neuIsingratioxx}) can be tested by extending the analytic results of \cite{Vaidya1976} for the Ising model to $T=T_c$ or by extending existing MC simulations  for  $G^{\text Ising}_b$ \cite{selke2009}.

We call attention to the fact
that the range of applicability of the $\varphi^4$ model is different from that of the Ising model. The eigenvalue $ \lambda_1 $
vanishes for $J_d=-J/2$ corresponding to $\det {\bf A} =0$ and $s=-1$ at which the matrix (\ref{33ax}) does not exist.
The discussion in Sec. II. C suggests an instability in the $\varphi^4$ model at $s_{LP}$ near $s=-1$ at which the line of critical points terminates.
Thus weak anisotropy no longer exists in this $\varphi^4$ model in the range  $s \leq s_{LP}$ and (\ref{3nbarAhNull2dx}) is not applicable to this range. By contrast, the Ising model  with the same couplings \cite{selke2005} is well behaved and has a smooth line of critical points down to $s=-\infty$ corresponding to $J_d=-J$ \cite{hout}, as shown in Fig. 1 of \cite{selke2009}, thus the $d=2$ $\varphi^4$ and Ising models have a fundamentally different phase diagram in the range $-\infty < s \lesssim -1$ corresponding to $-J<J_d\lesssim-J/2$.

This affects, of course,  the finite-size behavior in this regime. In \cite{dohm2008} the goal was to use $\varphi^4$ theory primarily in the range $s>0$  in order to explain the MC data \cite{selke2005} of the critical Binder cumulant of the anisotropic Ising model for $s>0$. In retrospect it is not surprising that later MC data for the Ising model \cite{selke2009} for negative $s$ agreed with the theoretical prediction \cite{dohm2008} for $s \gtrsim -0.6$
but not in the range $-1.5 \lesssim s \lesssim -0.6$ where $\varphi^4$ theory is not applicable.
The instability of the $\varphi^4$ theory at $s_{LP}\approx -1$ was not taken into account in the interpretation \cite{selke2009,kastening2013} of the disagreement
for $s \lesssim -0.6$. In particular it was not recognized  \cite{kastening2013}
that the matrix ${\bf \bar A}_2$, (\ref{33ax}), correctly describes the exact long-distance behavior of $G_b$, (\ref{3nbarAhNull2dx}), of the $\varphi^4$ model for  $s > s_{LP}$, including the asymptotic shape of the correlation length ellipse of the $\varphi^4$ model determined by (\ref{quadratic}). After substitution of (\ref{s}) into  ${\bf \bar A}_2$, (\ref{33ax}), the matrix  ${\bf  Q}_2(q)$, (\ref{33qx}), with $q=\xi_{0 \pm}^{(1)}/\xi_{0 \pm}^{(2)}$
is obtained which has the same form as ${\bf \bar A}_2(r)$ in  Eq. (12) of \cite{kastening2013,footnoteratio}.
We predict that the Binder cumulant of the $d=2$ $\varphi^4$ model differs significantly from that of the Ising model for $ s \lesssim -0.6$. Thus
we expect improved agreement of our  prediction \cite{dohm2008}  for $s<0$  with MC data of the Binder cumulant {\it for the $\varphi^4$ model  rather than for the Ising model}. Corresponding MC simulations for the $d=2$ $\varphi^4$ model would be desirable.
An improved version of the prediction of \cite{dohm2008} is obtained by replacing the matrix ${\bf \bar A}_3(s)$ of Eq. (8.19) of \cite{dohm2008} by the matrices (\ref{33a}) and (\ref{Q33a}) derived in Secs. IV. B and VII where this issue is further discussed in the context of multiparameter and two-scale-factor universality.

\renewcommand{\thesection}{\Roman{section}}
\renewcommand{\theequation}{3.\arabic{equation}}
\setcounter{equation}{0}
\section{Finite-size RG approach}
In a previous finite-size study of the anisotropic $\varphi^4$ theory \cite{dohm2008} with finite $n$,
the analysis was restricted to the case $n=1$ in a hypercubic geometry. In the following we extend this work to general $ n $ in a finite rectangular
$ L_1 \times L_2 \cdot \cdot \cdot \times L_d$ block geometry
with finite aspect ratios, (\ref{aspect}), with applications to  $L_\parallel^{d-1} \times L$ slab geometry including the limit $L_\parallel \to \infty$  to film geometry at finite $L$.
Eq. (\ref{xj})  implies that the rectangular block with a volume $V = \prod^d_{\alpha = 1} L_\alpha = N \tilde a^d$ is transformed into a parallelepiped shape  with the volume
\begin{eqnarray}
\label{volumex}
&&V'= N v'  = V/ (\det {\bf A})^{1/2}  = \prod^d_{\alpha = 1} L'_\alpha,\\
\label{Lalphastrich}
&&L_\alpha' = L_\alpha / (\det {\bf A})^{1/(2d)}, \;\;\; \alpha= 1, 2, ..., d,
\end{eqnarray}
with the transformed lengths $L_\alpha'$. The shear transformation yields the exact relation for the finite system
\begin{eqnarray}
\label{2vvx}
f_s(t, h,\{L_\alpha\}, {\bf A})& =& (\det {\bf A})^{- 1/2} f_s'(t,h',\{L_\alpha'\},  {\bf \bar A})\;\;\;\;
\end{eqnarray}
as a generalization of the hypercubic case \cite{dohm2008}. The aspect ratios (\ref{aspect})
\begin{eqnarray}
\label{rhoLprime}
&&\rho_\alpha=L/L_\alpha=L'/L'_\alpha , \; \rho_d\equiv1,\;\;\;\;\;\; \;\;\;\\
\label{Lprimex}
&&L' = L / (\det {\bf A})^{1/(2d)},
 \end{eqnarray}
with $L_d' \equiv L'$ are left invariant by the shear transformation. We define the geometric mean of the aspect ratios
\begin{equation}
\label{rhobarx}
\bar\rho= \Big[\prod^{d-1}_{\alpha = 1} \;\rho_\alpha\Big]^{1/(d-1)}.
\end{equation}
\subsection {Unrenormalized free energy density}
The perturbation approach for the model I (2.1) at $ h=0$ is based on the decomposition
${\bm\varphi}_j = {\bm\Phi} + {\bm\sigma}_j$
into the lowest-mode (${\bf k}={\bf 0}$) amplitude ${\bm\Phi}$
and into higher-mode contributions ${\bm\sigma}_j = V^{-1} {\sum_{\bf k\neq0}} e^{i{\bf k} \cdot
{\bf x}_j} \hat {\bm \varphi}({\bf k})$.
Since the details are similar to those presented in \cite{dohm2017I} for isotropic systems  we directly start from the partition function
\begin{eqnarray}
\label{2fx}
Z= V^{n/2}\tilde a^{-n} \int  d^n
{\bm\Phi} \exp \left\{- [H_0({{\bm\Phi}}^2)  + \bare{\Gamma}({{\bm\Phi}}^2) ] \right\}\;\;\;
\end{eqnarray}
with the lowest-mode Hamiltonian
\begin{eqnarray}
\label{lowHamilton}
H_0 ({{\bm\Phi}}^2) = V[(r_0/2) {{\bm\Phi}}^2
 + u_0 ({{\bm\Phi}}^2)^2  ]
\end{eqnarray}
and the higher-mode contribution
\begin{eqnarray}
\label{4hneu} {\bare{\Gamma}} ({{\bm\Phi}}^2) &=&  \big[- \;n (N-1) \ln(2 \pi) +VS_0\big(\bar r_{0{\rm L}}({{\bm \Phi}}^2),\{L_\alpha\},{\bf  A}\big) \nonumber\\ &+& (n-1)V S_0\big(\bar r_{0{\rm T}}({{\bm \Phi}}^2),\{L_\alpha\},{\bf  A}\big) \big]/2 \;
\end{eqnarray}
[see I (4.1)-I (4.12)]. The sum $S_0$ over the higher modes can be written as
\begin{eqnarray}
\label{Snullx}
S_0(r,\{L_\alpha\},{\bf  A})&=&\frac{1}{V} {\sum_{\bf k\neq0}} \ln \left\{\left[ r + \delta \widehat K (\mathbf k)\right] \tilde a^2\right\} \\
\label{Snullxdelta}
&=&\Delta(r,\{L_\alpha\},{\bf  A}) -V^{-1}  \ln( r  \tilde a^2)\nonumber\\ &+& \int_{\bf k} \ln \{[r + \delta \widehat K
(\mathbf k)] \tilde a^2\}, \\
\label{bb9xy}
\Delta(r, \{L_\alpha\},{\bf  A}) &=& V^{-1}
{\sum_{\bf k}} \ln
\{[r + \delta \widehat K (\mathbf k)] \tilde a^2\}\nonumber\\
&- &\int_{\bf k} \ln \{[r + \delta \widehat K (\mathbf k)]
\tilde a^2\}. \;
\end{eqnarray}
Here $\delta \widehat K (\mathbf k)$ has the long-wavelength form (\ref{2h}) with the anisotropic matrix ${\bf A}$, (\ref{2i}). The summations ${\sum_{\bf k\neq0}}$ run over the ${\bf k} \equiv (k_1, k_2, \ldots, k_d)$ vectors of the block geometry, with Cartesian components $k_\alpha = 2 \pi m_\alpha / L_\alpha, m_\alpha = 0, \pm 1, \pm 2, \cdots, \alpha = 1,2, \cdots, d$ in the range $ -\pi / \tilde a \leq k_\alpha < \pi / \tilde a$. An asymptotically exact calculation of the function $\Delta(r,  \{L_\alpha\},{\bf  A})$ for $L_\alpha/\tilde a\gg 1$, $0 < r\tilde a^2 \ll 1$,  $0 < r L_\alpha^2 \lesssim O(1)$,  and for finite $0 < \rho_\alpha < \infty$ can be carried out in a way similar to that in \cite{dohm2008}. The result is expressed in terms of $L'$ and the reduced anisotropy matrix ${\bf \bar A}$. It reads
\begin{eqnarray}
\label{F.155yy}
&&\Delta (r,\{L_\alpha\},{\bf  A} ) = \;  L^{-d} {\cal G}_0
(r L'^2, \{\rho_\alpha\}, {\bf \bar A}),\\
\label{calG0x}
&&{\cal G}_0(x, \{\rho_\alpha\}, {\bf \bar A}) \; = \;
\int^\infty_0 dz \; z^{-1} \exp \left[-xz/(4\pi^2)\right] \nonumber\\
&&\;\;\;\;\;\;\;\;\;\;\;\;\;\;\;\;\;\;\times \left\{\left(\pi/z\right)^{d/2} \; - \;{\bar \rho}\;^{d-1}\; K_d (z,
{\bf C}) \right\}.\;\;\;\;\;\;
\end{eqnarray}
The function $K_d (y, {\bf C})$ is defined for $y>0$ by
\begin{eqnarray}
\label{Kd}
K_d (y, {\bf  C}) =\sum_{\bf n} \;\exp (- y
{\bf n} \cdot {\bf  C n})\;
\end{eqnarray}
where the symmetric $d \times d$ matrix ${\bf C}= {\bf C}(\{\rho_\alpha\}, {\bf \bar A})$ has the elements
\begin{eqnarray}
\label{matrixC}
C_{\alpha\beta}=\rho_\alpha \rho_\beta \bar A_{\alpha\beta}.
\end{eqnarray}
The sum $\sum_{\bf n}$ runs over ${\bf n} = (n_1, n_2, ..., n_d) \; , n_\alpha = 0, \pm 1,
..., \pm \infty$.
The function ${\cal G}_0( x, \{\rho_\alpha\}, {\bf \bar A})$ decays exponentially for large $x$ and is logarithmically divergent for $x\to 0_+$.
This divergent part can be separated as
\begin{eqnarray}
\label{calG0decom}
&&{\cal G}_0(x, \{\rho_\alpha\}, {\bf \bar A}) = \;{\bar \rho}\;^{d-1}\;\big[\ln\big(\frac{x}{4\pi^2}\big)+ {\cal J}_0(x, \{\rho_\alpha\}, {\bf \bar A} )\big],\;\;\;\;\nonumber\\\\
\label{calJ3x}
&&{\cal J}_0(x, \{\rho_\alpha\}, {\bf \bar A} )=  \int_0^\infty
dy y^{-1}
  \big\{\exp \left[-xy/(4\pi^2)\right]
  \nonumber\\ &&\times \big[\;{\bar \rho}\;^{1-d}\;
  (\pi/y)^{d/2}
  - \; K_d (y,{\bf C}) + 1 \big]   - \exp (-y)\big\}
\end{eqnarray}
where the function ${\cal J}_0$  has a finite limit ${\cal J}_0(0, \{\rho_\alpha\}, {\bf \bar A} )$ for $x \to 0_+$.
From (\ref{Snullxdelta}), (\ref{F.155yy}), and (\ref{calG0decom}) we obtain
\begin{eqnarray}
\label{c1111u}S_0(r,\{L_\alpha\},{\bf  A}) = \int_{\bf k} \ln \{[r + \delta \widehat K
(\mathbf k)] \tilde a^2\} \nonumber\\
  + V^{-1}\big\{ \ln[L'^2/(4\pi^2\tilde a^2)]+ {\cal
J}_0(r L'^2, \{\rho_\alpha\}, {\bf \bar A})\big\}.
\end{eqnarray}
Note that $L'$, (\ref{rhoLprime}), rather than $L$ appears in the arguments of $\ln$ and of ${\cal J}_0$. The next steps are parallel to those of \cite{dohm2017I}. This leads to the unrenormalized free energy density of the anisotropic system in $2<d<4$ dimensions
\begin{eqnarray}
\label{free prime}
&&f(t,\{L_\alpha\})= f'^{(1)}_{b,ns} (t)\;\nonumber\\ && + (\det {\bf A})^{- 1/2}\;\delta f'(r_0 - r_{0c}, u'_0,L',\{\rho_\alpha\}, {\bf \bar A})\;\;\;\;\\
\label{free bulk nonsing}
&&f'^{(1)}_{b,ns}(t)= \frac{n}{2} \Big\{ -\frac{\ln (2 \pi)} {v'} +\int_{\bf k'} \ln \{[ \delta
\widehat K (\mathbf k')]  (v')^{2/d}\} \nonumber\\&&+ \;\; (r_0 - r_{0c}) \;
\int_{\bf k'} [\delta \widehat K (\mathbf k')]^{-1}\Big\}+ O(u'_0),\;\;\;\;\;\;\;\;\;\;\;\;\;\;\\
\label{deltafprime6}
&&\delta f'(r_0 - r_{0c}, u'_0,L',\{\rho_\alpha\}, {\bf \bar A})= \nonumber\\&&- \frac{1}{V'}\ln  \Big\{ \Big( \frac{2\pi V'}{L'^2}\Big)^{n/2}  \int  d^n {\bm\Phi'}  \exp \big[ - H^{eff}({{\bm\Phi}'}^2)  \big]\Big\},\;\;\;\;\;\;\;\;\;\\
\label{2fff}
&&\int  d^n {\bm\Phi'} \equiv 2 \pi^{n/2} \Gamma \bigg(\frac{n}{2}\bigg)^{-1}\int_0^\infty  d |{\bm\Phi}'||{\bm\Phi}'|^{n-1},
\end{eqnarray}
with ${\bm\Phi}'=(\det {\bf A})^{1/4}{\bm\Phi}$ and a nonsingular bulk part $f'^{(1)}_{b,ns}$ where $v'=(\det {\bf A})^{- 1/2}\tilde a^d$. The effective Hamiltonian in (\ref{deltafprime6}) reads
\begin{eqnarray}
\label{effHam}
H^{eff}({{\bm\Phi}'}^2)&=& \widehat H'_0({{\bm\Phi}'}^2) +{\bare{\Gamma}'}_{\rm L}({{\bm\Phi}'}^2) + (n-1){\bare{\Gamma}'}_{\rm T}({{\bm\Phi}'}^2) ,\nonumber\\\\
\label{lowHamilton1}
\widehat H'_0({{\bm\Phi}'}^2)&=&V'\left[(1/2) (r_0 - r_{0c}){{\bm\Phi}'}^2 + u'_0 {({{\bm\Phi}'}^2)}^2  \right],\;\;\;\;\;\;\;\;\;\\
\label{Gammalongbare}
{\bare{\Gamma}'}_{\rm L}({{\bm\Phi}'}^2)&=& -[V'A_d/(d\varepsilon)]{r_{0{\rm L}}'({{\bm\Phi}'}^2)}^{d/2} \nonumber\\&+& (1/2){\cal J}_0( r_{0{\rm L}}'({{\bm\Phi}'}^2){ L'}^2, \{\rho_\alpha\}, {\bf \bar A}),\;\;\;\\
\label{Gammatransbare}
{\bare{\Gamma}'}_{\rm T}({{\bm\Phi}'}^2)&=& -[V'A_d/(d\varepsilon)]{r_{0{\rm T}}'({{\bm\Phi}'}^2)}^{d/2} \nonumber\\&+& (1/2){\cal J}_0( r_{0{\rm T}}'({{\bm\Phi}'}^2){ L'}^2, \{\rho_\alpha\}, {\bf \bar A}),\;\;\;
\end{eqnarray}
with the ${{\bm\Phi}'}^2$ dependent parameters
\begin{subequations}
\label{rwidetilde}
\begin{align}
\label{rlongprime}
 r_{0{\rm L}}'({{\bm\Phi}'}^2)&=&r_0 - r_{0c}+12u'_0{{\bm\Phi}'}^2,
\\
\label{rtransprime}
r_{0{\rm T}}'({{\bm\Phi}'}^2)&=&r_0 - r_{0c}+4u'_0{{\bm\Phi}'}^2.
\end{align}
\end{subequations}
\subsection{Renormalized free energy density}
The quantity $\delta f'$ in (\ref{deltafprime6}) is expressed entirely in terms of quantities of the transformed isotropic system. Its multiplicative and additive renormalizations are the same as for the corresponding bulk quantity  $\delta f'_b(r_0 - r_{0c}, u'_0)$ in Sec. II since $L',\rho_\alpha$, and  ${\bf \bar A}$ are not renormalized. This ensures that the confined anisotropic system has the same critical exponents as the isotropic bulk system. Thus we employ the minimal subtraction scheme at fixed dimension for isotropic systems \cite{dohm1985} and define the renormalized counterpart of $\delta f'$ in $2<d<4$ dimensions as
\begin{eqnarray}
\label{5ddfinite}  && f'_R(r', u',L',\{\rho_\alpha\},\mu, {\bf \bar A}) = \nonumber\\ && \delta f'(Z_{r'}r', \mu^{\varepsilon}Z_{u'}Z_{\varphi'}^{-2}A_d^{-1}u',L',\{\rho_\alpha\}, {\bf \bar A}) \nonumber\\ && - (1/8)\mu^{-\varepsilon} r'^2 A_d\; A(u',\varepsilon) ,
\end{eqnarray}
where $u',r'$ are defined in (\ref{renormalized parameters}) and the renormalization constants $Z_{r'}(u',\varepsilon)$, $Z_{u'}(u',\varepsilon)$, $Z_{\varphi'}(u',\varepsilon)$, and $A(u',\varepsilon)$ are the same as in I (3.25) with $u$ replaced by $u'$. We take the same choice (\ref{my}) for the inverse reference length $\mu^{-1}$ as for the bulk theory. The subsequent treatment is parallel to that in \cite{dohm2017I}. This leads to
\begin{eqnarray}
\label{5aaaL} f'_R(r', u',L',\{\rho_\alpha\}, \mu, {\bf \bar A}) &&  = f'_R \big(r'(l),u'(l),L',\{\rho_\alpha\}, l\mu, {\bf \bar A} \big) \nonumber\\ && + \;A_d r'(l)^2{\cal B}'(l)/[2(l\mu)^\varepsilon],
\end{eqnarray}
with
\begin{eqnarray}
\label{approx renorm fprime-s-l}   && f'_R(r'(l), u'(l),L',\{\rho_\alpha\}, {\bf \bar A},l\mu) \nonumber\\&&= \frac{n}{2V'} \ln\Bigg\{\frac{{(l\mu L')}^{\varepsilon/2} {[\Gamma(n/2)]}^{2/n}{u'(l)}^{1/2}  }{2 \pi^2 A_d^{1/2}}\bar \rho^{(d-1)/2} \Bigg\}\nonumber\\&&-\frac{A_d}{{L'}^d }\Bigg\{\frac{(l\mu L')^d}{4d} -\frac{(n-1)}{\varepsilon}\Big[\frac{l_{\rm T}^2}{4(l\mu L')^\varepsilon}-\frac{l_{\rm T}^{d/2}}{d}\Big]\Bigg\}\nonumber\\&&+\frac{1}{2V'}\Big[{\cal J}_0( l^2\mu^2 {L'}^2, \{\rho_\alpha\}, {\bf \bar A})+ (n-1){\cal J}_0( l_{\rm T}, \{\rho_\alpha\}, {\bf \bar A})\Big]\nonumber\\&&- \frac{1}{V'}\; \ln  \Bigg\{  2 \int_0^\infty  ds s^{n-1} \exp \Big[ -\frac{1}{2} y'(l) s^2-s^4    \Big]\Bigg\},\;\;\;\; \;\;\;\;\;\;\;
\end{eqnarray}
\begin{eqnarray}
\label{ypsilonflowx}
&&y'(l)= \frac{r'(l)\; {(l\mu L')}^{d/2}  A_d^{1/2}}{(l\mu)^2 \;u'(l)^{1/2}}\;\;\bar \rho^{(1-d)/2} ,\\
\label{lT}
&&l_{\rm T}(t,L',\bar \rho) = l^2\mu^2 {L'}^2\nonumber\\&&-\;8\;\Big[(l\mu L')^\varepsilon\; u'(l) A_d^{-1}\;\bar \rho^{(d-1)/2}\Big]^{1/2}\;\vartheta_{2,n} (y'(l)),\;\;\;\;\;\;\;\;\;\;\\
\label{thetax}
&&\vartheta_{2,n} (y) = \frac{\int_0^\infty d s \;s^{n+1} \exp (-
\frac{1}{2} y s^2 - s^4)} {\int_0^\infty ds s^{n-1} \; \exp (-
\frac{1}{2} y s^2 - s^4)} \;,\\
\label{calB}
&&{\cal B}'(l) \equiv
\int_1^l B(u'(l')) \nonumber\\
&&\times \Big\{\exp\int_l^{l'}\Big[2\zeta_{r'}(u'(l'')) -
\varepsilon\Big]\frac{dl''}{l''}\Big\}\frac{dl'}{l'}.
\end{eqnarray}
The flow parameter $l$ is determined implicitly by
\begin{eqnarray}
\label{5v} \mu^2l^2&=& r'(l) +12\;\big[r'(l)}/{y'(l)\big]\;\vartheta_{2,n} (y'(l)),\\
\label{r}
r'(l) &=& r' \exp\Big[\int_1^l\zeta_{r'}(u'(l'))\frac{dl'}{l'}\Big],
\end{eqnarray}
with $r'(1)=r'=a't$, $a'=Z_{r'}(u',\varepsilon)^{-1}a_0$, $u'(1)=u'$.
In the regions $|y'| \gg 1$ of low and high temperatures far from $T_c$, the flow parameter approaches the bulk choices $l_-(t)$ and $l_+(t)$, respectively
\begin{eqnarray}
\label{bulk flow parameter}  \mu^2l^2 = \left\{
\begin{array}{r@{\quad \quad}l}
                         \mu^2 l_+^2 = r'(l_+)& \mbox{for} \;T > T_c ,\\
                         \mu^2 l_-^2 = -2r'(l_-) & \mbox{for} \;T <
                 T_c .
                \end{array} \right.
\end{eqnarray}
The result (\ref{5aaaL})-(\ref{r}) describes the crossover of the renormalized free energy density $f'_R$
at finite volume $V'$ for general $n$ from far below to far above $T_c$ including the critical regime, in agreement with I (4.52) and I (4.81) for the isotropic case, ${\bf \bar A}={\bf 1}$.
In the bulk limit we obtain
\begin{eqnarray}
\label{bulklimit}
f'_R(r'(l), u'(l),L',\{\rho_\alpha\}, {\bf \bar A},l\mu) \longrightarrow f'^\pm_{R,b} \big(l_\pm \mu,u'(l_\pm)\big)\;\;\;\;\;
\end{eqnarray}
where $f'^\pm_{R,b}$ is given by the one-loop bulk expressions I (3.40) and I (3.41), with $u$ replaced by $u'$.
\renewcommand{\thesection}{\Roman{section}}
\renewcommand{\theequation}{4.\arabic{equation}}
\setcounter{equation}{0}

\section{Results and applications}
\subsection{Finite-size scaling function and multiparameter universality}
In order to derive the asymptotic scaling forms of  $f'_{s}$ and of  $f_{s}$
we consider the limit of  a small flow parameter $l \ll 1$ or $ l\rightarrow 0$. In this limit we obtain
\begin{eqnarray}
\label{ustern}
&&u'(l) \rightarrow\;u'(0)\equiv {u'}^*=  u^*,\\
\label{6jx}
&&{\cal B}'(l) \rightarrow\; - \;(\nu/\alpha) \;B(u^*) , \;\;\\
\label{Qstern}
&& r'(l)/(\mu^2l^2)\rightarrow \;Q^*\; t
\;l^{-1/\nu}=Q^*\tilde x (\mu l L')^{-1/\nu},\\
\label{6cc}
&&y'(l) \rightarrow \; \tilde y = \tilde x \;Q^*\; \big[(\mu
l L')^{- \alpha /\nu} A_d \;{u^*}^{-1}\bar\rho\;^{1-d}\big]^{1/2}\;\;\;\;\;\;
\end{eqnarray}
where $\bar\rho$ is defined in (\ref{rhobarx}) and where
$\tilde x = t\;(L'/\xi'_{0+})^{1/\nu}$
is the scaling variable. Eqs. (\ref{ypsilonflowx}) and (\ref{5v}) imply asymptotically
$\mu l L' \rightarrow \; \tilde l = \tilde l(\tilde x,\bar \rho)$
where the scaling function $\tilde l(\tilde x,\bar \rho)$ is determined implicitly by
\begin{subequations}
\label{implicit}
\begin{align}
\label{6ffx} \tilde y + 12 \vartheta_{2,n}(\tilde y) =
\big[\tilde l^{d} A_d \;{u^*}^{-1}\bar \rho\;^{1-d}\big]^{1/2},
\\
\label{6ggx} \tilde y =\; \tilde x\;Q^*\;\big[\tilde l^{- \alpha /\nu} A_d \;{u^*}^{-1}\bar \rho\;^{1-d}\big]^{1/2}.
\end{align}
\end{subequations}
These two equations also determine the scaling function $\tilde y = \tilde
y(\tilde x,\bar\rho)$. The quantity $l_{\rm T}$, (\ref{lT}), becomes asymptotically $l_{\rm T}(t,L',\bar \rho)\to \tilde l_{\rm T}(\tilde x,\bar\rho)$ with
\begin{eqnarray}
\label{lTasymx}
\tilde l_{\rm T}(\tilde x,\bar \rho) = \tilde l^2-\;8\;\big[\tilde l^\varepsilon\; u^* A_d^{-1}\;\bar \rho\;^{d-1} \big]^{1/2}\;\vartheta_{2,n} (\tilde y).\;\;\;\;\;\;\;\;
\end{eqnarray}
Eqs. (\ref{6ffx})-(\ref{lTasymx}) have the same form as I (5.5a)-I (5.6) which implies that  $\tilde l(\tilde x,\bar\rho)$ is the same function as in \cite{dohm2017I}, but with a different definition of $\tilde x$ [see (\ref{tildexx})].
From (\ref{5aaaL}), (\ref{approx renorm fprime-s-l}), and (\ref{ustern}) - (\ref{lTasymx}) we derive the asymptotic scaling form for the {\it isotropic} (parallelepiped) system
\begin{eqnarray}
\label{6kx}  f'_R(r, u',L',\{\rho_\alpha\}, \mu, {\bf \bar A}) \;\,\; \longrightarrow \nonumber\\ f'_{s}(t,L',\{\rho_\alpha\}, {\bf \bar A})  = L'^{-d}
F(\tilde x,\{\rho_\alpha\}, {\bf \bar A})
\end{eqnarray}
with the scaling function $F$ given in (\ref{scalfreeaniso}) below. In order to obtain
$f_{s}$ for the {\it anisotropic} system in block geometry we use (\ref{2vvx}), (\ref{Lprimex}), and (\ref{ximean}).
This leads to
\begin{eqnarray}
\label{6k}  f_{s}(t,L,\{\rho_\alpha\}, {\bf  A})  = L^{-d}
F \Big(t
(L/\bar \xi_{0 +})^{1/\nu},\{\rho_\alpha\}, {\bf \bar A}\Big)\;\;\;
\end{eqnarray}
with the same scaling function $F$ as in  (\ref{6kx}) and (\ref{scalfreeaniso}) where now the scaling variable is expressed in terms of the mean correlation-length amplitude $\bar \xi_{0 +}$ and the physical length $L$. The equivalence of both representations is due to the fact that the  scaling variable $\tilde x$ can be expressed in two different ways,
\begin{eqnarray}
\label{tildexx}
t\;(L'/\xi'_{0+})^{1/\nu}&=&\;t\;(L/\bar \xi_{0+})^{1/\nu}\;\equiv\;\tilde x,
\end{eqnarray}
which follows from (\ref{ximean}) and (\ref{Lprimex}). The latter form $\tilde x =t(L/\bar \xi_{0+})^{1/\nu}$ is more appropriate for $f_s$, (\ref{6k}), since  $\bar \xi_{0 +}$  and $L$ are observable quantities of the anisotropic system whereas  the quantities $\xi'_{0 +}$ and $L'$ of the transformed isotropic system are not measurable \cite{diehl-cham}. The explicit result for $F$ reads for $2<d<4$
\begin{eqnarray}
\label{scalfreeaniso} &&F(\tilde x,\{\rho_\alpha\}, {\bf \bar A})= -\; A_d \;\Big\{
\frac{\tilde l^d}{4d} \;- \frac{(n-1)}{\varepsilon}\Big[\frac{{\tilde l}^2_{\rm T}}{4 \tilde l^\varepsilon }-\frac{{\tilde l_{\rm T}}^{d/2}}{d}\Big] \nonumber\\ && + \;\nu\;{Q^*}^2 \tilde x^2 \tilde
l^{- \alpha/\nu}\;B(u^*)/2\alpha \Big\}  \nonumber\\ && +\;\bar\rho\;^{d-1} \Big\{\frac{n}{2}\ln\Big[\frac{\tilde l^{\varepsilon/2} {[\Gamma(n/2)]}^{2/n}{u^*}^{1/2}  }{2 \pi^2 A_d^{1/2}} \;\bar\rho^{(d-1)/2}\Big]\nonumber\\ &&   -\ln  \Big[  2 \int_0^\infty  ds s^{n-1} \exp \big( -y(\tilde x, \bar\rho) s^2/2-s^4    \big)\Big]\nonumber\\ &&  + \frac{1}{2}{\cal J}_0( {\tilde l}^2, \{\rho_\alpha\}, {\bf \bar A})+  \frac{n-1}{2}{\cal J}_0( \tilde l_{\rm T},\{\rho_\alpha\}, {\bf \bar A})\Big\}\;\,\;\;\;\;\,\;\;\;
\end{eqnarray}
where $\bar\rho$ and ${\cal J}_0$ are defined by (\ref{rhobarx}) and (\ref{calJ3x}). The crucial information on the anisotropy is contained in ${\cal J}_0$ via the sum (\ref{Kd}) with the matrix ${\bf C}(\{\rho_\alpha\}, {\bf \bar A})$, (\ref{matrixC}). Eq. (\ref{scalfreeaniso}) is valid for general $n\geq 1$. It
describes the entire crossover from far below $T_c$ to far above $T_c$ and permits us to predict the effect of lattice anisotropy on the free energy and the Casimir force in the Goldstone-dominated low-temperature region as well as in the critical and high-temperature regions. For finite $L$ and $\rho_\alpha$, $F(\tilde x, \{\rho_\alpha\}, {\bf \bar A})$  is an analytic function of  $\tilde x$ near $\tilde x=0$, in agreement with general analyticity requirements.

The bulk part $F^\pm_b(\tilde x)$ of  $F(\tilde{x},\{\rho_\alpha\}, {\bf \bar A})$  is obtained from (\ref{scalfreeaniso}) in the limit of large $|\tilde x|$ as
\begin{eqnarray}
\label{3jjbulk}
F^\pm_b(\tilde{x}) =\left\{
\begin{array}{r@{\quad \quad}l}
                         \; Q_1 \tilde x^{d \nu}\quad          & \mbox{for} \;T > T_c\;, \\
                         \; (A^-/A^+)Q_1\mid\tilde{x}\mid^{d\nu}& \mbox{for} \;T <
                 T_c ,\;
                \end{array} \right.
\end{eqnarray}
where $Q_1$ and $A^-/A^+$ are the same as those for the isotropic system given in I (3.56) and I (3.57). This leads to the scaling form of the excess free energy density and the critical Casimir force of the anisotropic system
\begin{eqnarray}
\label{3k}
&&f^{ex}_{s}(t,L,\{\rho_\alpha\}, {\bf A})= L^{-d}F^{ex}(\tilde{x},\{\rho_\alpha\}, {\bf \bar A}),\;\;\;\;\;\;\;\;\;\;\;\;\;\;\;\;\;\;\;\;\;\;\\
\label{3kkx}
&&F^{ex}(\tilde{x},\{\rho_\alpha\}, {\bf \bar A})= F(\tilde{x},\{\rho_\alpha\}, {\bf \bar A}) - F^\pm_b(\tilde{x}),\\
\label{3l}
&&{F}_{\text Cas}(t,L,\{\rho_\alpha\}, {\bf A})=L^{-d}X(\tilde x, \{\rho_\alpha\}, {\bf \bar A}),\\
\label{3nn} &&X(\tilde{x},\{\rho_\alpha\}, {\bf \bar A}) =(d-1)  F^{ex} (\tilde{x},\{\rho_\alpha\}, {\bf \bar A})\nonumber \\ && -
\frac{\tilde{x}}{\nu}
\frac{\partial F^{ex}(\tilde{x},\{\rho_\alpha\}, {\bf \bar A})}{\partial\tilde{x}}- \sum_{\alpha=1}^{d-1}\rho_\alpha
\frac{\partial F^{ex}(\tilde{x},\{\rho_\alpha\}, {\bf \bar A})}{\partial \rho_\alpha}.
\end{eqnarray}
From previous studies at finite external field \cite{cd-1997,cd-2000} we infer that the extension of (\ref{6kx}) and (\ref{6k}) to finite $h'$ and $h$  has the structure
\begin{eqnarray}
\label{4a2}
&&f'_s (t,h', L',\{\rho_\alpha\},{\bf  \bar A})  = L'^{-d} \;F\big( \tilde x, \tilde x_h,\{\rho_\alpha\}, {\bf \bar A}\big),\;\;\;\;\;\;\;\;\\
\label{1cc}
&&f_s (t, h, L,\{\rho_\alpha\}, {\bf A})=  L^{-d} \;  F \big(\tilde x, \tilde x_h,\{\rho_\alpha\}, {\bf \bar A}\big),
\end{eqnarray}
where we have used
\begin{eqnarray}
\label{hprime}
 h'\; (L'/\xi_c')^{\beta\delta/\nu}
=\; h\;( L/\bar \xi_c)^{\beta\delta/\nu}\equiv  \tilde x_h
\end{eqnarray}
as follows from (\ref{2oox}) and (\ref{meancorampcx}).  Thus the same finite-size scaling function $F$ determines both $f_s'$ and $f_s$, i.e., the shear transformation does not change the structure of the scaling function but changes the form of its scaling arguments according to (\ref{tildexx}) and (\ref{hprime}). The dependence on $\rho_\alpha$ and ${\bf \bar A}$ remains unchanged.
As noted for the {\it bulk} properties in Sec. II, there exist two parameterizations (i) and (ii) of ${\bf \bar A}$  in terms of
the couplings $K_{i,j}$ or of the ratios
of the correlation-length amplitudes $\xi_{0+}^{(\alpha)}$.

Comparison with the Privman-Fisher scaling form (\ref{1b}) shows that $f_s$ as given by (\ref{6k}), (\ref{scalfreeaniso}), and (\ref{1cc}) violates two-scale-factor universality, as noted already in \cite{cd2004,dohm2006,chen-zhang,dohm2009PJ,dohm2008,DG}. In addition to the two nonuniversal metric factors,
\begin{eqnarray}
\label{constantsC}
C_1=(\bar\xi_{0+})^{-1/\nu}, \;\;\;C_2=(\bar \xi_c)^{-\beta\delta/\nu},
\end{eqnarray}
the scaling function $F$ depends on the $d(d+1)/2-1$  independent anisotropy parameters contained in ${\bf \bar A}$, contrary to the requirement that  ``the metric factors $C_1$ and $C_2$ are the {\it only} nonuniversal system-dependent parameters entering" \cite{pri}. For the violation of two-scale-factor universality it does not matter which parameterizations of ${\bf \bar A}$ are used.

Nevertheless,
some degree of universality is maintained in the finite-size scaling function $F$, (\ref{scalfreeaniso}), in analogy to multiparameter universality for {\it bulk} properties discussed in Sec. II.  It is natural to expect that this feature can be extended to weakly anisotropic {\it confined} systems, similar to the earlier extension of two-scale-factor universality from isotropic bulk systems \cite{aharony1974,ger-1,hohenberg1976,weg-1} to isotropic confined systems \cite{priv,pri}. In \cite{dohm2006,dohm2008} it was already hypothesized that
the scaling function $F$ has universal features in that it is valid not only for $\varphi^4$ lattice models but also for all weakly anisotropic systems with the same geometry, the same BC, and the same  matrix ${\bf \bar A}$ but no proof for the validity of this hypothesis was given.
As suggested by the exact confirmation through the results for the $(d=2,n=1)$ universality class in   (\ref{3oxicratiox}) and (\ref{xiratioisingx}) and for reasons of consistency with the universal validity of the critical bulk relations (\ref{3u})-(\ref{principalratio}) in terms of correlation lengths,
the parametrization (ii) in the form of ${\bf \bar A}\big(\{\xi_{0+}^{(\alpha)},{\bf e}^{(\alpha)}\}\big)$, (\ref{Anonunix}),  is required when $F$, (\ref{scalfreeaniso}), is applied to systems other than $\varphi^4$ models. Employing correlation lengths rather than couplings in the description of finite-size effects is in line with earlier representations of $F$ and $X$ for anisotropic Gaussian, Ising, and spherical models in film geometry \cite{Indekeu,kastening-dohm, DG,kastening2012}, see also Secs. V-VII. With this specification of ${\bf \bar A}$, the previous hypothesis of restricted universality for confined anisotropic systems \cite{dohm2006,dohm2008} becomes identical with our present unified hypothesis of multiparameter universality for {\it both} bulk and confined anisotropic systems. As long as  MC simulations only of $\varphi^4$ lattice models \cite{hasenbusch2010,Hasenbuschgesamt} are used for testing the validity of $F$, (\ref{scalfreeaniso}), both parametrizations (i) and (ii) of ${\bf \bar A}$ can be employed. Support for the validity of multiparameter universality comes also from the approximate agreement between $\varphi^4$ theory \cite{dohm2008,kastening2013} and MC data for the critical Binder cumulant of an anisotropic  Ising model  \cite{selke2005,selke2009}, as discussed in Sec. VII.

For systems with unknown bulk properties a meaningful analysis of finite-size data is possible only after a study of the bulk correlation function (Sec. II). The latter provides the information on the principal correlation lengths $\xi_{0 +}^{(\alpha)}$ and $\bar \xi_{0 +}$ needed for plotting of the finite-size data as a function of the scaling variable depending on $\bar \xi_{0 +}$. For a comparison with the theoretical prediction (\ref{scalfreeaniso}), the knowledge of the principal axes ${\bf e}^{(\alpha)}$ is required in order to construct the anisotropy matrix ${\bf \bar A}\big(\{\xi_{0+}^{(\alpha)},{\bf e}^{(\alpha)}\}\big)$, (\ref{Anonunix}), through ${\bf U}(\{{\bf e}^{(\alpha)}\}$) and ${\bf \bar{\mbox {\boldmath$\lambda$}}} \big(\{\xi_{0 +}^{(\alpha)}\}\big)$, (\ref{lambdaquerdef}), entering the scaling function $F$. Examples will be given below.

Similar to the bulk case,  we emphasize that the finite-size scaling form (\ref{scalfreeaniso}) satisfying multiparameter universality
still has a nonuniversal character as it is not only a function of nonuniversal correlation-length amplitudes through ${\bf \bar{\mbox {\boldmath$\lambda$}}} \big(\{\xi_{0 +}^{(\alpha)}\}\big)$ but also exhibits a nonuniversal directional finite-size dependence through ${\bf U}\big(\{{\bf e}^{(\alpha)}\}\big)$ similar to that of the  bulk case discussed in Sec. II, no matter what kind of representation of  ${\bf \bar{\mbox {\boldmath$\lambda$}}}$  is used.
The critical Casimir amplitude $X(0,\{\rho_\alpha\}, {\bf \bar A})$ is the most prominent example of a nonuniversal quantity of weakly anisotropic systems (see Fig. 2 below). In conclusion, the customary claim
that the scaling function of the critical Casimir force depends only on a few general and global properties is not valid for the subclass of anisotropic systems. The numerical confirmation \cite{hucht,vasilyev2009,hasenbusch2010} of the "universal" scaling function $X$ observed experimentally in isotropic $^4$He \cite{garcia} was possible only because {\it isotropic} $XY$ and $\varphi^4$ lattice models \cite{commentiso} were chosen for the MC simulations. For anisotropic models  of the same universality class this scaling function is changed significantly (Sec. V).

Our results
for a finite block geometry with $0 <\rho_\alpha < \infty$ include the case of a finite $L_\parallel^{d-1}\times L$ slab geometry  by setting $\rho_\alpha = \rho \equiv L/L_\parallel=L'/L'_\parallel>0, \alpha = 1, ..., d-1$ and replacing $\bar \rho$ by $\rho$. The necessary substitution is given in (\ref{calG3rhox}) and (\ref{calJ3rhox}). The case of film geometry is then obtained by letting $\rho \to 0$ at finite $L$ where (\ref{scalfreeaniso}) becomes singular at $\tilde x=0$ corresponding to an unshifted film critical point. This case will be separately discussed in Secs. V and VI.
Also the case of cylindrical geometry can be described by setting $L_\alpha= L_\parallel, \alpha = 1, ..., d-1$ and letting $L_d \to \infty$ at finite $L_\parallel$; this case was treated in \cite{dohm2011,chen-zhang} and will not be considered further in this paper.

We briefly comment on the result in Eq. (6.10) of \cite{dohm2008} which was restricted to the case $n=1$ in hypercubic $L^d$ geometry in $2<d<4$ dimensions. This result was expressed in terms of the scaling variable $\tilde x = t(L'/\xi'_{0+})^{1/\nu}$ of the transformed system which is equivalent to $\tilde x = t(L/\bar \xi_{0+})^{1/\nu}$ according to (\ref{tildexx}). As discussed in Sec. V. H of \cite{dohm2017I} for the isotropic case, our present result (\ref{scalfreeaniso}) applied to $n=1$ in a hypercubic geometry is a simplified version of Eq. (6.10) of \cite{dohm2008}.  Our present result avoids unsystematic terms which lead to an unreliable temperature dependence well below $T_c$ as mentioned in Sec. X. A of \cite{dohm2008}, compare Fig. 17 of \cite{dohm2008}, Fig. 11 of \cite{dohm2011}, and Fig. 11 of \cite{dohm2017I} for slab geometry, see also \cite{dohm2010erratum}.
\subsection {Two-dimensional anisotropy}
At present it is not known how to perform quantitative finite-size calculations for the $\varphi^4$ model in {\it two} dimensions. It was suggested \cite{dohm2008}, however, to incorporate a two-dimensional anisotropy of the type shown in Fig. 11 (a) of \cite{dohm2008} in a three-dimensional $\varphi^4$ model. Here we generalize and improve this suggestion. Consider a $d=2$ $\varphi^4$ lattice model with an anisotropy in the $x$-$y$ plane described by the matrix
${\bf A}_2
=
\left(\begin{array}{ccc}
 a & c \\
 c & b \\
\end{array}\right)$
where $\det{\bf A}_2>0$ and ${\bf \bar A}_2= {\bf A}_2/(\det{\bf A}_2)^{1/2}$. As an appropriate extension to $d=3$ we define the $d=3$ anisotropy matrix
\begin{align}
\label{extmatrix3}
{\bf A}_3
=
\left(\begin{array}{ccc}
 {\bf A}_2 & 0 \\
 0 & (\det{\bf A}_2)^{1/2} \\
\end{array}\right)
\end{align}
with $\det{\bf A}_3=(\det{\bf A}_2)^{3/2}$. The reduced matrix is
\begin{align}
\label{barextmatrix3}
{\bf \bar A}_3= \frac{{\bf A}_3}{(\det{\bf A}_3)^{1/3}}
=
\left(\begin{array}{ccc}
 {\bf \bar A}_2 & 0 \\
 0 & 1 \\
\end{array}\right)
\end{align}
which describes a three-dimensional system where the two-dimensional anisotropy is incorporated without an anisotropy in the $z$-direction. Now $d=3$ $\varphi^4$ theory can be employed in describing the anisotropy effects in the $x$-$y$ planes of this system  which are expected to be similar to those of the $d=2$ system. This strategy was successfully employed previously \cite{dohm2008,kastening2013} for the example of the Binder cumulant of the anisotropic $d=2$ model of Sec. II. E, (\ref{simplematrixx}). This issue is further discussed in Sec. VII.
\subsection {Three-dimensional anisotropy}
We consider two types $(I)$ and $(II)$ of anisotropic $\varphi^4$ models on a three-dimensional simple-cubic lattice which have a diagonal and a nondiagonal
anisotropy matrix ${\bf A}$, respectively. According to (\ref{lambdaalphaxix}), the diagonal matrix $  {{\mbox {\boldmath$\bar \lambda$}}}$ can be represented in both cases as
\begin{eqnarray}
 {{\mbox {\boldmath$\bar \lambda$}}}
\label{lambdaxi}
& =& \; 
\left(\begin{array}{ccc}
 \bar \xi_1^2 & 0 & 0 \\
  0 & \bar\xi^2_2 & 0 \\
  0 & 0 & \bar\xi^2_3\\
\end{array}\right) \;
\end{eqnarray}
with  $\bar \lambda_i =\bar \xi^2_i\equiv (\xi^{(i)}_{0+}/\bar \xi_{0+})^2$.
The simplest realization of type $(I)$ is a model with NN couplings $K_x=K_y\equiv K_\parallel=J_\parallel/\tilde a^2$ and $K_z \equiv K_\perp=J_\perp/\tilde a^2$
with an anisotropy matrix
\begin{align}
\label{simplematrix}
{\bf A}_{(I)}
&=
2\left(\begin{array}{ccc}
J_\parallel & 0 & 0 \\
0 & J_\parallel & 0 \\
0 & 0 & J_\perp \\
\end{array}\right),
\end{align}
with eigenvalues $\lambda_x= \lambda_y=2J_\parallel >0, \lambda_z=2J_\perp >0$ and $\det {\bf A}_{(I)}=8J_\parallel^2 J_\perp$. The eigenvectors are obviously parallel to the Cartesian axes. This implies ${\bf U}_{(I)} = {\bf 1}$ and ${\bf \bar A}_{(I)}=  {{\mbox {\boldmath$\bar \lambda$}}}$.
We must distinguish two different bulk correlation lengths
$\xi_{+\parallel}(t)=\xi^\parallel_{0+} t^{-\nu}$ and $\xi_{+\perp}(t)=\xi^\perp_{0+} t^{-\nu}$ above $T_c$. The mean correlation-length amplitude entering the scaling variable $\tilde x$, (\ref{tildexx}), is
$\bar \xi_{0+}=\big[(\xi^\parallel_{0+})^2\;\xi^\perp_{0+}\big]^{1/3}$.
The reduced anisotropy matrix is
\begin{align}
\label{simplematrixquer1}
{\bf \bar A}_{(I)}
=
\left(\begin{array}{ccc}
(J_\parallel/J_\perp)^{1/3} & 0 & 0 \\
0 & (J_\parallel/J_\perp)^{1/3} & 0 \\
0 & 0 & (J_\perp/J_\parallel)^{2/3} \\
\end{array}\right)\\
\label{simplematrixquer2}
=
\left(\begin{array}{ccc}
(\xi^\parallel_{0+}/\xi^\perp_{0+})^{2/3} & 0 & 0 \\
0 & (\xi^\parallel_{0+}/\xi^\perp_{0+})^{2/3} & 0 \\
0 & 0 & (\xi^\perp_{0+}/\xi^\parallel_{0+})^{4/3} \\
\end{array}\right)
\end{align}
according to (\ref{3qx}). This kind of anisotropy was studied in \cite{shenoy1995,DG,kastening-dohm} where both representations were used. There exists a large variety of different realizations of this matrix structure if
one allows for additional pair interactions beyond nearest neighbors along the
Cartesian axes. In this case the couplings  $J_\parallel$ and $J_\perp$ are
replaced by a sum of couplings without changing the structure of ${\bf A}_{(I)}$.

An example of type $(II)$ is a model with the same NN couplings
as in model $(I)$
and a NNN coupling $K_d=J_d/\tilde a^2$ along the diagonals in the $x$-$y$ planes, as illustrated in Fig. 11 of \cite{dohm2008}.
The anisotropy matrix is
\begin{equation}
 \label{33c}
 {\bf A}_{(II)}  = \; 2 \left(\begin{array}{ccc}
  J_\parallel+J_d & J_d & 0 \\
  J_d & J_\parallel+J_d & 0 \\
  0 & 0 & J_\perp \\
\end{array}\right) \; .
\end{equation}
The eigenvalues $ \lambda_1  =  2
(J_\parallel + 2 J_d)$, $ \lambda_2  =  2 J_\parallel $, $\lambda_3  = 2J_\perp$ are positive in the range  $ -\frac{1}{2} < J_d/J_\parallel <
\infty$, $J_\parallel>0$, $J_\perp>0$. There are three different principal bulk correlation lengths $\xi^{(i)}(t)=\xi_i t^{-\nu}$, $i=1,2,3$ above $T_c$ where we use the abbreviation $\xi_i\equiv \xi^{(i)}_{0+}$. The mean correlation-length amplitude entering the scaling variable $\tilde x$, (\ref{tildexx}), is
$\bar \xi_{0+}=(\xi_1\xi_2\xi_3)^{1/3}$.
The eigenvectors of ${\bf A}_{(II)}$ and the ensuing matrices ${\bf U}$ and ${\bf \bar A}_{(II)}$ are
\begin{eqnarray}
 \label{33h}
 &&{\bf e}^{(1)} = \frac{1}{\sqrt{2}}\left(\begin{array}{c}
  1 \\
  1 \\
  0 \\
\end{array}\right) , {\bf e}^{(2)} = \frac{1}{\sqrt{2}}\left(\begin{array}{c}
  -1 \\
  1 \\
  0 \\
\end{array}\right) ,{\bf e}^{(3)} = \left(\begin{array}{c}
  0 \\
  0 \\
  1 \\
\end{array}\right) ,\nonumber\\
 \label{orthomatrix}
 &&{\bf U}  = \; \frac{1}{\sqrt 2} \left(\begin{array}{ccc}
  1 & 1 & 0 \\
  -1 & 1 & 0 \\
  0 & 0 & \sqrt 2\\
\end{array}\right) ,\;\;\;\;\;\;\;\;\;\;\\
 \label{33dxxy}
 &&{\bf \bar A}_{(II)}  = \;[R(1-s^2)]^{-1/3}\;  \left(\begin{array}{ccc}
  1 & s & 0 \\
  s & 1 & 0 \\
  0 & 0 & R \\
\end{array}\right) ,\\
\label{Acorry}
 &&= \;
 \frac{1}{2(\xi_1\xi_2\xi_3)^{2/3}}\left(\begin{array}{ccc}
   \xi^2_1+  \xi^2_2\;\;\;& \xi^2_1-  \xi^2_2 & 0 \\
  \xi^2_1-  \xi^2_2 \;\;\;& \xi^2_1+\xi^2_2 \;\;& 0 \\
  0 & 0 & 2\xi^2_3\\\;\;\;\;\;
\end{array}\right)\;\;\;
\end{eqnarray}
with the anisotropy parameter $s$, (\ref{s}) with $J\equiv J_\parallel$, and \cite{footnote2008R}
\begin{eqnarray}
 \label{33e}
 R& = &\frac{J_\perp}{J_\parallel + J_d} =  \frac{2\lambda_3}{\lambda_1 \; +  \lambda_2}=\frac{2\xi^2_3}{\xi^2_1+\xi^2_2} \;.
\end{eqnarray}
In deriving (\ref{Acorry}) we have used (\ref{reducedA}), (\ref{lambdaxi}), and (\ref{orthomatrix}). Similar to type $(I)$ models , there exists a large variety of different realizations of type $(II)$ models, without changing the structure of (\ref{33h})-(\ref{Acorry}).
The representation (\ref{33dxxy}) is appropriate for  MC simulations of the $\varphi^4$ model whereas (\ref{Acorry}) is appropriate for other models for which $\xi_{0+}^{(i)}/\bar \xi_{0+}$ and ${\bf e}^{(i)}$  need to be identified. Possible candidates are  $d=3$ fixed-length spin models with the same couplings on the same lattice as for the $\varphi^4$ model ($II$). We conjecture that they have the same principal directions and orthogonal matrix ${\bf U}$ as in (\ref{33h}). Our conjecture is in conformity with the fact that it is valid for the analogous $d=2$ $\varphi^4$ and Ising models discussed in Sec. II. E.
\subsection{Quantitative predictions}
We consider a finite $L_\parallel^{d-1} \times L$ slab geometry with an aspect ratio $\rho=L/L_\parallel=L'/L'_\parallel$. Then the substitutions $ \bar \rho \to \rho$, ${\cal G}_0(x, \{\rho_\alpha\}, {\bf \bar A} ) \to {\cal G}_0(x,\rho, {\bf \bar A})$, and ${\cal J}_0(x, \{\rho_\alpha\}, {\bf \bar A} ) \to {\cal J}_0(x,\rho, {\bf \bar A})$ must be made with
\begin{eqnarray}
\label{calG3rhox}
&&{\cal G}_0(x,\rho, {\bf \bar A})= \int_0^\infty
dy y^{-1}
  \Big\{\exp {\left[-xy/(4\pi^2)\right]}
  \nonumber\\ &&\times \left[ (\pi/y)^{d/2}
    -\rho^{d-1} K_d (y, {\bf  C}_\rho) \right] \Big\},\;\;\;\\
\label{calJ3rhox}
&&{\cal J}_0(x,\rho, {\bf \bar A})= \int_0^\infty
dy y^{-1}
  \Big\{\exp {\left[-xy/(4\pi^2)\right]}
  \nonumber\\ &&\times \left[ \rho^{1-d}(\pi/y)^{d/2}
    - K_d (y, {\bf  C}_\rho) + 1 \right]  -  e^{- y}\Big\}\;\;\;
\end{eqnarray}
where the symmetric matrix ${\bf  C}_\rho$ has the elements
\begin{eqnarray}
\label{matrixCrho}  ({\bf  C}_\rho)_{\alpha \beta}= \left\{
\begin{array}{r@{\quad \quad}l}
                         \rho^2 \bar A_{\alpha \beta}& \mbox{for}\;\;\; \alpha,\beta \neq d ,\\
                         \rho \;\bar A_{\alpha d}& \mbox{for} \;\; \alpha \neq d, \beta=d,\\
                         \bar A_{d d}& \mbox{for} \;\; \alpha =d, \beta=d.
                \end{array} \right.
\end{eqnarray}
For models $(I)$ and $(II)$ we have the structure
 ${\bf  A} =  \left(\begin{array}{ccc}
  \bf B &  0  \\
  0 & A_{dd} \\
\end{array}\right)$
with $ A_{dd} > 0$ and with a $(d-1) \times (d-1)$ symmetric (positive definite) submatrix {\bf B} describing the anisotropies in the ``horizontal planes''. Correspondingly, $\delta \widehat K (\mathbf k)=\delta \widehat K (\mathbf q, p)$, has the long-wavelength form
\begin{equation}
\label{2hhAPP} \delta \widehat K (\mathbf q, p) = \sum_{\alpha,
\beta=1}^{d-1} B_{\alpha \beta} \; q_\alpha q_\beta \;\; + \;\; A_{dd}\;p^2 \; + O (k^4).
\end{equation}
The reduced anisotropy matrix ${\bf \bar A}$ is
\begin{eqnarray}
  \label{redaniso}
&& {\bf \bar A} = {\bf A} / (\det {\bf
A})^{1/d}= \left(\begin{array}{ccc}
  \bf \bar B &  0  \\
  0 & \bar A_{dd}  \\
\end{array}\right),\\
\label{Bbar}
&&{\bf \bar B} = {\bf B} / (\det {\bf A})^{1/d},\;\;\;\bar A_{dd}= A_{dd}/ (\det {\bf A})^{1/d}.\;\;\;\;
\end{eqnarray}
This case is described by (\ref{scalfreeaniso}) and (\ref{calJ3rhox}) with the replacements
\begin{eqnarray}
\label{substK}
&&K_d (y, {\bf  C_\rho})\to K_{d-1} (\rho^2 y,
{\bf \bar B}) \;K (\bar A_{dd} \;y), \\
\label{K1}
&&K(y) \equiv K_1(y,{\bf 1})= \sum^{\infty}_{m= -\infty} \;\exp (- y m^2),\\
\label{substisoF}
&&F(\tilde x,\{\rho_\alpha\}, {\bf \bar A})\to F(\tilde x,\rho,{\bf \bar A}),
\\
\label{substisoX}
&&X(\tilde x,\{\rho_\alpha\}, {\bf \bar A})\to X(\tilde x,\rho,{\bf \bar A})=(d-1)  F^{ex} (\tilde x,\rho,{\bf \bar A})\nonumber\\ && -
\frac{\tilde x}{\nu}
\frac{\partial  F^{ex}(\tilde x,\rho,{\bf \bar A})}{\partial\tilde x}- \rho
\frac{\partial F^{ex}(\tilde x,\rho,{\bf \bar A})}{\partial \rho}\;\;\;.
\end{eqnarray}
In the following we predict the effect of lattice anisotropy on the Casimir force scaling function $X$ of models $(I)$ and $(II)$ for several cases.
\subsubsection*{{\bf 1. Nonuniversal crossover from below to above ${\bf T_c}$ }}
\begin{figure}[!h]
\vspace{0cm}
\includegraphics[width=80mm]
{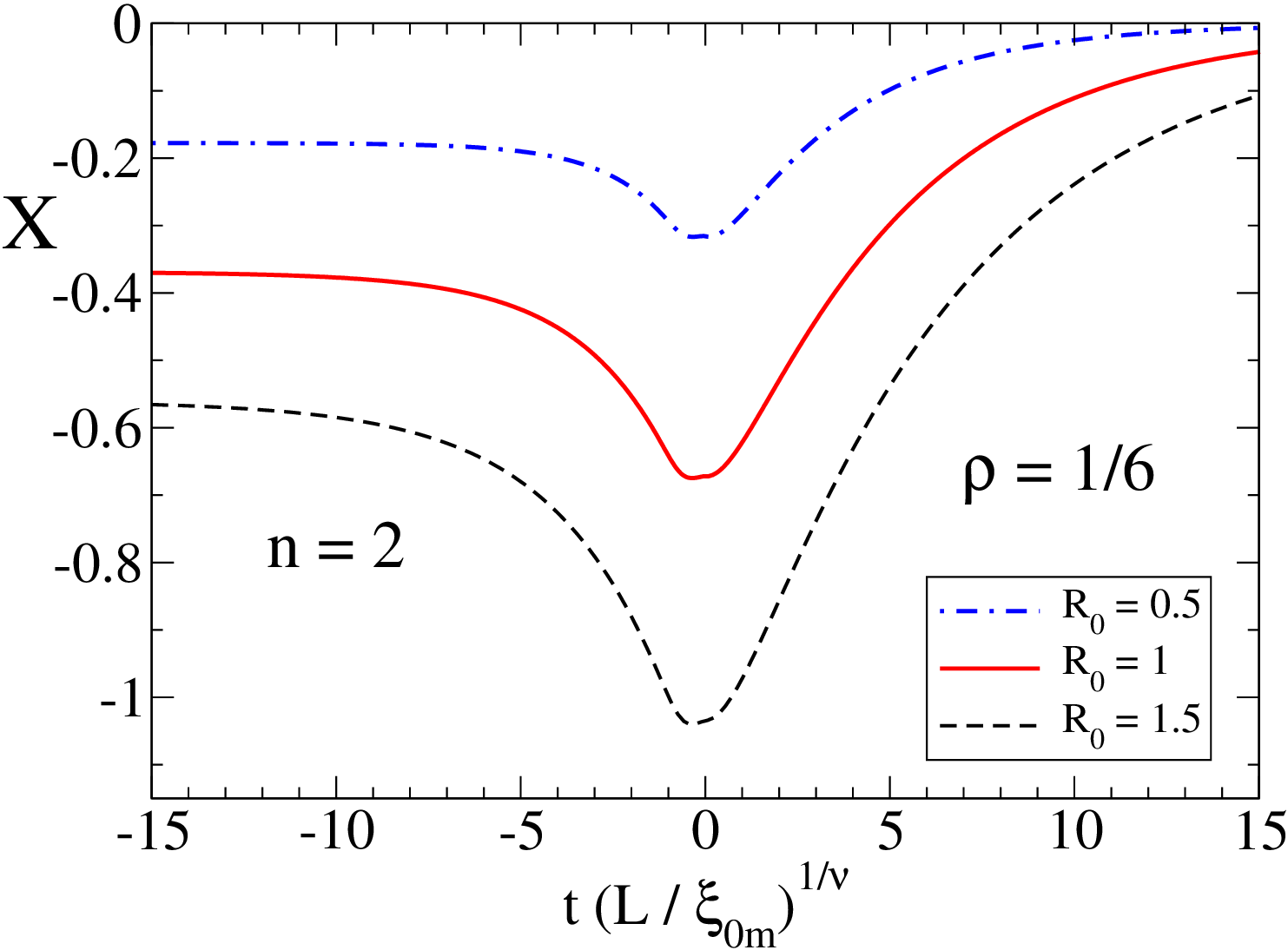}
\vspace{0.2cm}
\caption{(Color online)  Casimir force scaling function $X$, (\ref{substisoX}), (\ref{scalfreeaniso}), (\ref{calJ3rhox}) for $d=3$, $n=2$  in a slab geometry with the aspect ratio $\rho=1/6$ as a function of $t (L/ \xi_{0m})^{1/\nu}$.
Dashed and dot-dashed lines: anisotropic systems with the  matrix  ${\bf \bar A}_{(I)})$, (\ref{simplematrixquer2}), and the anisotropy parameter (\ref{333}) with $ R_0=1.5$ and $ R_0=0.5$, respectively, with $\xi_{0m}\equiv \bar \xi_{0+}$, (\ref{ximean}). Solid line: isotropic systems with $ R_0=1$ as shown in Fig. 3 (b) of \cite{dohm2017I}, with $\xi_{0m}\equiv  \xi_{0+}$. }
\end{figure}
For model $(I)$ we use the abbreviation
\begin{equation}
\label{333}
 R_0\equiv J_\perp/J_\parallel=(\xi_{0+}^\perp/\xi_{0+}^\parallel)^2
\end{equation}
corresponding to
$K_\perp/K_\parallel=\gamma_0^{-2}=(\xi_\perp/\xi_\parallel)^2$ in \cite{wil-1,shenoy1995}.
We calculate the  Casimir force scaling function from (\ref{scalfreeaniso})-(\ref{3nn}), (\ref{calJ3rhox}), and (\ref{substK})-(\ref{substisoX}), of a system in the ($n=2,d=3$) universality class with the anisotropy matrix $\bar {\bf A}_{(I)}$. The anisotropy effect for $R_0=0.5$ and $R_0=1.5$ on the crossover  from far below to far above $T_c$ is shown in Fig. 1 as a function of $\tilde x= t (L/\bar \xi_{0+})$ in the range $-15 < \tilde x < 15$ for $\rho=1/6$. For comparison the isotropic result for $R_0=1$ taken from Fig. 3 (b) of \cite{dohm2017I} is also shown. The finite low-temperature values of $X$ are due to the Goldstone modes \cite{dohm2017I}. While the position of the minima remains close to that for the isotropic case $R_0=1$, the depth of the minima as well as the magnitudes of the low-temperature Casimir forces are significantly changed by the anisotropy. Similar anisotropy effects are predicted by our theory for $n=1$ and $n=3$. This demonstrates the nonuniversality of the Casimir force scaling function due to lattice anisotropy both in the critical and noncritical regions.
\subsubsection*{{\bf 2. Change of sign of the critical Casimir amplitude due to anisotropy }}
From (\ref{6ffx})-(\ref{scalfreeaniso}) and (\ref{substisoX}) we obtain the Casimir amplitude in slab geometry at $T_c$
\begin{eqnarray}
\label{CasimiratTc}
&&X(0,\rho, {\bf \bar A}) =(d-1)  F (0,\rho, {\bf \bar A}) - \rho \partial F(0,\rho, {\bf \bar A})/\partial \rho,\;\;\;\;\;\;\;\;\;\;
\end{eqnarray}
\begin{eqnarray}
\label{scalfreeanisoTc} &&F(0, \rho, {\bf \bar A})= \rho^{d-1} \nonumber\\&&\times\;\Bigg\{-
u^* \left[\vartheta_{2,n} (0)\right]^2 \Bigg[\frac{36}{d} + 144\frac{(n-1)}{d\varepsilon}\;\Big(\frac{1}{3^{d/2}}-\frac{d}{36}\Big)\Bigg]\nonumber\\ && +\frac{n}{2}\ln\Bigg[\frac{\tilde l_c^2 \;{[\Gamma(n/2)]}^{2/n}  }{24 \pi^2\; \vartheta_{2,n} (0) }\Bigg] -\;  \ln \Big[\frac{1}{2}\;\Gamma(n/4) \Big] \nonumber\\ && + \frac{1}{2}{\cal J}_0( {\tilde l_c}^2, \rho, {\bf \bar A})  + \; \frac{n-1}{2}{\cal J}_0( {\tilde l_c}^2/3, \rho, {\bf \bar A})\Bigg\},\\
\label{ltildec}
&&\tilde l_c^{d/2} =12 {u^*}^{1/2}  A_d^{-1/2}
\vartheta_{2,n}(0)\;\rho^{(d-1)/2},\\
\label{theta2n}
&&\vartheta_{2,n} (0)= \;\Gamma\big((n+2)/4\big)/\Gamma\big(n/4\big) .
\end{eqnarray}
\vspace{0cm}
\begin{figure}
\vspace{0cm}
\begin{center}
\subfigure{\includegraphics[clip,width=7.2cm]{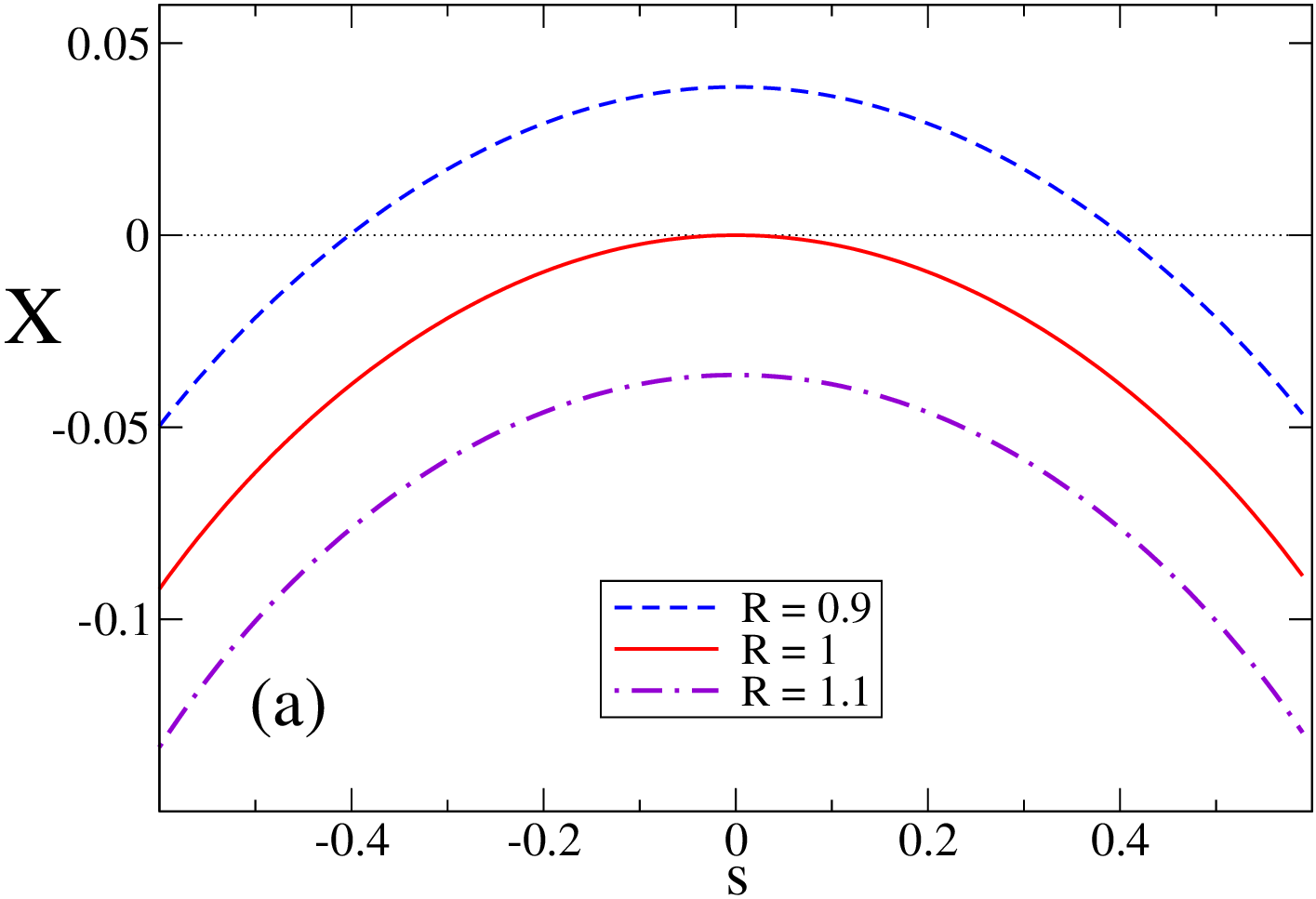}}
\subfigure{\includegraphics[clip,width=7.2cm]{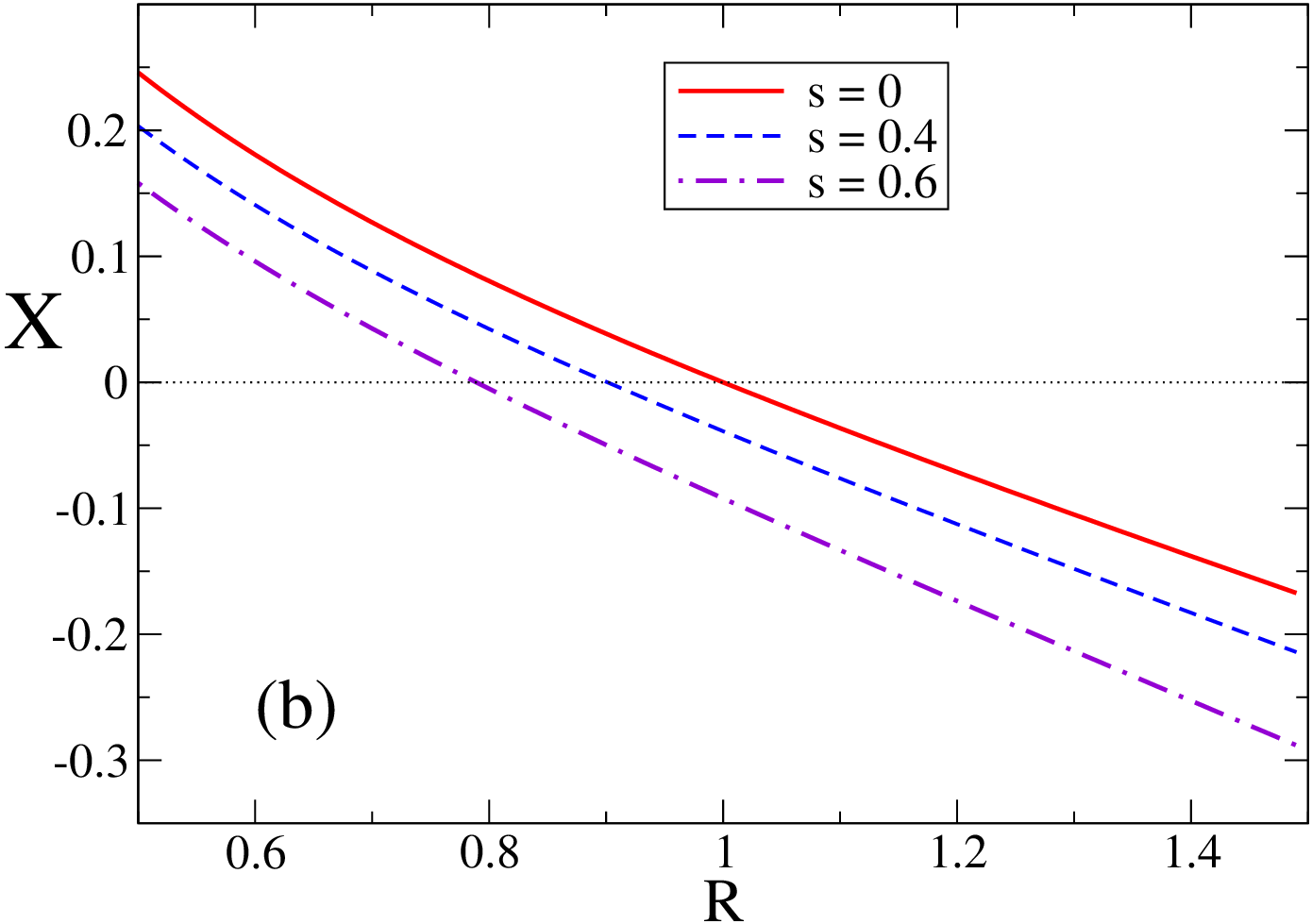}}
\end{center}
\vspace{1.3cm}
\caption{(Color online)  Casimir amplitude at $T_c$ for $d=3$, $n=1$,  (\ref{CasimiratTc})-(\ref{theta2n}), with the
anisotropy matrix  ${\bf \bar A}_{(II)}$, (\ref{33dxxy}), in a cubic $(\rho=1)$
geometry,  (a) as a function of
$s$, (\ref{s}), for $R = 0.9, 1, 1.1$ and (b) as a function of  $R$, (\ref{33e}), for $s=0, 0.4, 0.6$. $X(0,1, {\bf \bar A}_{(II)})$ vanishes for the isotropic case $s=0,R=1$ but not for the anisotropic case $s \neq 0,R > 1$.}
\end{figure}
\vspace{0cm}
From previous analytic and numerical work \cite{dohm2011,hucht2011} it is known that the Casimir amplitude at $T_c$ of {\it isotropic} systems vanishes in a cubic geometry, i.e., for  $\rho=1$. The Casimir amplitude at $T_c$ has received particular attention in the literature because of its alleged universality. It was claimed that it depends only on a few general and global properties of the system such as the bulk and surface universality classes of the phase transition, the system shape, and the boundary conditions.
In \cite{hucht2011}  "a general argument" was given that the Casimir force vanishes at $T_c$ for $\rho = 1$ and becomes repulsive in periodic systems for $\rho > 1$. In the following we show that this is not generally valid for weakly anisotropic systems.

For the example of model $(II)$, we show in Fig. 2 the effect of the anisotropy on X, (\ref{CasimiratTc})-(\ref{theta2n}), in the $n=1$ universality class in a cubic geometry for two cases:
(i) fixed $R=0.9,1,1.1$ in the range $-0.6 < s< 0.6$ for $\tilde x =0,\rho=1$,
(ii) fixed $s=0,0.4,0.6$  in the range $0.5< R < 1.5$ for $\tilde x =0,\rho=1$.
We see that both anisotropy parameters $s$ and $R$ affect the sign of the critical Casimir amplitude in a cubic geometry. In particular, it does not vanish but becomes attractive (negative) for $R>1, \rho=1,T=T_c$ and, by continuity, {\it remains attractive} for $R>1, \rho \gtrsim 1,T=T_c$. Clearly the absence or presence of a NN or NNN coupling is a microscopic detail. We conclude that our results invalidate the claim that the critical Casimir amplitude does not depend on microscopic details. For this conclusion it does not matter which kind of representation  is chosen for the anisotropy parameters $s$ and $R$.
\subsubsection*{{\bf 3. Change of sign of the low-temperature Casimir amplitude due to anisotropy }}
It is of interest to present the analytic form of  $F^{ex}$ and $X$ based on (\ref{scalfreeaniso}) in the region well below $T_c$ ($-\tilde x \gg 1$).
The derivation is sketched in Appendix C of \cite{dohm2017I} for the isotropic case. For the anisotropic case, the corresponding result can be obtained from I (5.14)-I (5.16) after replacing  ${\cal G}_0(x, \rho)$ and ${\cal J}_0(x,\rho)$ by ${\cal G}_0(x, \rho, {\bf \bar A})$, (\ref{calG3rhox}) and ${\cal J}_0(x,\rho, {\bf \bar A})$, (\ref{calJ3rhox}), respectively.

As discussed in \cite{dohm2017I} for the isotropic case, the low-temperature behavior differs fundamentally depending on wether  Goldstone modes are absent ($n=1$) or  present ($n>1$).
For $n=1$ we obtain the leading behavior for large negative $\tilde x$
\begin{eqnarray}
\label{Xbelowngleich1} &&X(\tilde{x},\rho, {\bf \bar A}) \approx \frac{1}{2}\;\Big[(d-1)   -(\tilde x/\nu)\;
\partial /\partial \tilde x \nonumber \\ &&- \rho\;
\partial /\partial \rho\Big]\;{\cal G}_0( (2|\tilde x| \;Q^*)^{2\nu }, \rho, {\bf \bar A})
\end{eqnarray}
with a vanishing low-temperature limit
\begin{eqnarray}
\label{lowXn1}
\lim_{\tilde x \to -\infty}X(\tilde x,\rho, {\bf \bar A})=0.
\end{eqnarray}
For $n>1$, $\rho > 0$, $X$ has the following {\it finite} low-temperature limit in a finite volume
\begin{eqnarray}
\label{lowXn2}
&& X(-\infty,\rho, {\bf \bar A})=\nonumber \\&& -[(n-1)/2]\;\rho^{d-1}\big[1+ \rho
\partial {\cal J}_0(0, \rho, {\bf \bar A})/\partial \rho\big].
\end{eqnarray}
where ${\bf \bar A}\big(\{\xi_{0{\rm T}}^{(\alpha)},{\bf e}^{(\alpha)}\}\big)$, (\ref{AnonunixlT}), may be used. The finite value of  $X(-\infty,\rho, {\bf \bar A})$ reflects the effect of the long-range fluctuations induced by the  Goldstone modes.
From (\ref{lowXn2}) we obtain the amplitude per component in the large-$n$ limit
in agreement with the exact result (\ref{X-large-nx-low-slab}). We note that (\ref{lowXn2}), divided by $n-1$, is  not identical with the Gaussian critical amplitude $X^G(0,\rho, {\bf \bar A})/n$, (\ref{cascalJcritslab}), unlike the case  of film systems for $n>1$ (Sec. V).

\vspace{1cm}
\begin{figure}[!h]
\includegraphics[width=80mm]
{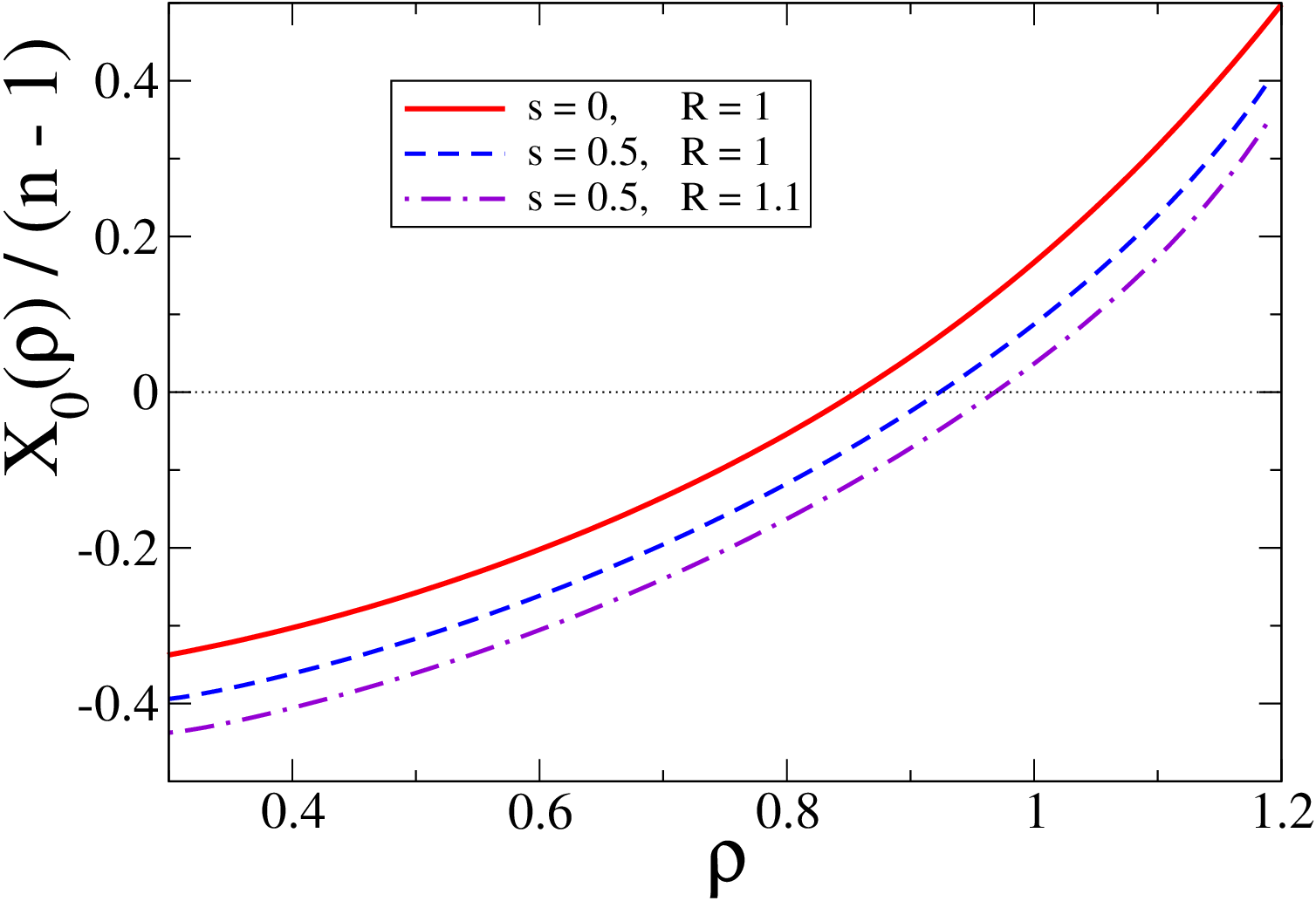}
\vspace{0cm}
\caption{(Color online) Low-temperature Casimir amplitude $X_-(-\infty,\rho, {\bf \bar A}_{(II)})\equiv X_0(\rho)$, (\ref{lowXn2}), divided by $n-1$, with the matrix (\ref{33dxxy}) and anisotropy parameters $s$, (\ref{s}), and $R$, (\ref{33e}), in a $d=3$ slab geometry  as a function of the aspect ratio $\rho$. Solid line: isotropic systems with $s=0, R=1$ (Fig. 5 of \cite{dohm2017I}). Dashed and dot-dashed lines: anisotropic systems with $s=0.5, R=1$ and $s=0.5, R=1.1$.  $X_0(\rho)$ vanishes at different values of $\rho$ depending on the anisotropy parameters. These curves are exact for $n\to \infty$, see (\ref{X-large-nx-low-slab}).}
\end{figure}

For the example of model $(II)$, a comparison of the amplitude (\ref{lowXn2}) divided by $n-1$ for isotropic (solid line) and anisotropic systems (dashed and dot-dashed lines) is shown in  Fig. 3 where the amplitude is plotted  for $d=3$ as a function of the aspect ratio $\rho$.
According to (\ref{lowXn2}) our theory predicts a vanishing of the low-temperature Casimir amplitude for general $1<n\leq\infty$ in a slab geometry if the condition
\begin{eqnarray}
\label{lowXzeroiso}
\rho\;\partial {\cal J}_0(0, \rho, {\bf \bar A})/\partial \rho=-1
\end{eqnarray}
is satisfied.  This condition is approximate for finite $n$ but becomes exact
in the limit $n\to \infty$ according to  (\ref{X-large-nx-low-slab}). For finite $n$, we expect model-dependent corrections at low temperatures
in a more complete theory.
The dependence on ${\bf \bar A}$ implies that the anisotropy has a significant macroscopic effect: it may affect the sign of the Casimir force in finite systems at low temperatures. The Casimir force vanishes in the isotropic case at $\rho=0.8567$ (solid line in Fig. 3) whereas in the anisotropic examples considered in Fig. 3 it vanishes at the nonuniversal values $\rho =0.9237 $ (dashed line) and $\rho =0.9683 $ (dot-dashed line), respectively.
We conclude that the low-temperature Casimir amplitude $X(-\infty,\rho, {\bf \bar A})$ is a nonuniversal quantity for which lattice anisotropy plays a significant role. Our quantitative predictions can be tested by MC simulations for spin models in the $XY$ ($n=2$) and Heisenberg ($n=3$) universality classes.
\renewcommand{\thesection}{\Roman{section}}
\renewcommand{\theequation}{5.\arabic{equation}}
\setcounter{equation}{0}
\section{Perturbation theory for film geometry}
Our lowest-mode separation approach is not applicable to  an $\infty^{d-1}\times L$ film geometry where no single lowest mode exists.
For this reason the scaling functions for {\it isotropic} film systems were derived in \cite{dohm2017I} by taking the film limit $ \rho \to 0$ at fixed $L$.
Here we show that the same result is obtained by renormalized perturbation theory  in the sense of an expansion around bulk mean field theory based on the decomposition
\begin{eqnarray}
\label{decompfilm}
{\bm \varphi}_j = {\bf M}_{mf} +  {\bm s}_j
\end{eqnarray}
where ${\bf M}_{mf}$ is the the bulk mean-field  order parameter.
The result is expected to be identical with the film limit $\rho\to 0$ of the results of Sec. IV.
In comparison to previous work for films at finite $n$ with periodic BC \cite{wil-1, KrDi92a, GrDi07,kastening-dohm}, the progress achieved here is the derivation of the scaling functions
for general $n$ below $T_c$ and the description of anisotropy effects above and below $T_c$.

Using the decompostion (\ref{decompfilm})  we obtain
the free energy density of the film system at $h=0$ in one-loop order as
\begin{eqnarray}
\label{free film one loop}&& f_{\text film}(t,L) = \frac{1}{2}r_0M_{mf}^2+u_0 M_{mf}^4  \nonumber\\&&- n\frac{\ln (2 \pi)} {2 \tilde a^d}+\frac{1}{2L} \sum_p
\int_{\bf q} \ln
\{[r^{mf}_{\rm L} + \delta \widehat K (\mathbf q , p)]\;  \nonumber\\&&+ \frac{n-1}{2L} \sum_p
\int_{\bf q} \ln
\{[r^{mf}_{\rm T} + \delta \widehat K (\mathbf q , p)]+ O(u_0)\;\;\;\;
\end{eqnarray}
with $\mathbf k\equiv(\mathbf q , p)$. Here  $\int_{\bf q}$ is a $(d-1)$ dimensional integral with finite lattice cutoff [see (\ref{3dd})]. The one-dimensional sum $\sum_p$ runs over $p=2\pi m/L, m = 0, \pm1, \pm2, ...$ up to $\pm\pi/\tilde a$. No spurious Goldstone divergence of the film free energy arises at our level of approximation due to the continuous mode spectrum with respect to the horizontal wave vector ${\bf q}$. At $h=0$, we have $M_{mf}^2=0$, $r^{mf}_{\rm L} =r^{mf}_{\rm T} =r_0$ for $r_0\geq0$ and $M_{mf}^2=-r_0/(4u_0)$, $r^{mf}_{\rm L} =-2r_0, r^{mf}_{\rm T} =0$ for $r_0\leq0$, respectively. This leads to the excess free energy density for $n\geq1,d>1$
\begin{eqnarray}
\label{gibbs-excess-film-above}
&&f^{ex,+}_{\text film}(t,L,{\bf A}) = \;\frac{n}{2 } \; \Delta_{\text film}(r_0, L,{\bf A}) + \;O(u_0),\;\;\;\;\;\;\\
\label{gibbs-excess-film-below1}
&&f^{ex,-}_{\text film}(t,L,{\bf A}) =\frac{1}{2 } \; \Delta_{\text film}(-2r_0 , L,{\bf A}) \nonumber\\&& \;\;\;\;\;\;\;\;\;\;\;\;\;\;\;\;\;\;\;\;\;\;\;\;+ \; \frac{n-1}{2 } \; \Delta_{\text film}(0, L,{\bf A})+\; O(u_0),\;\;\;\;\;\;
%
\end{eqnarray}
where the function $\Delta_{\text film}(r_0, L,{\bf A})$ is defined in I (6.5). These expressions are singular at $r_0=0$ for finite $L$ corresponding to an unshifted film critical point. In an exact theory, an $L$-dependent critical value $r_{0c,\text film}(L)$ corresponding to  $0<T_{c,\text{film}}(L) < T_c$ should occur for $ n=1,d>2$, for $n=2,d\geq 3$, and for $n>2, d>3$.
Our one-loop approximation does not capture this $L$-dependent shift. Similarly, no such a shift was captured  in earlier work  \cite{KrDi92a,wil-1,GrDi07,kastening-dohm}.
For an exact calculation of this shift in the large-$n$ limit
for $d>3$ see Sec. VI.

The shear transformation discussed in Sec. II preserves
the film geometry except that the thickness $L$ is transformed to a different thickness $\bar L$ of the transformed isotropic film which is given by Eq. (2.48) of \cite{kastening-dohm}.
Contrary to naive expectation, it is not $\bar L$ but rather \cite{cd2004}
\begin{align}
\label{Ltilde}
\widetilde L=[({\bf A}^{-1})_{dd}]^{1/2}L
\end{align}
that will appear in all results for the excess free energy for film geometry. Here  $({\bf A}^{-1})_{dd}$ denotes the $d$th diagonal element of the inverse of the  matrix ${\bf A}$. The length $\widetilde L$
represents the distance between those points on the opposite surfaces in the transformed  film system that are connected via the periodicity requirement \cite{cd2004,kastening-dohm}.

It is of interest to first consider the low-temperature limit $r_0 \to - \infty$ which has nothing to do with critical phenomena and for which no RG theory  is necessary. Since the longitudinal part of  (\ref{gibbs-excess-film-below1}) decays exponentially, $f^{ex,-}_{\text film}$ vanishes for $n=1$ in this limit. For $n>1$ the Goldstone modes yield a finite amplitude [compare (\ref{Deltafilm}), I (5.32]
\begin{eqnarray}
\label{gibbs-free-low}
&&\lim_{r_0 \to - \infty}f^{ex,-}_{film}(t,L,{\bf A}) = \; \frac{n-1}{2 } \; \Delta_{film}(0,L,{\bf A})\;\;\;\;\;\;\;\\
\label{fexfilmlow}
&&=- ({\det {\bf A}})^{-1/2}{\widetilde L}^{-d}(n-1)\; \pi^{-d/2}\Gamma(d/2)\zeta(d),
\end{eqnarray}
where $\widetilde L \gg \tilde a$ is assumed.
Employing the physical length $L$ of the original anisotropic film we obtain the  low-temperature  Casimir amplitude
\begin{eqnarray}
&&X_{\text{film},-\infty}({\bf \bar A } )
=  - L^d\lim_{r_0 \to - \infty}\partial[L
f^{ex,-}_\text{film} ]/\partial L \nonumber \\&&=
\label{casimirlowx}
-[({ \bf \bar A^{-1}})_{dd}]^{-d/2}\;(n-1)(d-1)\pi^{-d/2}\Gamma(d/2)\zeta(d)\;\;\;\;\;\;\;\;\;\;
\end{eqnarray}
for $n>1,d>1$ with
$({ \bf \bar A^{-1}})_{dd}= (\det {\bf A})^{1/d} ({\bf A^{-1}})_{dd}$.
We hypothesize, in the spirit of multiparameter universality,
that the result (\ref{casimirlowx}) derived within the $\varphi^4$ theory is more generally valid for weakly anisotropic O$(n)$-symmetric film systems with periodic BC in the low-temperature limit after the substitution of ${\bf \bar A}\big(\{\xi_{0{\rm T}}^{(\alpha)},{\bf e}^{(\alpha)}\}\big)$, (\ref{AnonunixlT}). We indeed obtain from (\ref{casimirlowx}) the correct Casimir force amplitude per component in the large-$n$ limit
in agreement with the exact result (\ref{Casimirfilmscaling}) and I (6.16).
We recall, however, that (\ref{casimirlowx}) has been derived within a one-loop approximation. For finite $n$ we expect model-dependent corrections to (\ref{casimirlowx}). Eq. (\ref{casimirlowx}), divided by $n-1$, is identical with the Casimir amplitude $X^G_{\text film}(0,{\bf \bar A})/n$, (\ref{CasimirGaussAmp}), of the Gaussian film system at $T_c$, compare I (5.33).

Now we turn to the temperature dependence. For  $L/\tilde a \gg 1$, $0\leq \tilde a r^{1/2}\ll 1$,  $0\leq L r^{1/2} \lesssim O(1)$
the function $\Delta_{\text film}(r,L,{\bf A})$ is given by
\begin{eqnarray}
\label{Deltafilmx1} &&\Delta_{\text film}(r,L,{\bf A})=  (\det {\bf A})^{-1/2}\widetilde L^{-d} \; {\cal G}_{0,\text film}
\big(r \widetilde L^2  \big), \;\;\;\;\;\;\;
\end{eqnarray}
as follows from (\ref{b7film}),(\ref{Deltafilm}), compare also I (6.6) for the isotropic case.
To obtain the correct scaling form
it is necessary to renormalize $f^{ex,\pm}_{\text film}(t,L,{\bf A})$.
First we may replace $r_0$
by $r_0-r_{0c}$  in the spirit of perturbation theory up to $O(1)$. The resulting quantity will be denoted by  $f^{ex,\pm}_{\text film}(r_0-r_{0c}, u_0,L,{\bf A})$. It is related to
$f'^{ex,\pm}_{\text film}(r_0-r_{0c}, u'_0,L',{\bf \bar A })$ of the transformed isotropic film system by
\begin{eqnarray}
\label{fexfilm}
&&f^{ex,\pm}_{\text film}(r_0-r_{0c}, u_0,L,{\bf A})\nonumber \\&&= (\det {\bf  A })^{-1/2} f'^{ex,\pm}_{\text film}(r_0-r_{0c}, u'_0,L',{\bf \bar A }).
\end{eqnarray}
The quantity  $f'^{ex,\pm}_{\text film}$ has no additive pole terms for $\varepsilon \to 0$ and is multiplicatively renormalizable with the same bulk $Z$ factors as used in Sec. II. C. The renormalized film excess free energy density of the transformed system is
\begin{eqnarray}
\label{fexfilmrenorm}
&&f'^{ex,\pm}_{R,\text film}(r', u',L',\mu,{\bf \bar A })\nonumber \\&&= f'^{ex,\pm}_{\text film}(Z_{r'}r',
       \mu^{\varepsilon}Z_{u'}Z_{\varphi'}^{-2}A_d^{-1}u',L',{\bf \bar A })
\end{eqnarray}
for $2<d<4$ [compare (\ref{5ddfinite})]. After integration of the RGE we choose the flow parameters $l_+$ and $l_-$
according to (\ref{bulk flow parameter}) which leads to
\begin{eqnarray}
\label{fexfilmrenormplus}
&&f'^{ex,+}_{R,\text film}(r', u',L',\mu,{\bf \bar A })
= \frac{n}{2}{\widetilde L}^{-d} {\cal G}_{0,film} \big(\mu^2 l_+^2 {\widetilde L}^2 \big), \;\;\;\;\;\;\;\;\;
\\
\label{fexfilmrenormminus}
&&f'^{ex,-}_{R,film}(r', u',L',\mu,{\bf \bar A })\nonumber \\&&=\frac{ {\widetilde L}^{-d}}{2}\Big[ {\cal G}_{0,film} \big(\mu^2 l_-^2 {\widetilde L}^2 \big) + (n-1)\; {\cal G}_{0,film} \big(0\big)\Big],\;\;\;\;\;\;\;
\\
\label{Lschlangeprime}
&&\widetilde L=[({\bf \bar A^{-1}})_{dd}]^{1/2}L',
\end{eqnarray}
with $L'= L / (\det {\bf A})^{1/(2d)}$. The choice $\mu^{-1}=\xi'_{0+}$ [see (\ref{my})] implies the scaling variable $t (\widetilde L/\xi_{0+}')^{1/\nu}$ for the transformed isotropic film system. Neglecting nonasymptotic corrections to scaling we obtain
the scaling form of the excess free energy of the transformed isotropic film  and of the original anisotropic film, respectively,
\begin{eqnarray}
\label{fexfilmscalingiso}
f'^{ex,\pm}_{R,\text film}& = & L'^{-d}F^{ex,\pm}_{\text film}(\tilde x_\perp,{\bf \bar A }),\\
\label{fexfilmscalinganiso}
f^{ex,\pm}_{R,\text film}& =& L^{-d}F^{ex,\pm}_{\text film}(\tilde x_\perp,{\bf \bar A }).
\end{eqnarray}
Here the scaling variable $\tilde x_\perp$ can be expressed in two different ways, similar to the case of block geometry,
\begin{eqnarray}
\label{scalfilm}
\tilde x_\perp
&=& t (\widetilde L/\xi_{0+}')^{1/\nu}
= t (L/ \xi^\perp_{0+})^{1/\nu},\\
\label{scalfilmxixneux}
 \xi^\perp_{0+}
&=&[({\bf A}^{-1})_{dd}]^{-1/2}\xi'_{0+},\;\;\;\;\;
\end{eqnarray}
where (\ref{scalfilmxixneux}) follows from (\ref{Ltilde}). The appropriate scaling variable for the anisotropic film is  $t (L/ \xi^\perp_{0+})^{1/\nu}$ whose natural reference length $\xi^\perp_{0+}$  turns out to be the amplitude of the bulk correlation length $\xi_{(d)}=\xi^\perp_{0+}t^{-\nu} $
above $T_c$ in the $d$th  direction, i.e., in the direction perpendicular to the film boundaries. This interpretation follows from considering the bulk correlation function (\ref{3nbar}) by choosing  ${\bf x}$ as  ${\bf x_{{\bf e}}}=x_{{\bf e}} {\bf e}_d$, ${\bf e}_d=(0,...,0,1)$ in the $d$th direction. Requiring
\be
\label{xibarplus}
|{\bf \bar{\mbox
{\boldmath$\lambda$}}}^{-1/2} {\bf U}{\bf x_{{\bf e}}}|/\bar \xi_+=| x_{{\bf e}}|/\xi_{(d)}
\ee
we obtain
\begin{eqnarray}
\label{xierechnung}
&&|{\bf \bar{\mbox
{\boldmath$\lambda$}}}^{-1/2} {\bf U}{\bf x_{{\bf e}}}|
= [({\bf \bar{\mbox
{\boldmath$\lambda$}}}^{-1/2} {\bf U}{\bf x_{{\bf e}}})\cdot ({\bf \bar{\mbox
{\boldmath$\lambda$}}}^{-1/2} {\bf U}{\bf x_{{\bf e}}})]^{1/2}
\nonumber\\
&&=[{\bf x_{{\bf e}}}\cdot ({\bf \bar A}^{-1}{\bf x_{{\bf e}}})]^{1/2}
=[{\bf e}_d\cdot ({\bf \bar A}^{-1}{\bf e}_d)]^{1/2} |x_{{\bf e}}|\;\;\;\;\;\;\;
\end{eqnarray}
where ${\bf e}_d\cdot ({\bf \bar A}^{-1}{\bf e}_d) =({\bf \bar A}^{-1})_{dd}$. Thus  (\ref{xibarplus}) yields $[({\bf \bar A}^{-1})_{dd}]^{1/2}/\bar \xi_+=1/\xi_{(d)}$. According to (\ref{ximean}) this is equivalent to
\begin{eqnarray}
\label{scalfilmxix}
 \xi^\perp_{0+}=[({ \bf \bar A^{-1}})_{dd}]^{-1/2} \;\bar \xi_{0+}=[({\bf A}^{-1})_{dd}]^{-1/2}\xi'_{0+},
\end{eqnarray}
which confirms (\ref{scalfilmxixneu}) and the interpretation given above. Obviously $\xi^\perp_{0+}$ is a nonuniversal quantity whose magnitude depends on the orientation of the film boundaries relative to the ellipsoidal shape of the correlation volume. The latter is determined by the intrinsic anisotropy which is totally unrelated to the film orientation. In contrast to $\widetilde L$ and
$\xi'_{0+}$, the physical lengths $L$ and $\xi^\perp_{0+}$ of the anisotropic systems are directly measurable quantities.

\vspace{0cm}
\begin{figure}[!ht]
\begin{center}
\subfigure{\includegraphics[clip,width=7.0cm]{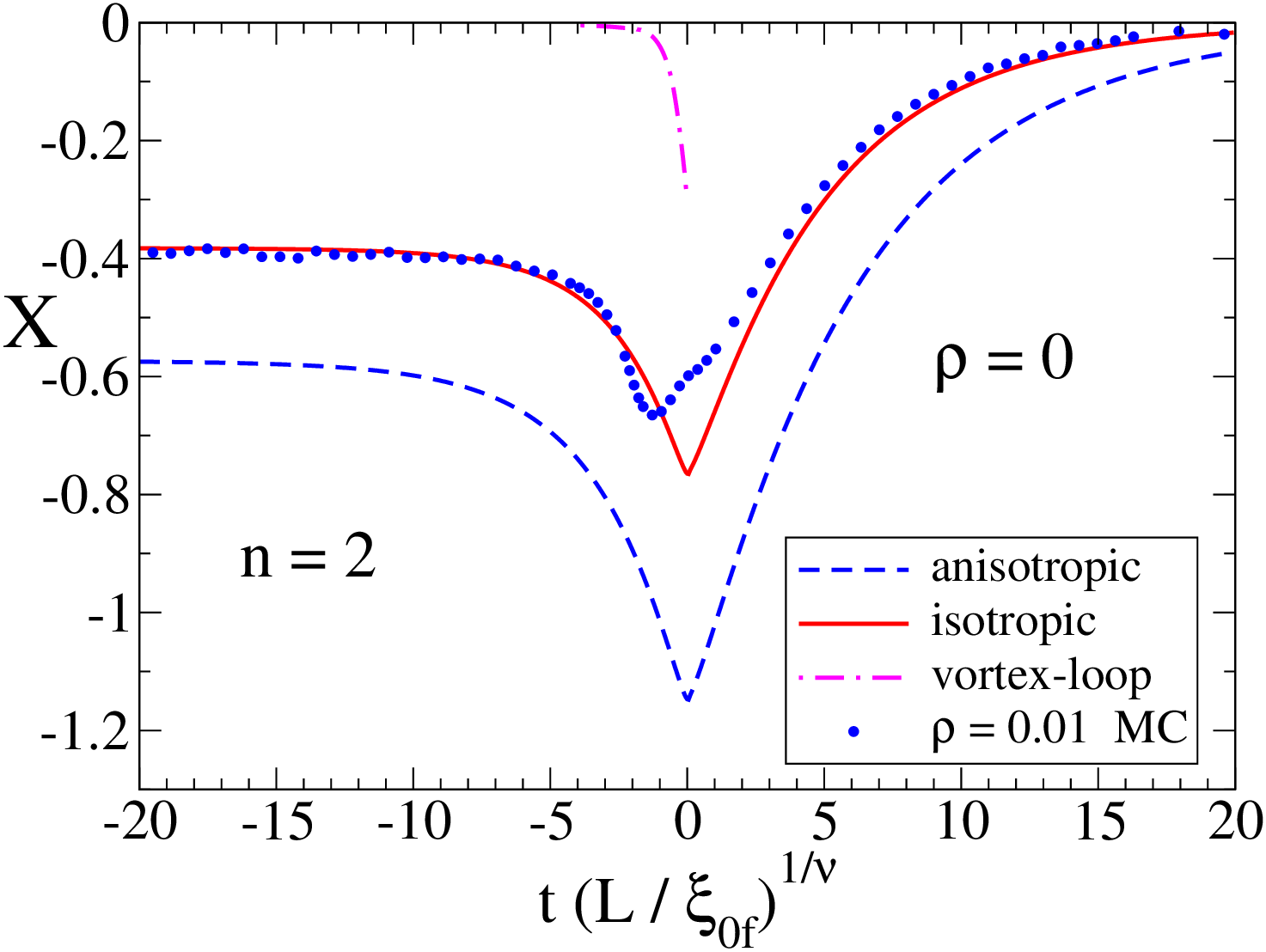}}
\end{center}
\vspace{0.5cm}
\caption{(Color online)  Scaling function $X_{\text film}$  in a film ($\infty^2\times L$) geometry ($\rho=0$) as a function of $t (L/ \xi_{0f})^{1/\nu}$
for $n=2$.  Solid line: isotropic system
from Fig. 8 (d) of \cite{dohm2017I} with $\xi_{0f}\equiv \xi_{0+}$.  Dashed line: from (\ref{fexfilmscalingplusxxx})-(\ref{fexfilmscalingminus}) with the matrix ${\bf \bar A}_{(I)}$, (\ref{simplematrixquer2}), and the anisotropy parameter $R_0=1.5$, (\ref{333}), with $\xi_{0f}\equiv \xi^\perp_{0+}$, (\ref{scalfilmxixneux}). Dot-dashed line: vortex-loop theory \cite{wil-1} without anisotropy effect. MC data for the $d=3$ isotropic XY model from \cite{hasenbusch2010} for $\rho=0.01$. }
\end{figure}
\vspace{0cm}

From (\ref{fexfilmrenormplus})-(\ref{fexfilmscalinganiso}) we obtain the  scaling functions of the {\it anisotropic} film system
\begin{eqnarray}
\label{fexfilmscalingplusxxx}
F^{ex,\pm}_{\text film}(\tilde x_\perp,{\bf \bar A })
&=&\;[({ \bf \bar A^{-1}})_{dd}]^{-d/2}\;F^{ex,\pm}_{\text film,iso}(\tilde x_\perp)\\\;\;\;
\label{Casimirfilmscalingxx}
F_\text{Cas,film}(t,L,{\bf A })&=&L^{-d}\;[({ \bf \bar A^{-1}})_{dd}]^{-d/2}\;X^{\pm}_{\text film,iso}
(\tilde x_\perp)\;\;\;\;\;\;\;\;\;
\end{eqnarray}
where  $F^{ex,\pm}_{\text film,iso}$ and $X^\pm_{\text film,iso}$  are
the scaling functions of the {\it isotropic} film given by I (5.29)-  I (5.31) which here, however, have a different scaling argument $\tilde x_\perp$ that is affected by anisotropy.
Eq. (\ref{Casimirfilmscalingxx}) contains (\ref{casimirlowx}) as the limiting case
$\tilde x_\perp \to -   \infty$. The different behavior of $F^{ex,\pm}_{\text film,iso}$ and $X^{\pm}_{\text film,iso}$ for $n=1$ and $n>1$ is discussed in \cite{dohm2017I}.
The scaling functions of the anisotropic film are described in terms of
the same function ${\cal G}_{0,\text film}$ that appears also in the isotropic case \cite{dohm2017I}.
This is the consequence of the fact that the shear transformation preserves  the film geometry,  but with a changed thickness determined by the anisotropy. In conclusion, even for the idealized $\infty^{d-1}\times L$ film geometry the Casimir force scaling function is affected by up to $d(d+1)/2-1$ nonuniversal anisotropy parameters, in addition to the thermodynamic length scale $\xi'_{0+}$ contained in $\tilde x_\perp$.
In particular, the critical Casimir amplitude $C_0X^{\pm}_{\text film,iso}
(0)$  with the nonuniversal prefactor
$ C_0=[({ \bf \bar A^{-1}})_{dd}]^{-d/2}$ is a nonuniversal quantity,
in contrast to the universal number $X_c$ in (\ref{casimirTc})
predicted by two-scale-factor universality \cite{pri}.
With the parametrization (ii) of ${\bf \bar A}\big(\{{\bf e}^{(\alpha)}\},\{\xi_{0+}^{(\alpha)}\}\big)$,
multiparameter universality replaces two-scale-factor universality for  anisotropic film systems.
As noted in \cite{dohm2017I}, our theory does not yield reliable results for $d \to 2$.  In particular it does not reproduce the results obtained for the $d=2$ anisotropic Ising model in a ($\infty\times L$) strip geometry \cite{Indekeu,kastening2012,rudnick}.

As an application to $d=3$ we consider model $(I)$ with the anisotropy matrix   ${\bf \bar A}_{(I)}$,  (\ref{simplematrixquer1}), and the anisotropy parameter $R_0$, (\ref{333}). Since
$[{\bf \bar A}_{(I)}^{-1}]_{33}=R_0^{-2/3}$ we obtain
the scaling functions
\begin{eqnarray}
\label{fexfilmscalingplus}
F^{ex,+}_{\text film}(\tilde x_\perp,{\bf \bar A}_{(I)})&=& \frac{n}{2}\;R_0\;{\cal G}_{0,\text film} \Big({Q^*}^{2\nu} \tilde x_{\perp}^{2\nu}\Big),\\
\label{fexfilmscalingminus}
F^{ex,-}_{\text film}(\tilde x_\perp,{\bf \bar A}_{(I)})&=& (1/2)\;R_0\;\Big\{{\cal G}_{0,\text film} \Big({Q^*}^{2\nu} |2\tilde x_{\perp}|^{2\nu}\Big)\nonumber\\&&+(n-1)\;{\cal G}_{0,\text film}(0)\Big\},
\end{eqnarray}
[compare I (5.29) and I (5.30)] with the critical and the low-temperature Casimir amplitudes, respectively,
\begin{eqnarray}
\label{CASfilmscalingc}
&&X_{\text film,c}= n R_0\;{\cal G}_{0,\text film} (0),\\
%
 %
\label{CASfilmscalinglow}
&&X_{\text film,0}= (n-1) R_0\;{\cal G}_{0,\text film}(0).
\end{eqnarray}
We illustrate these results by comparing in Fig. 4 the Casimir force scaling function $X^{\pm}_{\text film}$ of isotropic systems (solid line) with that of anisotropic systems (dashed line) obtained from (\ref{fexfilmscalingplus}) and (\ref{fexfilmscalingminus}) for $R_0=1.5$ and $n=2$. Also shown are MC data for the isotropic $d=3$ XY model from \cite{hasenbusch2010} for $\rho=0.01$. The anisotropy causes a sizable nonuniversal effect on the shape of the scaling function $X$ that should be detectable in MC simulations.  Correspondingly we predict that the "universal" Casimir force scaling function observed in isotropic superfluid $^4$He films \cite{garcia} and calculated in isotropic $XY$ and $\varphi^4$ models with Dirichlet BC \cite{hucht,vasilyev2009,hasenbusch2010,dohm2014,biswas2010,KrDi92a,GrDi07} is different from the scaling function of weakly anisotropic models with the same BC in the same bulk universality class.

The structure of (\ref{fexfilmscalingplusxxx}) and (\ref{Casimirfilmscalingxx}) becomes exact in the large-$n$ limit for $d\leq 3$ (see \cite{cd2004} and (\ref{fexfilmscalingplusx}), (\ref{Casimirfilmscaling}) below). The results (\ref{fexfilmscalinganiso})-(\ref{CASfilmscalinglow}) are expected to be reliable for $|t| (L/ \xi^\perp_{0+})^{1/\nu} \gtrsim O(1)$ for $d=3$ as supported by the comparison of our isotropic theory with the MC data for $\rho\ll 1$ in Fig. 4. Close to $T_c$, however, they exhibit the following shortcomings: (i) for  $n=1$ and $n=2$ they do not capture an $L$-dependent shift of the  film critical temperature $T_{cf}(L)$; (ii)) for $n=1$ and $n=2$, they do not reproduce the correct analytic form of the weak singularities that are expected \cite{vasilyev2009} for Ising- and $XY$-like film systems at $T_{cf}$; (iii) a singular (cusp-like) behavior of $F^{ex,\pm}_{\text film}$ at $\tilde x_\perp=0$ is present in (\ref{fexfilmscalingplus}) and (\ref{fexfilmscalingminus}) even for $n>2$, $d=3$ where no singularity at all should exist at finite $T$ in film geometry. These shortcomings also exist in isotropic theories of the Casimir force in film geometry with periodic BC \cite{wil-1, KrDi92a,GrDi07,kastening-dohm}. They are avoided in our lowest-mode separation approach where no artificial singularities are present (Fig. 1).
For a comment on the dot-dashed curve in Fig. 4 see Sec. VII.
\renewcommand{\thesection}{\Roman{section}}
\renewcommand{\theequation}{6.\arabic{equation}}
\setcounter{equation}{0}

\section{ Large-${\bf n}$  limit }
The exact treatment of the $\varphi^4$ theory in large-$n$ limit provides an important test for approximate results at finite $n$. Here we extend
previous exact results
\cite{cd1998,cd2004,dohm2008,dohm2009,dohm2011,dohm2017I}
by incorporating the results of the preceding sections with regard to anisotropy effects.
The singular part $\hat f_s$ of the exact free energy density per component $\hat f = \lim_{n \to \infty}\;f/n$ of the $\varphi^4$ lattice model in the limit $n \to \infty$ at fixed $u_0 n$  in a block geometry reads
\begin{eqnarray}
\label{s1}
&&\hat f_s(t,\{L
_\alpha\}, {\bf A})=\frac{(r_0- r_{0c})\hat \chi^{-1}}{8 u_0 n}- \frac{\hat\chi^{-2}}{16 u_0 n} \nonumber\\&&+(1/2)\;\widehat{\cal I}_3(\hat\chi^{-1}, {\bf A})+ (1/2)\;\Delta(\hat\chi^{-1},\{L
_\alpha\}, {\bf A}),\\
\label{4bx1x}
&&\hat\chi(t, \{L_\alpha\}, {\bf A})^{-1}= r_0 -r_{0c} + \; 4 u_0n  \partial\widehat{\cal I}_3(\hat\chi^{-1}, {\bf A})/\partial \hat\chi^{-1}\; \nonumber\\&&+ \; 4 u_0 n \partial\;\Delta(\hat\chi^{-1},\{L_\alpha\}, {\bf A})/\partial \hat\chi^{-1},
\end{eqnarray}
where $\widehat{\cal I}_3$ is defined in I (3.5d).
For $0\leq \hat\chi^{-1} \tilde a^2 \ll 1$, $2<d<4$, and $\det {\bf A}>0$
it is evaluated as
\begin{eqnarray}
\label{intsx}
\widehat{\cal I}_3(\hat\chi^{-1}, {\bf A})=-2(\det {\bf A})^{- 1/2}  A_d\; \hat\chi^{-d / 2}/(d\varepsilon),
\end{eqnarray}
compare I (3.6). For the finite-size contribution $\Delta$ see (\ref{bb9xy}) and (\ref{F.155yy}). For small $\hat\chi^{-1}$, the terms $\hat\chi^{-1}$
in (\ref{4bx1x}) and $\propto \hat\chi^{-2}$ in (\ref{s1}) are negligible. The remaining terms can be expressed in terms of $P(\hat x, \{\rho_\alpha\}, {\bf \bar A})=L\hat \chi^{-1/2}$ and $\hat F(\hat x, \{\rho_\alpha\}, {\bf \bar A})= L^d \hat f$ for $2<d<4$  as
\begin{eqnarray}
\label{3.9xx1}
\hat F(\hat x, \{\rho_\alpha\}, {\bf \bar A})\; &=& \;  \frac{A_d}{2 \varepsilon}
\Big[\hat x P^2 \; - \; \frac{2}{d} \; P^d \Big] \nonumber\\&+& \frac{1}{2}\;{\cal G}_0( P^2, \{\rho_\alpha\}, {\bf \bar A}),\\
\label{3.11xxx1}
P^{d-2} \;& =& \; \hat x  -   \frac{\varepsilon}{A_d}\;{\cal G}_1( P^2,\{\rho_\alpha\}, {\bf \bar A}),\\
\label{3.10primexxy1} {\cal G}_1( P^2, \{\rho_\alpha\}, {\bf \bar A}) \;  &=&-\partial {\cal G}_0(P^2, \{\rho_\alpha\}, {\bf \bar A})/\partial P^2\;\;\;\,
\end{eqnarray}
with the scaling variable (compare (\ref{tildexx}))
\be
\label{hatx}
t (L'/\xi_{0}')^{1/\nu_\infty}=\;t\;(L/\bar \xi_{0})^{1/\nu_\infty}\;\equiv\;\hat x, \;\;
\ee
where $\nu_\infty= (d-2)^{-1}$, $L'=L/(\det {\bf A})^{ 1/(2d)},L\equiv L_d $, and $\bar \xi_{0}
= (\det{\bf A})^{1/(2d)} \xi_{0}'$ with [compare I (6.12)]
\begin{eqnarray}
\label{corrinf}
\xi_{0}'=[4u_0'nA_d/(a_0\varepsilon)]^{1/(d-2)}.
\end{eqnarray}
For ${\cal G}_0$
see (\ref{calG0x})-(\ref{matrixC}).
The scaling function $  \hat F^{ex}=\hat F- \hat F_{b\infty}$ of the excess free energy density is obtained by subtracting the bulk part $\hat F_{b\infty}$ given in I (6.14).
The Casimir force scaling function $\hat X$ follows from (\ref{3nn}) as
\begin{eqnarray}
\label{X-large-nx}&& \hat X(\hat x,\{\rho_\alpha\}, {\bf \bar A}) =   \frac{A_d}{\varepsilon}\Big[\frac{1}{2}\hat x P^2- \frac{d-1}{d}P^d\Big] \nonumber\\&&+\hat F_{b\infty}(\hat x) - \frac{\bar\rho\;^{d-1}}{2}\sum_{\alpha=1}^{d-1}\rho_\alpha
\frac{\partial {\cal J}_0(P^2, \{\rho_\alpha\}, {\bf \bar A})}{\partial \rho_\alpha} \;\;\;\,\;\;\;\,\;\;\;
\end{eqnarray}
where ${\cal J}_0$ is given by (\ref{calJ3x}). For an application to slab geometry the substitutions of Sec. IV. D should be made. For an isotropic system  this yields (A8) and (A9) of \cite{dohm2017I}. The critical Casimir amplitude in a slab geometry is
\begin{eqnarray}
\label{X-large-nx-Tc}&& \hat X(0,\rho, {\bf \bar A}) =   -\frac{A_d(d-1)}{d\varepsilon} P_c^d - \frac{\rho^d}{2}\;
\frac{\partial {\cal J}_0(P_c^2, \rho, {\bf \bar A})}{\partial \rho} \;\;\;\,\;\;\;\,\;\;\;
\end{eqnarray}
with ${\cal J}_0(P^2_c, \rho, {\bf \bar A})$ given by (\ref{calJ3rhox}) where $P_c\equiv P_c(\rho,{\bf \bar A})$ is determined by
\begin{eqnarray}
\label{8i} P^{d-2}_c \; = \;  - \; (\varepsilon/A_d) \; \; {\cal
G}_1 (P^2_c,\rho, {\bf \bar A}) \; .
\end{eqnarray}
For  $-\hat x \gg 1$ we find the low-temperature behavior
\begin{eqnarray}
\label{4hhxx}
&&\hat F^{ex,-}(\hat x, \{\rho_\alpha\}, {\bf \bar A})\approx  (1/2) \bar\rho\;^{d-1} \Big\{-\ln|2\hat x | -1 \nonumber\\&&
+ \ln \Big[\varepsilon \bar\rho\;^{d-1}/(2 \pi^2 A_d)\Big]+ {\cal J}_0(0, \{\rho_\alpha\}, {\bf \bar A} ) \Big\}.\;\;\;\;\;
\end{eqnarray}
Unlike $\hat F^{ex,-}$, $\hat X$ has a finite limit for $\hat x \to - \infty$. In a slab geometry it is
\begin{eqnarray}
\label{X-large-nx-low-slab}
\hat X(-\infty,\rho, {\bf \bar A})& =&  - \frac{1}{2} \rho^{d-1}\big[ 1 +\rho
\partial {\cal J}_0(0, \rho, {\bf \bar A})/\partial \rho\big].\;\;\;\;\;\;\;\;
\end{eqnarray}
Both the critical and the low-temperature Casimir amplitudes are nonuniversal as they depend on the nonuniversal anisotropy matrix  ${\bf \bar A}$,
as demonstrated in Fig. 3 for model $(II)$. For isotropic systems (\ref{4hhxx}) and (\ref{X-large-nx-low-slab}) agree with Eqs. (3.17) and (3.21) of \cite{dohm2011}.

In an anisotropic ${\infty^{d-1} \times L}$ film system, (\ref{s1}) and  (\ref{4bx1x}) are replaced by I (6.3) - I (6.5), where now the quantities $\hat f_{\text film}(t,L, {\bf A}),\hat \chi_{f}(t, L, {\bf A}), \widehat{\cal I}_3(\chi_{f}^{-1}, {\bf A})$, and $\Delta_{\text film}(\hat \chi_{f}^{-1},L, {\bf A})$ depend on  ${\bf A}$.
The function $\Delta_{film}$ is evaluated in (\ref{Deltafilmx1}), thus
\begin{eqnarray}
\label{AbleitDeltafilm}
&&\partial\;\Delta_{\text film}(r,L, {\bf A})/{\partial r}= - \frac{\widetilde L^{2-d}}{(\det {\bf A})^{ 1/2}}   \; {\cal G}_{1,\text film}(r\widetilde L^2),\;\;\;\;\;\;\;\;\;
\end{eqnarray}
compare I (6.6) - I (6.8) and  Eq. (3.11) of \cite{dohm2011}. The treatment is parallel to that for the isotropic case in Sec. VI of \cite{dohm2017I} and in Sec. II of \cite{dohm2011}, except that the length $\widetilde L$ appears instead of $L$. As a consequence, the appropriate scaling variable for the anisotropic film system is
\begin{eqnarray}
\label{scalfilmlargen}
&&\hat x_\perp
= t (\widetilde L/\xi_{0}')^{1/\nu_\infty}
= t (L/ \xi^\perp_{0})^{1/\nu_\infty},\\
\label{scalfilmxi}
&& \xi^\perp_{0}
 \label{scalfilmxiprime}
=[({\bf A}^{-1})_{dd}]^{-1/2}\xi'_{0},\;\;\;\;\;\;\;\;\;\;
\end{eqnarray}
where $\xi^\perp_{0}$ is the amplitude of the bulk correlation length
above $T_c$  in the direction perpendicular to the film boundaries, see  (\ref{scalfilm}) and the subsequent discussion.  The results for $\hat F_{\text film}= L^d\hat f_{\text film}$ and $\hat X_{\text film}= L^d \hat F_{\text Cas,film}$ of the {\it anisotropic} film system read
\begin{eqnarray}
\label{fexfilmscalingplusx}
\hat F_{\text film}(\hat x_\perp,{\bf \bar A })
&=&[({ \bf \bar A^{-1}})_{dd}]^{-d/2}\hat F_{\text film,iso}(\hat x_\perp)\\\;\;\;
\label{Casimirfilmscaling}
\hat X_{\text film}(\hat x_\perp,{\bf \bar A })&=&[({ \bf \bar A^{-1}})_{dd}]^{-d/2}\hat X_{\text film,iso}
(\hat x_\perp)\;\;\;\;\;\;\;\;\;
\end{eqnarray}
where  $\hat F_{\text film,iso}$ and $\hat X_{\text film,iso}$  are
the scaling functions of the {\it isotropic} film given by I (6.9) -  I (6.26) which here, however, have a different scaling argument $\hat x_\perp$ that is affected by anisotropy through the correlation length $\xi^\perp_{0}$.

As shown in \cite{dohm2017I} it is necessary to distinguish the case $d\leq3$ from the case $d>3$ where a finite shift of the reduced film critical temperature $t_{cf}<0$ exists. In the anisotropic system for $d>3$, the film transition occurs at $\hat x_\perp=x^*$ where $x^*= (\varepsilon/A_d)G_{1,{\text film}}(0)$ is the same as given in I (6.18) for the isotropic system, in accordance with multiparameter universality.
Nonuniversality enters, however, the definition of the scaling variable
$\hat x_\perp$, (\ref{scalfilmlargen}), and the fractional shift
\begin{eqnarray}
\label{shiftfilmaniso}
t_{cf}= [T_{cf}(L)-T_c]/T_c &=&x^*(L/\xi^\perp_{0})^{-1/\nu_\infty}<0\;\;\;\;\;\;\;\;\;\;\;\;
\end{eqnarray}
of the film critical temperature which differs from the shift I (6.19) in the isotropic case by the correlation length $\xi^\perp_{0}$.  It is obvious that the same kind of nonuniversal anisotropy effect should occur in film systems of the ($n=1,d>2$) and ($n=2,d\geq3$) universality classes with realistic BC. Thus the comments on Figs. 4, 6, 13, and 15 in \cite{vasilyev2009} on the "universal values" of $x^*$ should be complemented by an information about the anisotropy-dependent fractional shift of $T_c$ in anisotropic films. In particular we predict that this effect occurs in superconducting films which should exhibit an anisotropy-dependent fractional shift of the Kosterlitz-Thouless transition different from that in isotropic superfluid $^4$He films. Also the location and depth of the minimum of the Casimir force scaling function should be be changed by anisotropy. This implies that it is not possible to predict the Casimir force scaling function of anisotropic superconducting films only on the basis of the known scaling function for superfluid $^4$He films, without additional experimental input with regard to the principal axes and correlation lengths. This clearly underscores the impact of nonuniversality on observable properties in weakly isotropic systems within the same bulk universality class.
\renewcommand{\thesection}{\Roman{section}}
\renewcommand{\theequation}{7.\arabic{equation}}
\setcounter{equation}{0}
\section{ Comparison with earlier results}
(i){\it Vortex-fluctuation theory}.
It was pointed out in \cite{wil-1} that a Casmir force should occur in anisotropic superconducting films. A quantitative prediction of the scaling function $X$ was presented for $T \leq T_c$ for the case of periodic BC but no  anisotropy effect on $X$ was found, in disagreement with our prediction in Sec. V. The result of \cite{wil-1} is shown in Fig. 4 as dot-dashed line. No derivation of this result was given but an earlier vortex-fluctuation theory for the anisotropic $XY$ model \cite{shenoy1995} was invoked where it was shown that the critical {\it exponents} are unchanged by anisotropy, as expected. It is not proven in \cite{shenoy1995}, however, that asymptotic {\it amplitudes}, in particular that of the excess free energy density, remain unaffected by anisotropy. A counterexample is the fully anisotropic $d=3$ Ising model in a $L\times L\times \infty$ geometry whose singular part of the free energy density was found to depend on lattice anisotropy \cite{Yurishchev}. Therefore we consider as unjustified the conclusion  \cite{shenoy1995} that "anisotropy is irrelevant". This conclusion also  contradicts the established RG classification \cite{pri} of lattice anisotropy as a {\it marginal} perturbation. Thus we consider as unfounded the claim that the Casimir force scaling function in anisotropic superconducting films is "essentially the same" \cite{wil-1} as in isotropic $^4$He films. More specifically,  it was noted in a Comment that the "noncritical" Goldstone part  was not taken into account in \cite{wil-1}. In our result for $F_{\text film}^{ex,-}$  below $T_c$, (\ref{fexfilmscalingminus}), the constant Goldstone part is represented by the transverse contribution proportional to $n-1$. Thus our remaining longitudinal contribution which carries the temperature dependence (the first term on the right-hand side of (\ref{fexfilmscalingminus})) is to be compared with the prediction of \cite{wil-1}. Although the shape of our longitudinal term resembles that of the result of \cite{wil-1} (dot-dashed line in our Fig. 4) its magnitude is strongly dependent on the anisotropy parameter $R_0$ as shown by the dashed line, in disagreement with the vortex-loop theory. Thus it would be highly desirable to test these predictions by MC simulations for the anisotropic $XY$ model.

(ii){\it Scaling functions in film geometry}. The structure of our scaling functions for finite $n$, (\ref {fexfilmscalingplusxxx}), (\ref{Casimirfilmscalingxx}), and in the large-$n$ limit, (\ref{fexfilmscalingplusx}),(\ref{Casimirfilmscaling}), agrees with the structure of exact results \cite{cd2004,kastening-dohm,DG} and of phenomenological results \cite{kastening2012} for general $d$. In \cite{cd2004,kastening-dohm,DG} this structure was asserted to violate two-scale-factor universality, in contrast to the opposite claim in \cite{kastening2012}. No comparison was made in \cite{kastening2012} with the earlier results \cite{cd2004,kastening-dohm,DG}.
In the following we show that the claim of \cite{kastening2012} is not correct.

Two representations of the free energy scaling form were given in \cite{kastening2012} for film geometry in $d$ dimensions: (a) Eqs. (110) and (134), (b) Eq. (117). Both forms (a) and (b) agree with multiparameter universality but violate two-scale-factor universality. The form (a) depends on the system-dependent quantity ${\bf \bar A}$ which violates the requirement \cite{pri} that the metric factors $C_1$ and $C_2$ are the {\it only} nonuniversal, system-dependent parameters entering (\ref{1b}). The form (b) has the system-dependent prefactor $C_0=({\bf \hat n}^T{\bf \bar A}^{-1}{\bf \hat n})^{-d/2}=[({ \bf \bar A^{-1}})_{dd}]^{-d/2}$, in disagreement with the  requirement  \cite{pri} that no such prefactor should exist. The claim after Eq. (134) of \cite{kastening2012} that no new nonuniversal factor needed to be introduced as compared to the isotropic scaling form (\ref{1b}) disagrees with Eq. (117) of \cite{kastening2012}  where the nonuniversal prefactor $C_0$ appears explicitly in front of ${\cal F}_{iso}(\tilde x)$.

If two-scale-factor universality is claimed to be valid for the {\it confined} anisotropic system it is necessary to verify the consistency of this claim with the scaling forms of the {\it bulk} system
\cite{hohenberg1976,pri,priv}. In \cite{kastening2012} no such bulk scaling forms for the free energy density and for the correlation function of anisotropic systems were given. Here we determine the singular part $\hat f_{b,s}$ of bulk free energy density that follows from our exact result (\ref{scalfilmlargen}), (\ref{fexfilmscalingplusx}), and I (6.9) - I (6.14) in the large-$n$ limit. Taking the bulk limit $L\to \infty$ at fixed $t>0$ we obtain for $2<d<4$
\begin{eqnarray}
\label{fexfilmscalingplusxbulklimit}
\hat f_{b,s}
&=& \lim_{L\to \infty}L^{-d}[({ \bf \bar A^{-1}})_{dd}]^{-d/2}\hat F_{\text film,iso}(\hat x_\perp)\\
\label{filmbulklimit}
&=&Y_\infty \Big[\prod^d_{\alpha = 1} \xi_{0}^{(\alpha)}\Big]^{-1}t^{d\nu_\infty}=A_1t^{d\nu_\infty}\;\;\;
\end{eqnarray}
where $\xi_{0}^{(\alpha)}$ are the amplitudes of the principal correlation lengths and $Y_\infty$ is a universal constant [see I (6.14)]. The form (\ref{filmbulklimit}) violates two-scale-factor universality for reasons given after (\ref{3n4}). This confirms that also Eq. (117) of \cite{kastening2012} in inconsistent with two-scale-factor universality for the bulk system. This is the consequence of the nonuniversal prefactor $C_0$.

In \cite{kastening2012} it is stated that the correlation lengths $\xi_{0+}^{(\alpha)}$ are nonuniversal quantities
but it is suggested to call a quantity  universal if it depends on $\xi_{0+}^{(\alpha)}$ through ${\bf \bar A}$. In this terminology, ${\bf \bar A}$  is a universal quantity if it is expressed in terms of correlation lengths.  We consider this terminology as selfcontradictory since a function of a nonuniversal quantity is a nonuniversal quantity. In particular, microscopic details, such as the presence or absence of a NNN coupling, cause a significant change of this quantity ${\bf \bar A}$ which is incompatible with the meaning of universality. In \cite{kastening2012} it is  argued that quantities depending on ${\bf \bar A}$ depend "only on macroscopic near-critical correlation lengths". The dependence on the principal axes, whose orientation is a   system-dependent, i. e., nonuniversal feature, is ignored in \cite{kastening2012}, and it is overlooked that quantities may be nonuniversal even if they are macroscopic and near-critical. For example, the  amplitude $A_1$ in (\ref{3u}) and (\ref{filmbulklimit}) determining the critical specific-heat amplitude is a function of macroscopic near-critical correlation lengths. It is well established that this amplitude is nonuniversal for both isotropic and anisotropic systems.

(iii) {\it Phenomenological conjectures}. In \cite{kastening2012} the critical behavior of anisotropic $d$-dimensional film systems is discussed. In  Eqs. (95) and (96) of \cite{kastening2012}, where  ${\bf S}$ and ${\bf R}$ correspond to our ${\bf \bar{\mbox
{\boldmath$\lambda$}}}^{-1/2} $ and ${\bf U}$,  the shear transformation of  the $\varphi^4$ theory \cite{cd2004,dohm2006,dohm2008} is adopted in a modified form which corresponds to  ${\bf x'} = {\bf \bar{\mbox
{\boldmath$\lambda$}}}^{-1/2} {\bf U} {\bf x}$, instead of (\ref{xj}), and  to $\delta \widehat K' ({\bf k'})=c'_0  {\bf k'}^2 +O(k'^4)$ with a modified coefficient $c'_0\neq 1$, instead of (\ref{Aprimematrix}) with $c'_0=1$. This modification has no impact on observable properties. The coefficient $c'_0$ is a dummy parameter of the primed theory since any coefficient $c'_0\neq1$ is canceled in all quantities of the anisotropic system, in particular in the  matrix ${\bf \bar A}$ and in the matrix ${\bf \Xi}$ in Eq. (97) of \cite{kastening2012} which corresponds to $(\bar\xi_\pm)^2 {\bf \bar A}$ of the $\varphi^4$ theory. Thus, within the $\varphi^4$ theory, the modified shear transformation of \cite{kastening2012} is essentially the same as the shear transformation with $c'_0=1$ and leads to the same matrix  (\ref{Anonunix}).
While it is established within  $\varphi^4$ theory  \cite{cd2004,dohm2006,dohm2008} that weakly anisotropic systems are related to isotropic systems by a shear transformation, no analytic foundation is given in \cite{kastening2012} for extending this property to weakly anisotropic systems in $d$ dimensions other than $\varphi^4$ models. More specifically, no Hamiltonian, partition function, or correlation functions are presented in \cite{kastening2012} that provide a basis for a definition of the correlation lengths, the principal axes, the matrices in Eqs. (95)-(103) of \cite{kastening2012}, and the free energy density $f_s(T,L)$ in Eqs. (110), (114), and (124) of \cite{kastening2012}, for systems other than $\varphi^4$ models. In particular, while our matrix ${\bf U}$ is precisely defined through the principal directions ${\bf e}^{(\alpha)}$ of the bulk correlation function $G_b$, (\ref{3nbarA}), of the $\varphi^4$ theory, the $d$-dimensional orthogonal matrix ${\bf R}$ of \cite{kastening2012} is an empty quantity that has no analytic definition. It is claimed that the long-distance correlations of $d$-dimensional anisotropic bulk systems are described by a correlation length ellipsoid with a $T$-independent orientation but no justificaton is given in \cite{kastening2012}. Thus we consider the  results of \cite{kastening2012} for general anisotropic systems in $d$ dimensions as phenomenological conjectures. These  conjectures are consistent with multiparameter universality but violate two-scale-factor universality. The finite-size scaling functions derived for the $d=2$ anisotropic Ising model in \cite{kastening2012} are claimed to be exact without justification. Their derivation is based on the bulk correlation lengths $\xi_i$ adopted from Eq. (A22) of \cite{Indekeu} which, for the anisotropic triangular lattice, are only a conjecture.  These correlation lengths have not been compared with the exact results of \cite{Vaidya1976} and no exact analytic determination is given for the principal axes (i.e., for the angle $\theta$)  in terms of the couplings $K_i$ and the lattice structure.

(iv) {\it Critical Binder cumulant}.
For given geometry and BC,  two-scale-factor universality implies that $U^*$, (\ref{1f}), is the same number for all systems in a given ($d,n$) universality class (see \cite{dohm2009PJ} and Sec. 10.1 in \cite{priv}). By contrast, a dependence of $U^*$ on anisotropy was discovered in \cite{cd2004} and confirmed  by MC data of \cite{selke2005} which demonstrate a dependence of $U^*$ on the NNN coupling $J_d$. This is a clear violation of two-scale-factor universality. An analytic  prediction of the nonuniversal deviation of  $U^* ({\bf \bar A})$ from its isotropic value $U^* ({\bf 1})$  \cite{dohm2008,kastening2013}  is in approximate agreement with MC data
for the anisotropic $d=2$ Ising model of Sec. II. E
\cite{kam,selke2005,selke2009}. The agreement in the range $J_d \lesssim -J/2$ was obtained in \cite{kastening2013} by replacing the matrix ${\bf \bar A}_3(s)$ of Eq. (8.19) of \cite{dohm2008} by the matrix ${\bf \bar A}_3(r)$ of Eq. (13) of \cite{kastening2013} containing the anisotropy matrix ${\bf \bar A}_2(r)$, Eq. (12) of \cite{kastening2013}, of the Ising model expressed in terms of correlation lengths.
In view of our results of Secs. II. E and Sec. IV, however, it is  reassuring that no {\it ad hoc} substitution
of a matrix of the $d=2$ Ising model into a  formula of the $d=3$ $\varphi^4$ RG theory is necessary but a coherent derivation of the anisotropy dependence of the Binder cumulant can be given entirely within the $d=3$ $\varphi^4$ RG theory, as proposed in Sec. IV. B. Application of (\ref{extmatrix3}) and (\ref{barextmatrix3}) to the present case yields a $d=3$ $\varphi^4$ model with a NN coupling $ J_z=[(J+2J_d)J]^{1/2}$ in the $z$ direction (rather than  $J +J_d$ chosen in \cite{dohm2008}), with
\begin{eqnarray}
 \label{neu33c}
 {\bf A}_3 & =& \; 2 \left(\begin{array}{ccc}
  J+J_d & \;\;J_d & 0 \\
  J_d &\;\; J+J_d & 0 \\
  0 & 0 & \;\;[(J+2J_d)J]^{1/2} \\
\end{array}\right) \; ,\;\;\;\;\;\;\\
%
%
  \label{33a}
 {\bf \bar A}_3(s) &=&\;\left(\begin{array}{ccc}
  {\bf \bar A}_2(s) & 0  \\
  0 & 1  \\
\end{array}\right)\\
\label{Q33a}&=&\;\left(\begin{array}{ccc}
  {\bf  Q}_2(q) & 0  \\
  0 & 1  \\
\end{array}\right)\equiv {\bf  Q}_3(q),
\end{eqnarray}
where the two-dimensional matrices  ${\bf \bar A}_2(s)$ and ${\bf  Q}_2(q)$ are given by (\ref{33ax}) and (\ref{33qx}) with $s$ and $q$ defined by (\ref{s}) and (\ref{ratioq}), respectively. We expect (\ref{neu33c})-(\ref{Q33a}) to provide a better description of the two-dimensional anisotropy in the $x$-$y$ plane than provided by ${\bf \bar A}_3(s)$ in Eq. (8.19) of \cite{dohm2008}. The application to the $d=2$ Ising model is then obtained, on the basis of the hypothesis of  multiparameter universality, by  substituting into ${\bf  Q}_2(q)$, the correlation-length ratio $q \to q^{\text Ising}=\xi_+^{(1){\text Ising}}/\xi_+^{(2){\text Ising}}$ of the $d=2$ Ising model.
The resulting ${\bf  Q}_3(q^{\text Ising})$ has the same form as ${\bf \bar A}_3(r)$ in Eq. (13) of \cite{kastening2013,footnoteratio}.
The correctness of this strategy was proven analytically for the example discussed in Sec. II. B,  (\ref{corrlambdaratiox})-(\ref{xiratioisingx}). The agreement
with the MC data \cite{kam,selke2005,selke2009,kastening2013}
supports the hypothesis of multiparameter universality for weakly anisotropic systems predicting that the dependence of $U^*$ on nonuniversal correlation-length ratios has a universal functional form.
A verification of this hypothesis can be provided by MC simulations for $U^*$  of the anisotropic $d=2$ $\varphi^4$ model where the MC data should be analyzed in terms of the correlation-length ratio $q=\xi_{0+}^{(1)}/\xi_{0+}^{(2)}=[(J+2J_d)/J]^{1/2}$ of the $\varphi^4$ model. The result should agree with the  MC data for $U^*$ of the Ising model shown in Fig. 2 of \cite{kastening2013}.

The $d=3$ $\varphi^4$ theory \cite{dohm2008} is, of course, not capable of predicting the isotropic value of $U^*$ of the $d=2$ Ising model. This implies that the latter value serves as an adjustable parameter in the comparison with the MC data
which leads to an adjusted function for $U^*$ in Fig. 2 of \cite{kastening2013}.
\section*{ACKNOWLEDGMENT}
I am grateful to M. Hasenbusch for providing the data of Ref. \cite{hasenbusch2010} in numerical form.
\renewcommand{\thesection}{\Roman{section}}
\setcounter{equation}{0} \setcounter{section}{1}
\renewcommand{\theequation}{\Alph{section}.\arabic{equation}}

\section*{ Appendix A: Gaussian model}
We consider the Gaussian model, i.e., I (2.13), for $u_0=0, h=0,r_0=a_0t,r_{0c}=0$ with an anisotropic interaction  $\delta \widehat K ({\mathbf k})$.
Although this model is defined only for $r_0\geq0$ and has no low-temperature phase, it plays an important role in our finite-size $\varphi^4$ theory. We employ the results of Sec. II.
For the transformed system
with the isotropic interaction (\ref{Aprimematrix}), a unique bulk correlation length $\xi'_G(t)= r_0^{-1/2} = \xi_{0G}' t^{-\nu_G}$ exists with $\xi_{0G}' = a_0^{-\nu_G} $,
$\nu_G=1/2$. The Gaussian principal correlation lengths are denoted by
$\xi_G^{(\alpha)} (t) = \lambda_\alpha^{1/2}
\xi'_G(t)= \xi^{(\alpha)}_{0G} t^{- \nu_G}$,
with $\xi_{0G}^{(\alpha)} = \lambda_\alpha^{1/2} \xi_{0G}'$,
and their geometric mean is
$\bar \xi_{0G}=\Big[\prod^d_{\alpha = 1} \xi_{0G}^{(\alpha)}\Big]^{1/d}
= (\det{\bf A})^{1/(2d)} \xi_{0G}'.$
First we consider the rectangular block geometry defined in Sec. I.
The Gaussian excess free energy density
is
\begin{eqnarray}
\label{b7}
f^{G,ex}(r_0,\{L_\alpha\},{\bf A})=  \frac{n}{2}\;\Delta(r_0,\{L_\alpha\},{\bf A})
\end{eqnarray}
where $\Delta(r_0, \{L_\alpha\},{\bf A})$ is defined in (\ref{bb9xy}).
Due to the ${\bf k}={\bf 0}$ term of $\sum_{\bf k}$,  $ \Delta(r_0, \{L_\alpha\},{\bf A})$ diverges for $r_0 \to 0$ at finite $V$. This divergence is an unphysical artifact of the Gaussian model which comes from the absence of the $\varphi^4$ term. The divergent ${\bf k}={\bf 0}$ term of  $ \Delta$, however, does not contribute to the Gaussian Casimir force $F_{\text Cas}^G = -\partial (L f^{G,ex})/\partial L$  since $L V^{-1} \ln (r_0 \tilde a^2)=  (\prod^{d-1}_{\alpha = 1} L_\alpha)^{-1} \ln (r_0 \tilde a^2)$ is independent of $L$, thus $F_{\text Cas}^G$ has a finite limit for $r_0\to 0$ at finite $V$ [see (\ref{cascalJcrit})].
The calculation of  $\Delta(r_0,\{L_\alpha\},{\bf A})$ is parallel to that in \cite{dohm2008}. The result is given in (\ref{F.155yy})-(\ref{matrixC}) for large $L_\alpha/\tilde a$, small $0 < r_0^{1/2}\tilde a \ll 1$, fixed $ 0 < L_\alpha r_0^{1/2}\lesssim O(1)$ and finite $0<\rho_\alpha < \infty$. This leads to the finite-size scaling form
\begin{eqnarray}
\label{b23}
 f^{G,{\text ex}}(r_0, \{L_\alpha\},{\bf A})&=&L^{-d} F^{G, {\text ex}}(\tilde x_G, \{\rho_\alpha\}, {\bf \bar A}),\;\;\;\;\\
\label{fexgauss}F^{G, {\text ex}}(\tilde x_G, \{\rho_\alpha\}, {\bf \bar A}) &=& (n/2) {\cal G}_0 (\tilde x_G, \{\rho_\alpha\}, {\bf \bar A}),
\end{eqnarray}
with the Gaussian scaling variable
\begin{eqnarray}
\label{scalGaussx}
r_0 L'^2 = t (L'/\xi_{0G}')^{1/\nu_G}=t (L/\bar\xi_{0G})^{1/\nu_G}\equiv \tilde x_G \;\;
\end{eqnarray}
where $L' = L / (\det {\bf A})^{1/(2d)}$.
The function ${\cal G}_0$ can be decomposed as given in (\ref{calG0decom}). This can be derived from (\ref{calG0x}) by adding and subtracting $(\prod^{d-1}_{\alpha = 1}\rho_\alpha )\ln [x/(4\pi^2)]$ and using the integral representation
\begin{eqnarray}
\label{b10x} \ln w = \; \int_0^\infty dy y^{-1} \left[ \exp
{\left(-y\right)} - \exp {\left(- w y\right)}\right]
\end{eqnarray}
with $w=x/(4\pi^2)$.  It is the function ${\cal J}_0$, (\ref{calJ3x}), rather than ${\cal G}_0$ that determines the Gaussian Casimir force
\begin{eqnarray}
\label{3l}
&&F^G_{\text Cas}(t,\{L_\alpha\},{\bf A})=L^{-d}X^G(\tilde x_G, \{\rho_\alpha\}, {\bf \bar A}),\\
\label{cascalJ}
&&X^G(\tilde x_G, \{\rho_\alpha\}, {\bf \bar A}) =-(n/2)\;\bar\rho\;^{d-1}\nonumber \\&& \times \Bigg\{2+
\Bigg[\frac{\tilde x_G}{\nu_G}
\frac{\partial}{\partial \tilde x_G}+ \sum_{\alpha=1}^{d-1}\rho_\alpha
\frac{\partial}{\partial \rho_\alpha}\Bigg]{\cal J}_0(\tilde x_G, \{\rho_\alpha\}, {\bf \bar A})\Bigg\},\;\;\;\;\;\;\;\;\;\;
\end{eqnarray}
as derived from (\ref{calG0decom}), (\ref{3nn}), and (\ref{fexgauss}). For $\tilde x_G \to 0$, ${\cal G}_0$ diverges logarithmically [see (\ref{calG0decom})].
This divergence is canceled in $X^G$ which yields a finite  Casimir amplitude
\begin{eqnarray}
\label{cascalJcrit}
&&X_c^G\equiv X^G(0, \{\rho_\alpha\}, {\bf \bar A})\nonumber \\&&=-\frac{n}{2}\;\bar\rho\;^{d-1}\Bigg\{2+ \sum_{\alpha=1}^{d-1}\rho_\alpha
\frac{\partial}{\partial \rho_\alpha}{\cal J}_0(0, \{\rho_\alpha\}, {\bf \bar A})\Bigg\}.\;\;\;\;\;\;\;\;\;\;
\end{eqnarray}
For a slab geometry it is given by
\begin{eqnarray}
\label{cascalJcritslab}
X^G(0, \rho, {\bf \bar A})=-(n/2)\;\rho^{d-1} \big[2+ \rho\;
\partial{\cal J}_0(0, \rho, {\bf \bar A})/\partial \rho\big]\;\;\;\;\;\;\;\;
\end{eqnarray}
with ${\cal J}_0$ given by (\ref{calJ3rhox}).
The amplitudes (\ref{cascalJcrit}) and (\ref{cascalJcritslab}) divided by $n$ differ from the exact low-temperature amplitudes in the large-$n$ limit [for slab geometry see
(\ref{X-large-nx-low-slab})], unlike the amplitude for film geometry (\ref{CasimirGaussAmp}) given below.
For cubic geometry ($\rho_\alpha=1$), our function
(\ref{calJ3x}) is related to  $J_0(x,{\bf \bar A})$ in Eq. (C2) of \cite{dohm2008} by
${\cal J}_0(x,\{1\}, {\bf \bar A})=J_0(x,{\bf \bar A})$.
For isotropic systems  in  slab geometry, our function (\ref{calJ3rhox})
is  related to  $J_0$  in Eq. (4.26) of \cite{dohm2011} by
${\cal J}_0(x,\rho,{\bf 1})= \rho^{1-d} J_0(x,\rho) +(\rho^{1-d} - 1)\ln [x/(4\pi^2)]$.

Now we consider the $\infty^{d-1} \times L$ film geometry. The Gaussian excess free energy density is for $d>1$ and $r_0\geq0$
\begin{eqnarray}
\label{b7film}  f_{\text film}^{G,ex}(r_0,L,{\bf A})= (n/2)\;\Delta_{\text film}(r_0,L,{\bf A}),
\end{eqnarray}
with $ \Delta_{\text film}(r_0,L,{\bf A})$ defined in I (6.5).
Using
(\ref{b10x}),
interchanging the integration $\int dy$ with $\sum_p
\int_{\bf q}$ and $\int _{\bf k}$  and using
$L^{-1} \sum_p \int_{\bf q} 1 = \int_{\bf k} 1 = \tilde a^{-d}$,
we rewrite $ \Delta_{\text film}$ as
\begin{eqnarray}
\label{b11} \Delta_{\text film}(r_0,L,{\bf A})& =&
\int_0^\infty dy y^{-1} e^{-r_0 \tilde a^2
y}\Big[\int_{\bf k} \exp \{-  \delta \widehat K (\mathbf k)
\tilde a^2 y \} \nonumber\\& -&  L^{-1} \sum_p
\int_{\bf q} \exp \{-
\delta \widehat K (\mathbf q , p) \tilde a^2 y \} \Big].
\end{eqnarray}
Assuming  large
$L/\tilde a$, small $0 \leq r_0^{1/2}\tilde a \ll 1$ and fixed $ 0 \leq L r_0^{1/2}\lesssim O(1)$, we may replace
$\delta \widehat K $ by its long - wavelength form and let the integration and summation limits of
$\int_{\bf k}$  and $L^{-1} \sum_p \int_{\bf q}$ go to $\infty$. Then we obtain
\begin{eqnarray}
\label{b11filmx} &&\Delta_{\text film} =
\int _0^\infty dy y^{-1} e^{-r_0 \tilde a^2
y}\Big[\int_{\bf k}^\infty \exp (-  {\bf k} \cdot {\bf
A k}
\;\tilde a^2 y )\nonumber\\&&-  L^{-1} \sum_p^\infty \exp (-
A_{dd} p^2 \tilde a^2 y )
\int_{\bf q}^\infty  \exp (-
\Psi (\mathbf q, p)\; \tilde a^2 y ) \Big], \;\;\;
\qquad
\end{eqnarray}
with $\Psi (\mathbf q, p)= {\bf q} \cdot {\bf B q} + 2p\; {\bf q} \cdot {\bf b}$
where ${\bf B }$ represents the $(d-1) \times (d-1)$ symmetric matrix which results from removing the $d$th row and $d$th column of
$ {\bf A}  = \;  \left(\begin{array}{ccc}
  {\bf B} & {\bf b}  \\
   {\bf \bar b} & A_{dd}  \\
\end{array}\right).$
Here ${\bf \bar b}$ and ${\bf b}$ denote the first $d-1$ elements $A_{d\alpha}=A_{\alpha d},\alpha=1,2,...,d-1$ of the $d$th row and $d$th column of ${\bf A }$. We assume $\det {\bf B}>0$, $A_{dd}>0$.
Integration over ${\bf q}$ yields
\begin{eqnarray}
\label{b15film}\int_{\bf q}^\infty \exp \{-
\Psi (\mathbf q, p)\; \tilde a^2 y \}&&
= (\det {\bf B})^{- 1/2} (4\pi \tilde a^2  y)^{(d-1)/2}\nonumber\\ && \times
\exp ( {\bf b} \cdot {\bf B}^{-1} {\bf b} \;p^2 \tilde a^2 y ) .\;\;\;\;\;\;\;\;\;
\end{eqnarray}
The matrix ${\bf A }$
can be decomposed as
\begin{equation}
 \label{34}
  \left(\begin{array}{ccc}
  {\bf B} & {\bf b}  \\
  \bar {\bf b} & A_{dd}  \\
\end{array}\right)= \;\left(\begin{array}{ccc}
  {\bf B} & {\bf 0}  \\
  \bar{ \bf b} & 1  \\
\end{array}\right)\left(\begin{array}{ccc}
  {\bf 1} & {\bf B^{-1} b}  \\
  {\bf 0} & \;\;\;\; A_{dd}- {\bf b}\cdot {\bf B^{-1} b} \\
\end{array}\right)\;\;\;\;\;\;\;\;\;\;\;\;\;\;\;\;\;\;
\end{equation}
which implies
$\det {\bf A}  = ( A_{dd}- {\bf b}\cdot {\bf B^{-1} b}) \det {\bf B}$,
thus the exponential arguments $\propto p^2$ can be rewritten as
\begin{eqnarray}
\label{p}
p^2\;[ -A_{dd}+ {\bf b}\cdot {\bf B^{-1} b}]\tilde a^2 y = - p^2\; ( \det {\bf A}/ \det {\bf B}) \;\tilde a^2 y.\;\;\;\;\;\;\;
\end{eqnarray}
The ratio $\det {\bf B}/ \det {\bf A}$ is expressed in terms of the matrix elements $({ \bf  A^{-1}})_{dd}$ and $({ \bf \bar A^{-1}})_{dd}$ as
\begin{eqnarray}
\label{AdurchB}
\det {\bf B}/\det {\bf A}=  ({\bf A^{-1}})_{dd}=  (\det {\bf A})^{-1/d}({ \bf \bar A^{-1}})_{dd}.\;\;\;\;\;
\end{eqnarray}
The Gaussian bulk integral in (\ref{b11filmx}) yields
\begin{eqnarray}
\label{anigaussx}
\int_{\bf k}^\infty \exp \{-  {\bf k} \cdot {\bf
A k} \tilde a^2 y \}=(\det {\bf A})^{- 1/2}(4\pi \tilde a^2  y)^{-d/2}.\;\;\;\;\;\;\;
\end{eqnarray}
Substituting into (\ref{b11filmx}) the integration variable
$z= 4 \pi^2 ( \det {\bf A}/ \det {\bf B})\tilde a^2 y / L^2$
and using (\ref{p})-(\ref{anigaussx}) we finally obtain for $d>1$ and $r_0\geq 0$
\begin{eqnarray}
\label{Deltafilm}  f_{\text film}^{G,ex}(r_0,L,{\bf A})= (n/2)  (\det {\bf A})^{-1/2}\widetilde L^{-d}\; {\cal G}_{0,\text film}
\big(r_0 \widetilde L^2 \big)\;\;\;\;\;\;
\end{eqnarray}
where $\widetilde L$ and ${\cal G}_{0,\text film}(x)$  are given by (\ref{Ltilde}), (\ref{Lschlangeprime}), and I (5.28), respectively.
The scaling forms are
 $F^G_{\text{Cas},{\text film}}$ $=L^{-d}X^G_{\text film}(\tilde x^\perp_{G},{\bf \bar A})$ and $ f_{\text film}^{G,ex}= L^{-d}F^{G,ex}_{\text film}(\tilde x^\perp_{G},{\bf \bar A})$
with
\begin{eqnarray}
&&F_{\text film}^{G,{\text ex}}(\tilde x^\perp_{G},{\bf \bar A})=(n/2)[({ \bf \bar A^{-1}})_{dd}]^{-d/2}  \; {\cal G}_{0,\text film}\big(\tilde x^\perp_{G}\big),\;\;\;\;\;\;\;\;\;\;\\
\label{CasimirfilmscalingGauss}
&&X^G_{\text film}(\tilde x^\perp_{G},{\bf \bar A })=(d-1)F^{G,ex}_{\text film}(\tilde x^\perp_{G},{\bf \bar A })\nonumber\\&&-\;
(\tilde x^\perp_{G}/\nu_G)\;
\partial F_{\text film}^{G,ex}(\tilde x^\perp_{G}, {\bf \bar A })/\partial\tilde x^\perp_{G},\;\;\;\;\;\;\\
\label{scalGauss}
&&\tilde x^\perp_{ G}=r_0 \widetilde L^2
= t (\widetilde L/\xi_{0G}')^{1/\nu_G}
= t (L/ \xi^\perp_{0G})^{1/\nu_G},\;\;\;\;\;\\
\label{scalfilmxixneu}
 &&\xi^\perp_{0G}
=[({\bf A}^{-1})_{dd}]^{-1/2}\xi'_{0G},\;\;\;\;\;
\end{eqnarray}
where $\xi^\perp_{0G}$ is the correlation length perpendicular to the boundaries [see (\ref{scalfilmxixneux})]. At $T_c$ the Casimir amplitude is
\begin{eqnarray}
\label{CasimirGaussAmp} X^G_{\text film}(0,{\bf \bar A})=  (d-1)\;[({ \bf \bar A^{-1}})_{dd}]^{-d/2}\;(n/2)\; {\cal G}_{0,\text film}(0).\;\;\;\;\;\;
\end{eqnarray}
The scaling functions of the {\it anisotropic} film system can be written as
\begin{eqnarray}
\label{fexfilmscalingplusxx}
F^{G,ex}_{\text film}(\tilde x^\perp_{G},{\bf \bar A })
&=&[({ \bf \bar A^{-1}})_{dd}]^{-d/2}F^{G,ex}_{\text film,iso}(\tilde x^\perp_{G})\;\;\;\;\;\;\;\;\;\\\;\;\;
\label{Casimirfilmscalingx}
X^{G}_{\text film}(\tilde x^\perp_{G},{\bf \bar A })&=&[({ \bf \bar A^{-1}})_{dd}]^{-d/2}X^G_{\text film,iso}
(\tilde x^\perp_{G})\;\;\;\;\;\;\;\;\;
\end{eqnarray}
where  $F^{G,ex}_{\text film,iso}$ and $X^G_{\text film,iso}$  are
the scaling functions of the {\it isotropic} film given by I (A.4), I (A.5) which here, however, have a different scaling argument $\tilde x^\perp_{G}$ that is affected by anisotropy. Thus anisotropy changes both the overall amplitude and the scaling argument, in structural agreement with (\ref{fexfilmscalingplusx}) and (\ref{Casimirfilmscaling}) for $n \to \infty$.

\end{document}